\documentclass[12pt,authoryear]{elsarticle}

\usepackage{amssymb}
\usepackage{amsmath}

\usepackage[dvipsnames]{xcolor}

\newif\ifmycolor
\mycolortrue   

\ifmycolor
   \definecolor{vertdc1}{RGB}{20,89,33}
   \definecolor{blood}{RGB}{193,41,41}
   \definecolor{viol}{RGB}{109,10,186}
   \definecolor{dgreen}{RGB}{9,95,29}
   \definecolor{DarkGreen}{RGB}{9,95,29}
   \definecolor{dorange}{RGB}{197,69,6}
   \definecolor{CNRSBlue}{RGB}{26,48,81}
   \definecolor{CNRSLightBlue}{RGB}{10,141,167}
   \definecolor{DarkRed}{RGB}{212,0,0}
   \definecolor{DarkRed2}{RGB}{150,0,0}
   \definecolor{Violet}{RGB}{122,47,214}
\else
   \definecolor{vertdc1}{RGB}{0,0,0}
   \definecolor{blood}{RGB}{0,0,0}
   \definecolor{viol}{RGB}{0,0,0}
   \definecolor{dgreen}{RGB}{0,0,0}
   \definecolor{DarkGreen}{RGB}{0,0,0}
   \definecolor{dorange}{RGB}{0,0,0}
   \definecolor{CNRSBlue}{RGB}{0,0,0}
   \definecolor{CNRSLightBlue}{RGB}{0,0,0}
   \definecolor{DarkRed}{RGB}{0,0,0}
   \definecolor{DarkRed2}{RGB}{0,0,0}
   \definecolor{Violet}{RGB}{0,0,0}

   \definecolor{Melon}{gray}{0}
   \definecolor{Emerald}{gray}{0}
   \definecolor{Brown}{gray}{0}
   \definecolor{purple}{gray}{0}
   \definecolor{Cerulean}{gray}{0}
   \definecolor{Orange}{gray}{0}
   \definecolor{BurntOrange}{gray}{0}
   \definecolor{CadetBlue}{gray}{0}
   \definecolor{teal}{gray}{0}
   \definecolor{violet}{gray}{0}
   \definecolor{magenta}{gray}{0}
   \definecolor{Gray}{gray}{0}
   \definecolor{red}{gray}{0}
   \definecolor{orange}{gray}{0}
\fi

\usepackage{hyperref}
\hypersetup{
colorlinks=true,
breaklinks=true,
urlcolor= RedViolet,
citecolor=blue,
linkcolor=blue,
breaklinks=true
}
\usepackage{breakurl}

\usepackage{url}

\newcommand{\vb}[1]{{\color{blue}\texttt{\detokenize{#1}}}}

\usepackage{etoolbox}
\AtBeginEnvironment{figure}{\centering}
\usepackage{adjustbox}

\journal{Icarus}

\begin{document}

\begin{frontmatter}



\title{Flow dynamics of cryolava or impact melt on Titan’s surface} 


\author[LPG]{Daniel Cordier} 
\ead{daniel.cordier@cnrs.fr}
\author[LEATP]{Bastien Bodin}
\author[LPG]{St\'{e}phane Le Mou\'{e}lic}
\author[JPL]{Ashley G. Davies}

\affiliation[LPG]{organization={LPG Nantes Universite, Univ Angers, Le Mans Universite, CNRS, UMR 6112},
             addressline={}, 
             city={Nantes},
             postcode={44000}, 
             state={},
             country={France}}
  
\affiliation[LEATP]{organization={Universite de Reims Champagne Ardenne, LEATP},
             addressline={}, 
             city={Reims},
             postcode={51100}, 
             state={},
             country={France}}

\affiliation[JPL]{organization={NASA Jet Propulsion Laboratory, California Institute of Technology},
             addressline={}, 
             city={Pasadena},
             postcode={91109}, 
             state={CA},
             country={USA}}

\begin{abstract}
   {Titan, a unique body in the solar system, resembles a life-size Miller–Urey experiment where
   pockets of liquid water might interact with abundant organic matter. Its young surface (less than a billion years) shows signs of material 
   exchange with the interior, and an isotopic signature is detected in the atmosphere. 
   Cryovolcanism and meteor impacts represent potential high energy mechanism for these exchanges.
   Although cryolava or impact melt properties are largely unknown, Titan's internal structure makes a significant water content in these 
   fluids highly probable. The temperature of these fluid mixtures at the onset of the flows is above 0$^{\rm o}$C.}
  %
   {The primary objective of this work is to investigate the dynamical and thermal properties of potential Titanian flows,
   while ensuring their spatial extents remain broadly consistent with the few available observations. We also aim to assess whether their 
   evolution could enable the hydrolysis of encountered organic material.}
  %
   {On Earth, in the context of lava flow risk assessment, numerical techniques based on cellular automata have been developed for many years. 
    This approach allows simulating the flow of Bingham fluids over distance scales of several tens of kilometers.}
  %
   {In the context of Titan's surface, and within the explored parameter space, flows with a spatial extent of several tens of kilometers
    can be produced with our model. The resulting flow thickness is about a meter. The flow extent is primarily controlled by the total erupted 
    volume and the terrain topology. As expected, the rheology and thermo-physical properties of the erupted fluid have significant influence 
    on the flow's extent and the cooling rate of the cryolava. We predict hydrolysis reactions between an aqueous cryolava and the organic 
    matter likely ubiquitous on Titan’s surface.  Our model code is publically available.}
  %
\end{abstract}





\end{frontmatter}



\section{\label{intro}Introduction}
%
\begin{figure*}[htbp]
\centering
\adjustbox{center}{%
\includegraphics[angle=0, width=15 cm]{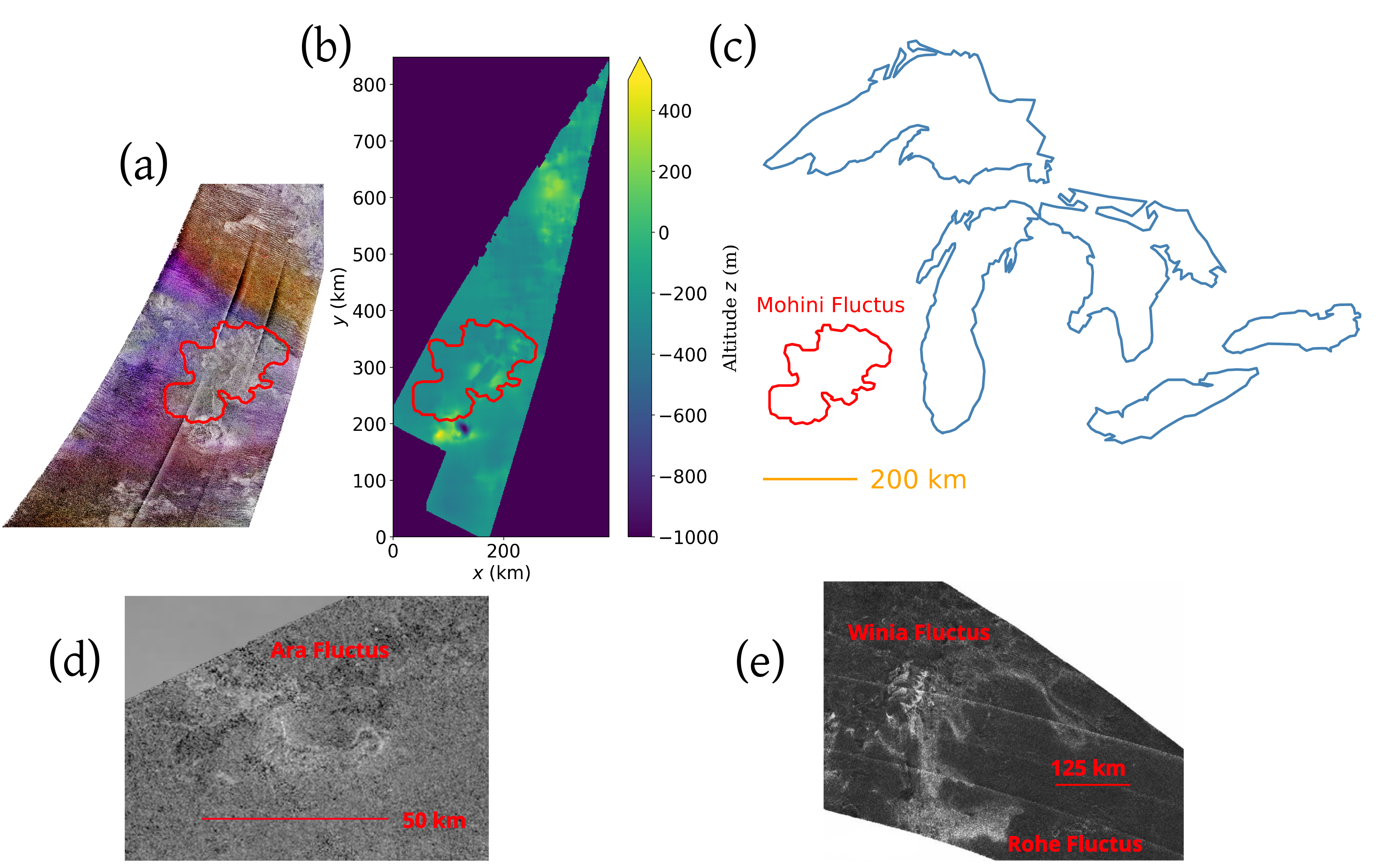}
}

\caption[]{\label{topoMohiniFluctus}Panel (a) Composite image showing the Sotra Patera/Mohini Fluctus area, Mohini Fluctus is outlined in red . 
           A RADAR Synthetic Aperture Radar (SAR) swath acquired 
           during the T25 flyby is colorized using the Visible and Infrared Mapping Spectrometer (VIMS)
           band ratio color composite mosaic described in \cite{lemouelic_etal_2019}, 
           where red is controlled by $1.59/1.27$~$\mu$m ratio, green by the $2.03/1.27$~$\mu$m ratio and blue by the $1.27/1.08$~$\mu$m.
           In this color scheme, brownish tones correspond to dune fields and bluish areas may be enriched in water ice compared to the 
           surroundings. Mohini Fluctus and its associated deposits appear as uniformly bright areas in VIMS, with no particluar color tint, 
           so with the same absence of spectral signature as the bright mainlands of Titan. 
           Panels (b) represents the digital elevation model of the same region, published by \cite{malaska_etal_2022}, the range of elevation 
           has been reduced to $[-1000 \, \mathrm{m}, +800 \, \mathrm{m}]$ in order to improve the visibility of low amplitue variations.
           The radar image in panel (a) has a spatial resolution of approximately $640$~m/px, whereas that in panel
           (b) is characterized by a coarser resolution of about $\sim 1400$~m/px. In the Digital Elevation Model (DEM) shown in panel (b), 
           the typical elevation difference between adjacent pixels is approximately $7$~m.
           Panels (c) shows a size comparison between Mohini Fluctus and Great American Lakes.
           This figure can be rebuilt using data publicly available\footnotemark.
           The maps displayed in the first three panels are plotted at the same scale in order to facilitate direct comparisons.
           Panels (d) and (e) present three additional flow-like features that are less well defined than Mohini Fluctus, namely 
           Ara Fluctus and Rohe Fluctus. Panel (e) also displays Winia Fluctus, which was not classified as a cryovolcanic feature by
           \cite{lopes_etal_2013}.}
\end{figure*}
\footnotetext{Fig.~\ref{topoMohiniFluctus} material: \url{https://gitlab.univ-nantes.fr/icy-moons/src/fig_titan_mohinifluctus}}
%
    Titan has long been recognised as having a strong exobiological potential \citep[see for instance][]{sagan_khare_1982}.
Titan's bulk composition contains a large amount of water, and  its surface is rich in organics.
This concomitance enables interactions between water and organic matter, opening the way to prebiotic chemistry potentially similar to
that of primordial Earth \citep{coustenis_1995}.
Researchers have also put forward ideas about forms of life as we don't know it.
Indeed, an exotic prebiological activity, harbored by liquid hydrocarbons, have been proposed \citep{mckay_smith_2005,strobel_2010,Lunine_2010,stevenson_etal_2015a,stevenson_etal_2015b}.
 In the present work, we focus on the first exobiological perspective, which involves the encounter of water with organic material.\\

    While the sedimentation of organic matter initially produced in Titan’s atmosphere toward the surface is commonly considered \citep{neish_etal_2018},
the reverse process may also occur. In this scenario, organic matter is sequestered in the interior during the moon's formation
\citep{reynard_sotin_2023} and subsequently subjected to physico-chemical
transformations over geological timescales. Such material may be transported to the surface via high energy processes like cryovolcanism
\citep[\textit{e.g.}][]{tobie_etal_2006,lopes_etal_2013} or meteorite impacts  \citep[\textit{e.g.}][]{kalousova_etal_2024}.
More subtle low energy mechanisms, such as bubble bursting at the surfaces of Titan's polar hydrocarbon seas \citep{cordier_etal_2024}, could
be also at work. Titan's surface appears to be a major site for the occurrence of a wide variety of organic species from diverse origins.\\

   Titan's low surface temperature, determined to be around $94$~K \citep{fulchignoni_etal_2005,jennings_etal_2009}, prevents
the existence of long standing bodies of liquid water at the moon surface, even in the presence of an anti-freezing agent like ammonia 
\citep{grasset_pargamin_2005,lunine_stevenson_1987}. Nonetheless, two scenarios have been advanced to account for the potential occurrence of 
liquid water at the surface: (1) a local melting of the crust by a meteorite impact \citep[\textit{e.g.}][]{kalousova_etal_2024}, and
(2) a cryovolcanic eruption \citep[\textit{e.g.}][]{kargel_1992,lopes_etal_2013,wood_radebaugh_2020}.\\

%
 In the present study, we consider the flow of a fluid referred to generically as a ``cryofluid'', which may correspond either to cryolava
stricto sensu or to impact-generated melt. 
 Within the framework we consider here, we do not regard a possible subsurface ocean on Titan as a direct source of cryolava. Indeed, when such 
an ocean was first proposed, it was thought to lie beneath an ice crust approximately $80$–$100$~km thick. This estimate was based on models of Titan's 
internal structure \citep[\textit{e.g.}][]{sohl_etal_2014}, as well as on Cassini measurements of Schumann resonances \citep{beghin_etal_2012, beghin_2015}. 
The ascent of liquid from such an ocean to the surface would clearly pose significant physical challenges. 
More recently, the very existence of this subsurface ocean has been called into question, based both on a reanalysis of gravitational data 
\citep{petricca_etal_2025} and on a reinterpretation of the electromagnetic activity recorded during the Cassini descent \citep{lorenz_legall_2020}.
Regarding cryolava, we favor a scenario similar to that 
proposed for Europa \citep{fagents_2003,lesage_etal_2020,lesage_etal_2022}, where a relatively shallow volume of ice ($1-10$~km below the surface) 
is melted by tectonic processes, as suggested by some simulations \citep{vilella_etal_2020}. 
Concerning the cryofluid generated by an impact, we envision a scenario in which part of the impact-melt liquid flows through a breach in the crater 
rim. In the case of cryolava, a geophysical process generates overpressure within a subsurface liquid reservoir, which drives the liquid to the surface.
In the case of impact melt, by contrast, the liquid flows through a breach under the sole effect of gravity.
Given the current observational and theoretical constraints, there is no indisputable basis for 
favouring one case over the other. 
 We emphasize that we do not model the behavior of the melt pool, as has been done by some authors \citep[\textit{e.g.}][]{hedgepeth_etal_2022}, 
but rather that of the fluid that may potentially escape from this type of reservoir.
Regardless of whether the fluid originates from cryovolcanic activity or from an impact event, 
both processes may produce flows propagating downslope under its own weight. Such cryofluid flows
may also have been widespread during Titan's earliest epochs \citep{raulin_2005}. However, identifying present-day remnants of these primordial 
activities is unlikely, given the relatively young age of Titan’s surface \citep[$0.2$–$1$~Gyr,][]{lorenz_etal_2007,neish_lorenz_2012}.\\
%

 Organic material is central to research on Titan. Consequently, the quantities present in its various potential reservoirs have
been estimated.
Recent studies addressing organic material derived from accretion and subsequently delivered to the hydrosphere have estimated the total
organic matter concentration within the putative subsurface ocean \citep{miller_etal_2026}. The global contribution of insoluble 
organic matter (IOM) and soluble organic matter (SOM) is estimated to range between $10^{-2}$~mol~L$^{-1}$ and $10^{-1}$~mol~L$^{-1}$, 
which exceeds the average concentration found in Earth’s oceans. 
  As previously mentioned, transposing such values to those relevant for
cryolava implies transport through an ice crust several tens of kilometers thick, if the ocean exists, 
a process that could be achieved by ice convection, except within the conductive layer near the surface. To traverse this latter layer, additional 
processes would need to be invoked, among which cryovolcanism or impacts are the most likely.
       Another aspect of the question regarding the organic matter content of cryolava (or impact melt) concerns the interaction 
between a flow and organic material already present at the surface. Indeed, organic matter produced by atmospheric
photochemistry settles on the surface and, over time, may potentially form layers several tens or even hundreds of meters thick,
as indicated by atmospheric chemistry models \citep[\textit{e.g.}][]{lavvas_etal_2008a,lavvas_etal_2008b,lavvas_etal_2010,lavvas_etal_2013,krasnopolsky_2009,krasnopolsky_2010,krasnopolsky_2014}.
  A strong line of evidence in favor of organic material having accumulated at the surface of Titan, in the form of particles, is provided 
by the observation of dunes in equatorial regions \citep[\textit{e.g.}][]{lorenz_etal_2006,barnes_etal_2015}. 
Based on radar observations from Cassini, dune height has been estimated to be in the range of $100$ to $150$~m.
   If we assume that Titan's cryofluid flows have final thicknesses broadly comparable to those of terrestrial lava flows, which typically 
range from about one to several tens of meters, and more rarely reach a few tens of meters \citep{rossi_1997,self_etal_2021,blasizzo_etal_2022}, 
we can then consider ratios
(at least in terms of volume)
of organic matter to aqueous fluid close to unity, if not significantly greater than one. The final outcome strongly depends on the 
physico-chemical mixing processes at play. This interaction may be mechanical in nature, involving entrainment of material 
({\it i.e}., erosion) or flow through a porous medium (such as dunes), but also chemical, through dissolution and/or hydrolysis of organic 
matter in granular form.
    It should be noted that the rheological properties of the initial fluid, whether cryolava or ``impact melt'', may be modified by 
the incorporation of sedimentary particles or by the dissolution of organic matter \citep{stickel_powell_2005,maffettone_greco_2015}.

In the prebiotic perspective of interaction between liquid water and organic matter, cryofluid flows, whatever their origin, have several
advantages over crater lakes: (1) the extent of the contact surface with organic matter (assuming it is evenly distributed on Titan's surface) 
is far greater in the case of flows. This large contact surface induces 3 other advantages: (2) a potentially very high ratio of quantity of 
organic matter to quantity of water per unit of Titan's surface (even greater than 1), (3) the nature of the organic matter encountered by 
the liquid can be more varied since the organic species accumulated at the surface may differ from one place to another, 
(4) greater ease of {\it in situ} analysis/exploration. The possible large ratio organic/water has to be compared to the situation found with impact 
crater formation, where only $\sim 0.1$--$1$\% of organic material initial present in the crater footprint could be mixed with water.
   For example, a $2$-km in size spherical impactor, made of non-porous ice, traveling at $7$~km~s$^{-1}$ can produce a
melt volume within the crater which of around $3\times 10^4$~Mt  \citep[see Fig.~4 in][]{artemieva_lunine_2003}.
The portion of material falling into the crater, and interacting with the liquid water
\citep[see Fig.~7 in][]{artemieva_lunine_2003} could be of the order of $200 - 300$~Mt leading in the most favorable case to a fraction organic
material/liquid water of around 1\%. In addition, $80$~s after an impact, organic material mixed with  the melt volume may be estimated 
to be around $30$~Mt \citep[see Sect.~5.3 in][]{artemieva_lunine_2003} leading to a ``concentration'' of organic material one order of magnitude smaller.\\

  The organic matter initially present on the surface of Titan can be hydrolyzed upon contact with ``warm''
({\it i.e.} with a temperature above 0$^{\circ}$C) water, such as that which is
hypothesized to exist in cryofluid. This hydrolysis is expected to lead to the formation of new compounds, following a chemical reaction kinetics
that depends on the cooling curve of the flow \citep{neish_etal_2009,neish_etal_2010}.  
Numerical simulations have shown that impact melt pools  remain liquid
for $10^2$ to $10^4$~years \citep{obrien_etal_2005,artemieva_lunine_2005,hedgepeth_etal_2022,wakita_etal_2023,kalousova_etal_2024}. 
Because of their possible shallowness, cryofluid flows may cool faster than melt trapped within the impact crater, assuming similar
composition and initial temperature.
    Laboratory experiments on tholin hydrolysis have been conducted around room temperature. 
The time allowed for chemical reactions before product analysis, has varied from one day \citep{khare_etal_1986,raulin_etal_2007} to a few months
\citep{poch_etal_2012,brasse_etal_2017} and up to a couple of years 
\citep{neish_etal_2010,cleaves_etal_2014}.
   Higher temperatures, according to the Arrhenius law, result in higher reaction rates, allowing hydrolysis reactions to proceed over 
shorter timescales. For example, \cite{farnsworth_etal_2024} showed that, for the hydrolysis of 2-aminopropanenitrile in a 15\% NH$_3$ solution, 
the reaction rate decreases by approximately two orders of magnitude as the temperature decreases from 21$^\circ$C to $-22^\circ$C. This therefore 
implies substantially longer timescales at lower temperatures to achieve a comparable extent of reaction.
Although a potentially high concentration of organic matter transported by a cryofluid flow is not expected to accelerate the reaction rates for a
specific reaction, contact with a large amount of sediment, and potentially with a wide variety of organic compounds, may result in a greater 
diversity of chemical species produced through interaction with liquid water.\\

   The first observational evidence for possible cryovolcanism on Titan emerged after 2005 with the arrival of the Cassini--Huygens spacecraft at Saturn.
Among the earliest lines of evidence was the detection of argon-40 in Titan's atmosphere by the Gas Chromatograph Mass Spectrometer (GCMS)
aboard the Huygens lander \citep{niemann_etal_2005}.
This isotope is produced by the radioactive decay of potassium-40, whose reservoir
is expected to reside within Titan's interior. Its detection therefore suggests the existence of a pathway connecting the interior to the atmosphere, 
a role that could potentially be fulfilled by cryovolcanic activity. A second line of evidence is based on geomorphological interpretations 
derived from Synthetic Aperture Radar (SAR) images of Titan's surface. These observations revealed features that may correspond to cryovolcanoes, 
several of which exhibit lobate structures suggestive of flow emplacement \citep{lopes_etal_2007}. This initial survey identified only three 
caldera-like structures from which a fluid appeared to have flowed. The analysis was later revisited using a much larger dataset 
\citep{lopes_etal_2013}, leading to the identification of approximately ten candidate cryovolcanic features, among which five were considered 
the most compelling candidates. Among these candidates, as recognized by the Cassini Radar Science Team \citep{CassiniFinalReport_RADAR_2019}, 
the best example of a flow on Titan's surface is probably ``Mohini Fluctus'' located in the Sotra Patera region \citep[40.0$^{\rm o}$W, 
14.5$^{\rm o}$S,][]{lopes_etal_2013}.
As shown in Fig.~\ref{topoMohiniFluctus}, this lobate geomorphological features is approximately $180$~km long. 
Using a method based on 
polygons\footnote{The Python class \texttt{Polygon} enables area computations for surfaces defined by polygonal shapes. This class is provided by the module \texttt{shapely.geometry} from the \texttt{shapely} package.}, 
we estimate the area covered by Mohini Fluctus to be around 33,000~km$^2$. Using an estimated maximum thickness of $100$~m
\citep{lopes_etal_2013}, this deposit would have a total volume of up to $3,300$~km$^3$ ($3.3\times 10^{12}$~m$^3$).
   To illustrate the range of sizes and morphologies exhibited by the observed flow features, we also present radar images of Ara Fluctus and
Rohe Fluctus in Fig.~\ref{topoMohiniFluctus}. Ara Fluctus is notably smaller than Mohini Fluctus, with a maximum extent of approximately $\sim 50$~km. 
Rohe Fluctus may be interpreted as more extensive than Mohini Fluctus, although in this case the possibility of multiple source regions cannot be excluded.
In comparison
with Mohini Fluctus, both Ara and Rohe Fluctus are significantly less well defined morphologically.\\

  A flow fed by a breach in the wall surrounding a molten pool produced by an impact represents another plausible
scenario. 
   Impact melt flows have been identified on Mars \citep[\textit{e.g.}][]{morris_etal_2010}, Venus
\citep[\textit{e.g.}][]{asimov_wood_1992}, and on the Moon \citep[\textit{e.g.}][]{neish_etal_2014}. 
   The situation is different for icy worlds that lack a substantial atmosphere, which is the case for all of them except Titan. Indeed, on moons 
such as Europa, Callisto, or Ganymede, the absence of an atmosphere means that a fluid composed predominantly of liquid water would rapidly 
evaporate and freeze. On such bodies, ice creep is considered a plausible process; however, this type of phenomenon is beyond the scope of 
the present study.
On Titan's surface, impact craters are far more common than candidate cryovolcanoes; the most recent census identifies a total of $90$~objects ranging from 
certain to possible \citep{hedgepeth_etal_2020}. These craters vary in diameter from approximately $3$~km to several hundred kilometers. Menrva, the largest, 
measures $400$~km across, while Selk and Sinlap are somewhat smaller, with a diameter of $\sim 80$~km \citep{soderblom_etal_2010,lemouelic_etal_2008}. 
Selk is of particular interest because its region has been 
selected for exploration by the {\it Dragonfly} rotorcraft \citep{lorenz_etal_2021}. The areas of potential flows on the slopes of Selk crater 
\citep[see, for instance, Fig.~6 in][]{lorenz_etal_2021} are comparable in size to those of Mohini Fluctus.
  The scenario we consider here involves a possible outflow of liquid from an impact-melt pool formed during an impact. More exotic scenarios, 
such as the deposition of a solid–liquid ejecta mixture \citep[see the possible composition of the ejecta in][]{artemieva_lunine_2003} over a 
relatively limited region, followed by subsequent flow, are also conceivable. We have chosen to exclude such scenarios from our study, as 
we consider them to be less likely.
   As we show above, both putative cryolava flows and impact-melt flows exhibit maximum extents ranging from a few tens to several hundreds of kilometres.
Our objective here is not to reproduce any specific flow feature. Indeed, the observational constraints required to achieve such a goal remain insufficient.
Instead, our approach is to model a flow representative of what might be considered a typical or average cryolava or impact-melt flow.\\

This paper is motivated by two main scientific questions: 
(1) to roughly reproduce typical cryolava or impact melt flows and to investigate the primary trends in their dynamical behavior
across the parameter space,
(2) to study the related thermal behavior of cryolava, or impact melt, and its compatibility with the hydrolysis time scales
of organics.
  We emphasize that the dynamical and thermal behaviors are coupled, and thus cannot, a priori, be studied independently of each other.
 Our methodology is based on the construction of a ``reference model''. This reference case then serves as the basis for a systematic
investigation of the influence of the various individual physical parameters involved. The aim is to identify the main trends governing the behaviour of a typical Titan flow.\\

 In Sec.~\ref{model}, we describe our reference model which will subsequently serve as a basis for comparison;
in Sec.~\ref{param}, we examine the influence of the physical parameters on the flow dynamics.; and in Sec.~\ref{subs},
we explore the some possible interactions between the flow and the underlying substrate. Finally, Sec.~\ref{concl} is dedicated to conclusions and perspectives.

%
\begin{figure*}[htbp]
\centering
\adjustbox{center}{%
\includegraphics[angle=0, width=14cm]{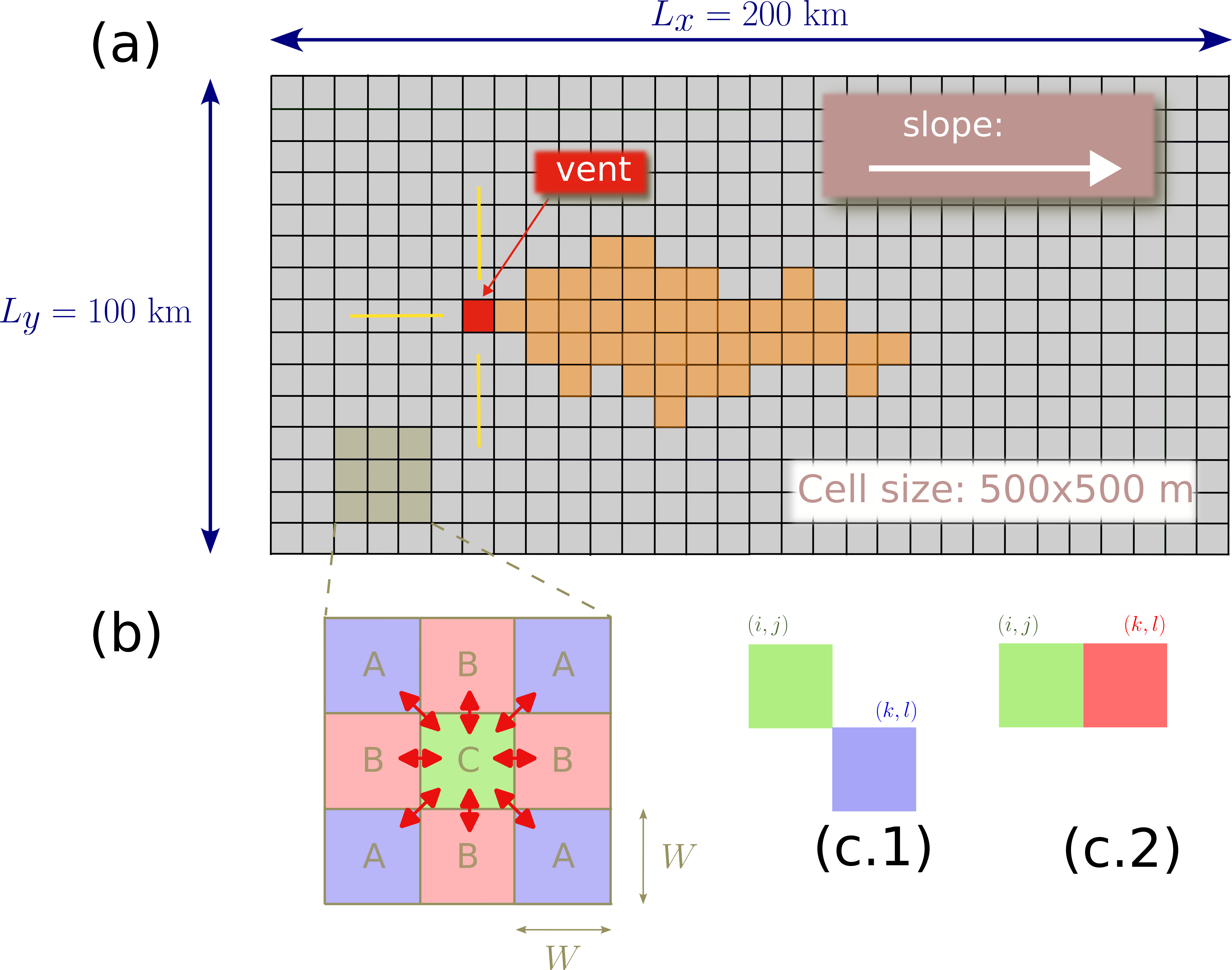}
}

\caption[]{\label{CAprinciple}(a) The cellular automata method is based on a spatial partitioning of the terrain into a grid of 
           square-shaped cells. Each individual cell has dimensions $W \times W$, while the full simulation domain spans $L_x \times L_y$.  
           In this sketch, the cryovolcanic vent is shown in red and outlined by yellow segments.  
           During the simulation, the fluid progressively occupies cells according to the physical rules implemented in the cellular 
           automata framework. An example of the system state at a given time is shown, with cells containing cryolava displayed in orange.
           (b, c.1, c.2) According to the principles of cellular automata, if we consider a $3 \times 3$ block of cells, the central cell 
           interacts with its neighbors through exchanges of mass and energy. Two types of neighbors are distinguished: cells that share an 
           edge with the central cell (type B) and cells that touch it only at a corner (type A). This distinction lies at the origin of the 
            anisotropy issues inherent to the cellular automata approach.}
\end{figure*}
%

\section{Construction of the reference model}
\label{model}

\begin{figure}[htbp]
\begin{center}
\includegraphics[angle=0, width=8cm]{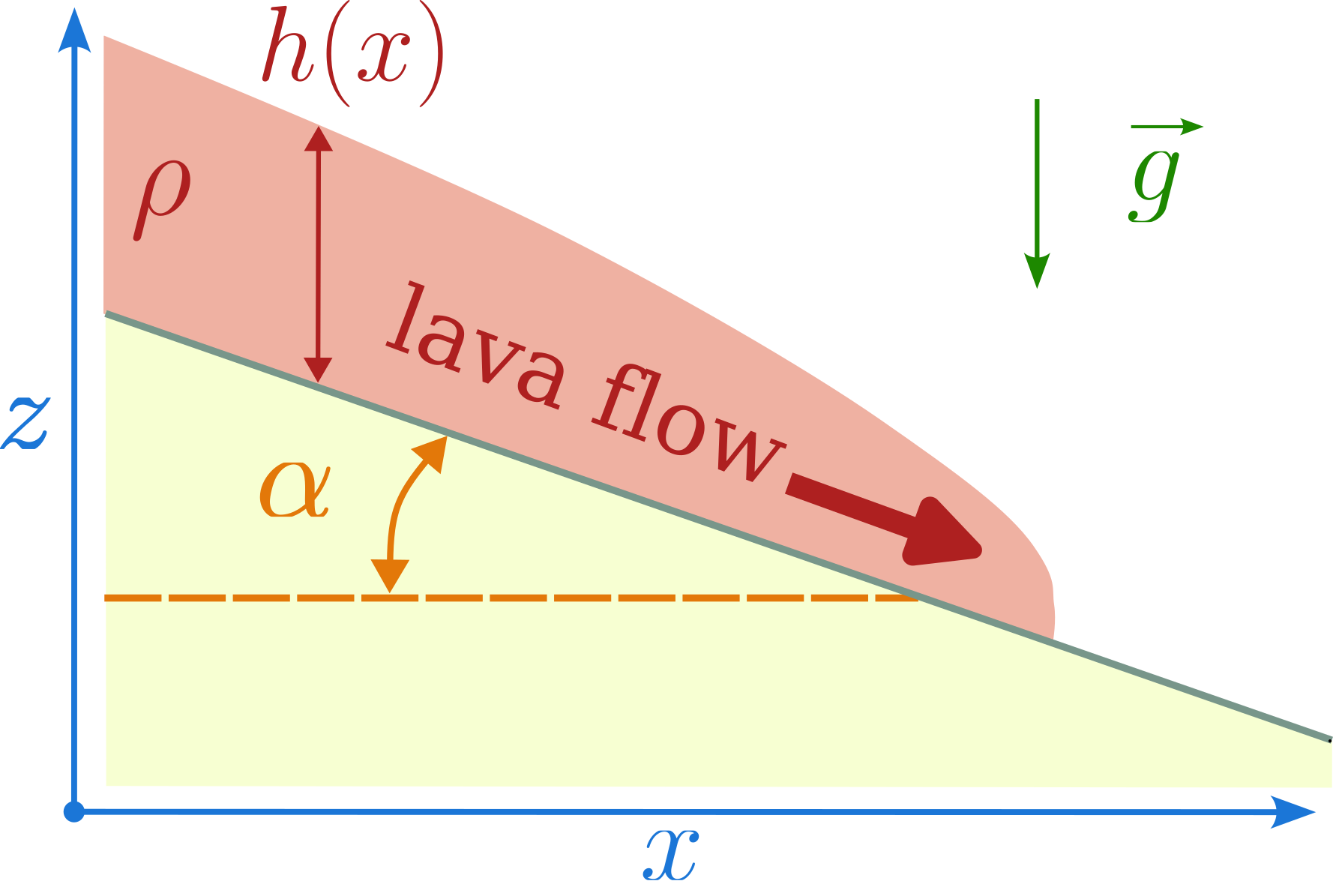}

\caption[]{\label{lavaslope}Conceptual sketch of a lava flow moving downslope. The local lava thickness is denoted by $h(x)$, and the angle of the 
           terrain with respect to the horizontal is $\alpha$. This figure is inspired by Fig.~1 of \cite{miyamoto_sasaki_1997}.}
\end{center}
\end{figure}

    Although the very first works that opened the research field of cellular automata date back to the early 1950s, 
the field gained serious momentum at the beginning of the 1980s with a seminal article by Stephen Wolfram \citep{wolfram_1983}, 
who later also published an overview of the cellular automata domain \citep{wolfram_2002}. This numerical approach has a wide 
variety of applications in fields as diverse as physics \citep{creutz_1986}, chemistry \citep{menshutina_etal_2020}, economics 
\citep{calero_2022}, and biology, ranging from tumour growth \citep{alarcon_etal_2003} to pattern formation \citep{deutsch_dormann_2005}, 
as well as social sciences \citep{hegselmann_1996}, traffic flow modelling \citep{feng_etal_2023}, and even wildfire propagation 
\citep{trucchia_etal_2020}. The geosciences are no exception, with applications in soil erosion \citep{tang_etal_2022}, dune field dynamics
\citep{werner_1995}, drainage network modelling \citep{coppola_etal_2007}, and, closer to our domain of interest, lava flow simulations \citep{ishihara_etal_1990,miyamoto_sasaki_1997,miyamoto_sasaki_1998,ciraudo_etal_2007,vicari_etal_2007,rongo_etal_2016}. 
These terrestrial models are primarily aimed at natural hazard prediction. Only a few planetary cases have been studied using a 
cellular automata approach: some Venusian lava flows \citep{miyamoto_sasaki_2000}, putative ice flows on Europa \citep{miyamoto_etal_2005}, 
Martian slope streaks \citep{miyamoto_etal_2004a}, and debris flows on Martian sand dunes \citep{miyamoto_etal_2004b}.\\
   Here, we propose an application of a cellular automata technique
to putative cryofluid flows on the surface of Titan. The method employed is primarily based on three papers: \cite{ishihara_etal_1990},
\cite{miyamoto_sasaki_1997}, and \cite{vicari_etal_2007}.
In this framework, space is represented by a simulation box with length $L_x$ and width $L_y$, 
which is divided into a grid of square cells with dimensions $W$ (see Fig.~\ref{CAprinciple}-a). Each cell $(i, j)$ has its own elevation $z_{i,j}$, 
which may follow the slope of an inclined plane, as exemplified in Fig.~\ref{CAprinciple}-a, or, in more sophisticated cases, be provided by
a digital elevation model derived from field observations. Cells containing material are also associated with lava properties, such as the 
local flow thickness $h_{i,j}$, fluid density $\rho_{i,j}$, temperature $T_{i,j}$, and so on. In the following, we will often omit 
the indices $(i,j)$ to simplify the notation.\\
    In the general case, the behavior of a fluid is governed by the Navier-Stokes equation \citep{navier_1823,stokes_1845}:
\begin{equation}\label{NSequa}
    \frac{\mathrm{D} u}{\mathrm{D} t} = K -\frac{1}{\rho} \mathrm{Grad} p + \frac{\eta}{\rho} \Delta u
\end{equation}
where $K$ represents the external forces (here, the weight), $p$ is the pressure (Pa), $\rho$ the density (kg~m$^{-3}$), 
$\eta$ the dynamic viscosity (Pa~s), and $u$ the local fluid velocity (m~s$^{-1}$).  
The inertial term can be neglected if the fluid is not subjected to strong accelerations, {\it i.e.} $\mathrm{D} u/\mathrm{D} t \simeq 0$.  
In this case, Eq.~\ref{NSequa} can be reformulated as
\begin{equation}\label{NSreformulated}
   \eta \frac{\partial^2 u}{\partial z^2} = -\frac{\partial h}{\partial x} \rho g \cos \alpha + \rho g \sin \alpha
\end{equation}
The local lava thickness $h$ and the terrain slope angle $\alpha$ are defined in Fig.~\ref{lavaslope}.  
The gravity $g$ is fixed at the Titanian value $1.352$~m~s$^{-2}$ throughout this work.  

Assuming a Bingham fluid with plastic viscosity $\eta$ (Pa~s) and yield stress $\tau_0$ (Pa),  
from Eq.~\ref{NSreformulated} we can define a critical thickness $h_c$ associated with a pair of adjacent cells $(i,j)$ and $(k,l)$ 
(see Fig.~\ref{CAprinciple} c.1 and c.2) as \citep[see, for instance,][]{miyamoto_sasaki_1997}:
\begin{equation}\label{hc}
  h_{c, (i,j) - (k,l)} = \frac{\tau_0}{\rho g \left(\sin \alpha - \frac{\partial h}{\partial x} \cos \alpha\right)}
\end{equation}
The liquid flows from cell $(i,j)$ to cell $(k,l)$ when $h_{\rm eff, (i,j) - (k,l)} > h_{c, (i,j) - (k,l)}$, where $h_{\rm eff}$ is the
effective lava thickness, which, for a flat and horizontal terrain, is given by the difference $|h_{(i,j)}-h_{(k,l)}|$.  
A comprehensive determination of $h_{\rm eff, (i,j) - (k,l)}$ for all conceivable cases can be found in the literature \citep{ishihara_etal_1990,vicari_etal_2007}.  

The volume of fluid $\Delta V_{(i,j)-(k,l)}$ (m$^3$) transported between two cells during the time step $\Delta t$ can be written as \citep[see, for instance,][]{miyamoto_sasaki_1997}:
\begin{equation}\label{DeltaV}
\Delta V_{(i,j)-(k,l)} = \pm \frac{\tau_0 h_c^2 W}{3 \eta} \, \left[\left(\frac{h_{eff}}{h_c}\right)^3 -\frac{3}{2} \left(\frac{h_{eff}}{h_c}\right)^2
          + \frac{1}{2}\right] \, \Delta t
\end{equation}
   The sign of the flux is determined by the direction of lava flow: it is positive if the flow goes from cell $(k,l)$ to cell $(i,j)$, and 
negative otherwise. The flow direction itself depends on the respective elevations of the lava surface in the two adjacent cells: the fluid
flows from the cell where this elevation is highest toward the one where it is lowest, subject to the previously mentioned condition on $h_{\rm eff}$.
This flux $\Delta V_{(i,j)-(k,l)}$ is consistent with the principle of mass conservation since
\begin{equation}\label{MatterConserv}
     \Delta V_{(i,j)-(k,l)} = - \Delta V_{(k,l)-(i,j)}
\end{equation}

Equation~\ref{DeltaV} warrants several comments:  
(1) For small values of the yield stress $\tau_0$, the fluid flows from one cell to its neighbor for any non-zero value of the effective 
    lava thickness $h_{\rm eff}$.  
(2) Conversely, the yield stress $\tau_0$ can, at a certain point, halt the liquid flow. The higher its value, the more rapidly the flow stops, and 
    the thicker the remaining lava becomes, all else being equal.  
(3) In the case of a flat and horizontal terrain, when $\partial h/\partial x \sim 0$, the critical thickness $h_{\rm c}$ becomes infinite, 
    and consequently the flow ceases.  
(4) In certain situations, the volume $\Delta V$ removed from a given cell may exceed the lava volume initially present in that cell, 
    leading to a negative volume at time $t+\Delta t$. To prevent this, the time step $\Delta t$ in Eq.~\ref{DeltaV} must be properly adjusted.  
    This last point relates to the challenge of time-step management, for which it is difficult to provide a universal prescription.
    One possible approach is to impose a maximum value on the relative variations in cryofluid thickness from one time to the next, for example by
    choosing not to exceed a $1/16$ increase or decrease, as done by \cite{bilotta_etal_2012}.\\
%
%

We would like to point out that, although at first glance one might think that Newtonian fluids ({\it i.e.}, corresponding to $\tau_0 \sim 0$) 
could be treated using the method described here, very small values of the critical thickness $h_{\rm c}$ may result in very large 
liquid flows (see Eq.~\ref{DeltaV}). This can produce a non-negligible inertial term in the Navier-Stokes equation, thereby violating the assumptions underlying 
the development of this cellular automata approach. However, the method remains valid if the viscosity $\eta$ is sufficiently large 
to keep $\Delta V_{(i,j)-(k,l)}$ within a range of relatively small values.\\
   During a time step $\Delta t$, the heat content of a given cell is modified by material exchanges with neighboring cells, 
cryofluid thermal radiation, and cooling due to wind. We denote $Q_t$ and $Q_{t+\Delta t}$ as the internal energy of the considered cell
at times $t$ and $t+\Delta t$, respectively. The principle of energy conservation requires that
\begin{multline}\label{QtDeltat}
Q_{t+\Delta t} = Q_t + \sum_k \, \rho_k \, \Delta V_k \, C_{p,k} \, T_{t, k} \\
                - (\epsilon \, \sigma_B \, T_t^4 + \Delta Q_{{\rm wind}, t}) \, W^2 \, \Delta t
\end{multline}
where the index $k$ runs over the 8 neighboring cells. Here, $C_{p, k}$ is the cryofluid specific heat (J~K$^{-1}$~kg$^{-1}$),
$\Delta V_k$ is the volume exchanged (see Eq.~\ref{DeltaV}), $T_{t, k}$ is the temperature of the $k$-th neighboring cell at time $t$, 
$\epsilon$ is the thermal emissivity of cryofluid, $\sigma_B$ is the Boltzmann constant, $T_t$ is the temperature of the considered cell at time $t$,
and $\Delta Q_{{\rm wind}, t}$ (W~m$^{-2}$) represents the cryofluid cooling due to wind. This latter term is provided in Eq.~3 of
\cite{davies_etal_2010}, which is reproduced here as Eq.~\ref{Qwind_Davies2010}
\begin{equation}\label{Qwind_Davies2010}
   \Delta Q_{\rm wind} = U_w \, f \, (T_{\rm cryofluid}-T_{\rm air}) \, \rho_{\rm air} \, C_{p, {\rm air}}
\end{equation}
where $U_w$ is the wind speed (m~s$^{-1}$), $f$ is the friction factor, $f= 0.0036$ \citep[see][for more details]{davies_etal_2010},
$T_{\rm cryofluid}$ and $T_{\rm air}$ are the temperature of the cryofluid, and the air temperature, respectively.
The adopted properties for Titan’s atmosphere are: density $\rho_{\rm air} = 5.2$~kg~m$^{-3}$,
temperature $T_{\rm air} = 94$~K \citep{fulchignoni_etal_2005}, and specific heat capacity $C_{\rm p, air} = 1004.0$~J~mol$^{-1}$~K$^{-1}$.
For the reference model, we assumed the absence of wind so
as to avoid introducing the additional uncertainty associated with the poorly constrained wind speed. The sensitivity of the results to this
parameter is examined later in the study.\\

At each time step, the cell temperature is updated according to
\begin{equation}\label{Tupdate}
   T_{t+\Delta t} = \frac{Q_{t+\Delta t}}{\rho \, C_p \, V_{t+\Delta t}}
\end{equation}
  we emphasize that, within the framework of the cellular automaton approach, the temperature assigned to each cell corresponds to a vertical average.
The heat generated by viscous dissipation is neglected in Eq.~\ref{QtDeltat}, as our estimates show that it is approximately ten orders of 
magnitude smaller than the initial thermal energy of the fluid. In addition, the surface pressure on Titan is around $1.5$~bar and the ambient 
temperature is close to $90-94$~K \citep{fulchignoni_etal_2005}, cryofluid vaporization can be safely neglected.
    As is well known, techniques based on cellular automata suffer from anisotropy issues. In the case of lava flow simulations,
several methods have been proposed to mitigate this problem, all of which are based on Monte Carlo-type algorithms.  
The first method \citep{miyamoto_sasaki_1997} is relatively simple and computationally efficient, but by its nature it violates the conservation of matter and energy, 
{\it i.e.}, Eq.~\ref{MatterConserv} is no longer strictly satisfied. Nonetheless, this drawback can be controlled by choosing a sufficiently small time step, 
so that the conservation laws are satisfied on average to within $\lesssim 1$\%.  
The second method \citep{vicari_etal_2007}, which rigorously enforces the conservation laws, is considerably more time-consuming, {\it e.g.} at least
by a factor of ten.  
We have implemented both approaches in our model. The cellular automata method allows for the accurate reproduction of actual basaltic lava flows 
under appropriate eruption scenarios, for instance on Earth \citep{vicari_etal_2007} or Venus \citep{miyamoto_sasaki_2000}.  
More details concerning the anisotropy correction are provided in Appendix~\ref{appAnisoCorr}.
We have verified that, under equivalent assumptions, our implementation produces results consistent with previously published studies 
\citep[\textit{e.g.}][]{miyamoto_sasaki_1997}.\\

%
\begin{figure}[htbp]
\begin{center}
\includegraphics[angle=0, width=9cm]{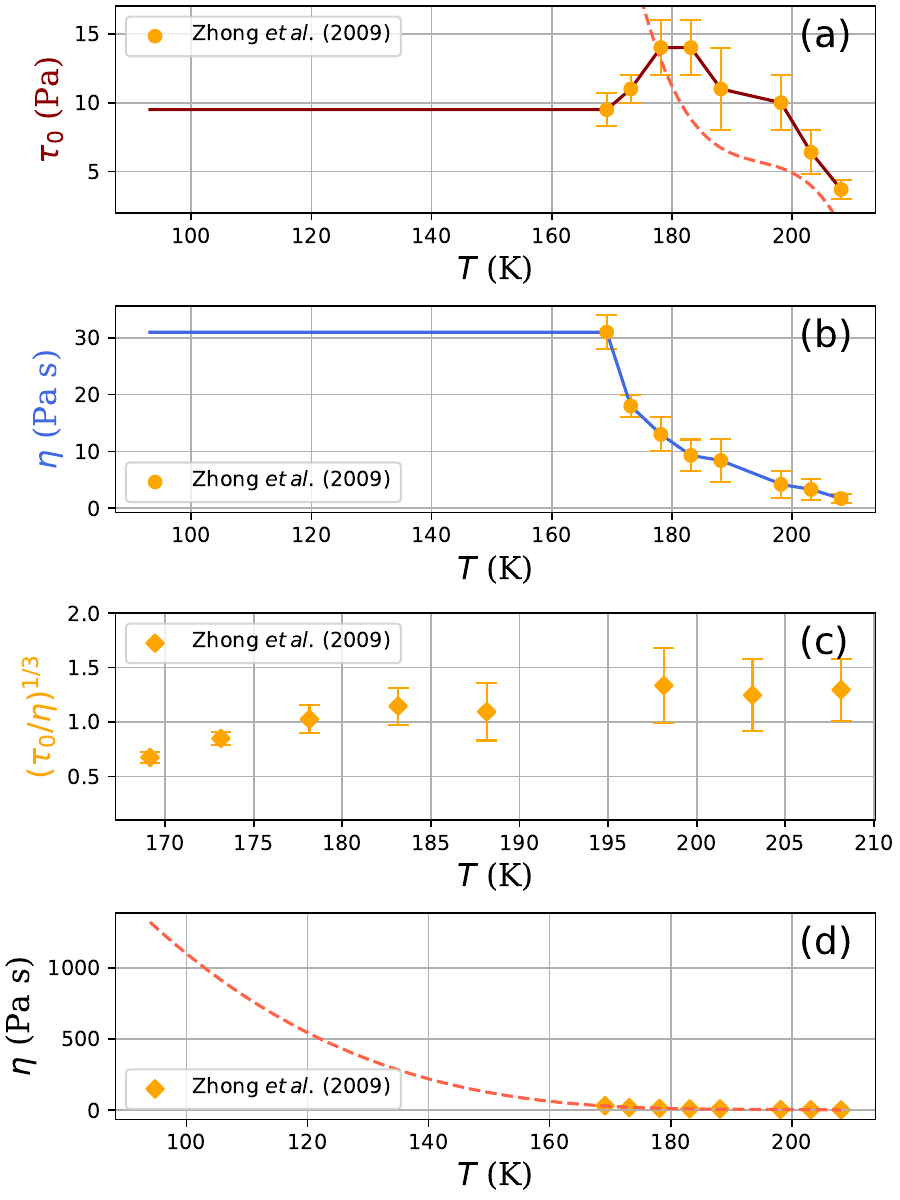}

\caption[]{\label{visZhong09_a}The rheology of a liquid mixture of 40\% methanol$+$water, determined by laboratory measurements performed
           by \cite{zhong_etal_2009}. The original Zhong~{\it et al.}'s data have been extrapolated for low temperatures, down to a value 
           representative for Titan's surface, {\it i.e.} $93.15$~K ($-180^{\rm o}$C), the maximum temperature of Zhong~{\it et al.}'s results
           domain $208.15$~K ($-65^{\rm o}$C) is adopted as the eruption temperature. 
           (a) The yield stress $\tau_0$ measured by 
           \cite{zhong_etal_2009} with constant extrapolation at low temperature, the redish dashed line corresponds to the interpolation/extrapolation
           of $\eta$ data (see also panel-d and maintext), 
           (b) the Bingham plastic viscosity from the same reference.
           (c) The ``ratio'' $(\tau_0/\eta)^{1/3}$, computed according to a suggestion from \cite{pinkerton_wilson_1994}. 
           (d) The Zhong {\it et al}.'s plastic viscosity with cubic interpolation, 
           and extrapolation down to temperatures consistent 
           with Titan's surface temperature (red dashed line).
           All the data have been read in Zhong~{\it et al.}'s Tab.~1 and error bars have been reported for $\tau_0$ and $\eta$.}
\end{center}
\end{figure}
%

   To define our simulation box, we considered horizontal dimensions representative of either Mohini Fluctus or Selk 
(see Fig.~\ref{topoMohiniFluctus}), {\it i.e.}, $200 \times 100$~km, in accordance with the description provided in Sect.~\ref{intro}.  
The average slope is set to $0.3^{\rm o}$, which is consistent with observations from Mohini Fluctus radar topography, 
where an elevation difference of $1500$~m over $300$~km corresponds to a mean slope of $0.3^{\rm o}$ 
\citep[see Fig.~\ref{topoMohiniFluctus}-b for the elevation map derived by][]{malaska_etal_2022}.  
Using the same data, together with the average pixel size (approximately $1400$~m) and the mean elevation difference between adjacent
pixels, we similarly derive an average slope of about $0.3^{\rm o}$.
Another example for which topography has been derived is the Ganesa Macula region. Although it was not classified as a cryovolcanic structure 
by \cite{lopes_etal_2013}, it exhibits prominent flow-like features. Examination of the topography of this region 
\citep[see Fig.~3 of][]{lopes_etal_2013} indicates elevation differences of approximately $\sim 600$~m over a distance of about $100$~km. 
These values correspond to a slope of roughly $0.3^{\rm o}$. Using a similar approach, \cite{lopes_etal_2013} derived a comparable estimate 
for Hotei Regio, obtaining a slope of approximately $0.09^{\rm o}$ \citep[see Fig.~10 of][]{lopes_etal_2013}.
On Earth, land slope distributions typically peak at a few degrees \citep{tang_etal_2020}, depending on the region considered.  
A gentler slope on Titan is plausible, given that the surface of this moon appears generally smoother than that of our planet.  
For this reason, we adopted a ``standard'' slope of $0.3^{\rm o}$, even though \cite{soderblom_etal_2010}, using photoclinometry, 
found locally much steeper slopes, reaching up to $30$~degrees. Indeed, the photoclinometry technique, which relies on VIMS data, yields
rather uncertain results, with a spatial resolution inferior to that of radar data. Furthermore, the available data focus on the crater 
rim rather than on features that could be interpreted as impact melt flows.

To keep computation times manageable, we set the cell size to $500 \times 500$~m, {\it i.e.} $W= 500$~m. 
This spatial resolution is reasonable when compared with Cassini SAR\footnote{Synthetic Aperture Radar} image resolutions, which range from $\sim 350$~m to over $1$~km \citep{lopes_etal_2013}. As illustrated in Fig.~\ref{topoMohiniFluctus}, a spatial resolution of $500$~m is amply sufficient for Mohini
Fluctus, given that the corresponding observations have resolutions of $640$~m/px$^{-1}$ and $1400$~m/px$^{-1}$. For a flow like Ara Fluctus, 
which is considerably less well defined morphologically, the use of a much higher spatial resolution would not be justified by the available 
observational constraints. With this first set of topographic assumptions, we are clearly addressing non-channelized flows.
The source of cryofluid is assigned to a cell of the computational grid, implying that the volcanic vent has a characteristic size smaller
than $500 \times 500$~m.\\

Furthermore, the substrate is assumed to be a perfect insulator and not subject to thermal erosion.  Insulating properties could be provided by a 
thick layer of organic material resulting from the sedimentation of particles produced in Titan's atmosphere.
Indeed, the literature provides estimates of the thermal conductivity of Titan’s organic sediments as low as
$\kappa_{\rm sedim} \simeq 0.025$~W~m$^{-1}$~K$^{-1}$ \citep{schurmeier_dombard_2018}. One may compare the thermal flux directed toward 
the crust, through a sediment layer of thickness $\Delta z_{\rm sedim}$, with the radiative flux from the upper external surface of a flow. 
These two fluxes are comparable when
\begin{equation}
\Delta z_{\rm sedim} \simeq \frac{\kappa_{\rm sedim} (T_{\rm cryofluid} - T_{\rm crust})}{\sigma_B (T_{\rm cryofluid}^4 - T_{\rm air}^4)}
\end{equation}
Taking $T_{\rm cryofluid} \sim 300$~K and $T_{\rm crust} = T_{\rm air} \sim 100$~K,
numerical application yields $\Delta z_{\rm sedim} \simeq 1$~cm. This thickness is significantly smaller than the estimated height--derived from radar
observations \citep{mastrogiuseppe_etal_2014b}--of $60$ to $120$~m of organic sediments. It goes without saying that, when cooling occurs via natural or
forced convection in the atmosphere above the flow, the situation becomes even more favorable for neglecting cooling by contact with the ground. Nevertheless,
we acknowledge that mechanical or thermal erosion could occur in certain locations; however, modeling such effects constitutes a research endeavor in its own
right, beyond the scope of the present article.\\
%

 Regardless of the origin of the cryofluid considered (cryolava or impact melt) it is highly probable that water
constitutes the major liquid component of the fluid, as it is widely accepted that the outermost layers of Titan's interior consist
of ice with the uppermost layers perhaps in the form of clathrates \citep[see for instance][]{sotin_etal_2021} with a layer of organic material on top.
This ice crust, approximately $100$~km thick \citep{sohl_etal_2014}, may overlie a
global ocean whose existence remains strongly debated \citep{lorenz_legall_2020,petricca_etal_2025}, as does its composition. For the latter, 
several solutes have been proposed, such as ammonium sulfate \citep{fortes_etal_2007}, sodium chloride, and magnesium sulfate 
\citep{vance_etal_2018}; however, it is currently not possible to discriminate between these competing hypotheses \citep{sotin_etal_2021}. 
Although infrared observations show spectral signatures that may be consistent with ice outcrops \citep{griffith_etal_2019, coutelier_etal_2021}, 
the surface composition remains poorly constrained. Spectral analyses of the Sotra Patera region conducted by \cite{solomonidou_etal_2016} 
revealed albedo variations consistent with possible cryovolcanic activity; however, it is impossible to derive precise information 
on the chemical composition of the surface from these data. 
%
   The results of \cite{solomonidou_etal_2016} suggest an increase in surface albedo at all the wavelengths considered, which could be indicative 
of a change in the surface state of the ground in the region under consideration. This change, associated with a brightening, is consistent with the formation of 
several types of crystals that could form during the cooling of a binary mixture.
%
Finally, owing to the suspected presence of alkanofers
\citep{hayes_etal_2017,horvath_etal_2016,faulk_etal_2020,cordier_etal_2021}, as well as possible forms of cryovolcanism in the polar
regions \citep{wood_radebaugh_2020}, the possibility of a cryofluid containing a significant fraction of liquid hydrocarbons
cannot be excluded. \\

%
%
   In a very general sense, rheology describes the transport of momentum within a fluid. This transport occurs first at the molecular scale,
such that variations in chemical composition alter the rheology; it also takes place at the mesoscopic scale ({\it i.e}. $\mu$m, mm, cm, etc.), where
suspended solid particles contribute to this transport through collisions with one another. This effect is described by the
Einstein--Roscoe law \citep[][]{einstein_1906,einstein_1911,roscoe_1952}, which is widely used in volcanology
\citep[see][]{mader_etal_2013}.\\

   Both water and liquid hydrocarbons may contain dissolved species, such as ammonia, methanol, or salts in the case of
water \citep[see, e.g.,][]{kargel_1990,kargel_etal_1991,morrison_etal_2023},
and organic compounds such as butane or acetylene in the case of hydrocarbons \citep{cordier_etal_2013b,cordier_etal_2016b}.
In both cases, the liquid may also be laden with solid particles: aqueous solutions undergo crystallization as cooling proceeds and
may additionally entrain solid organic sediments encountered during surface flow; liquid hydrocarbons may likewise transport
such materials. The latter could also, for example following fracturing or an impact event, become enriched in ice particles,
since ice is insoluble in liquid hydrocarbons.
  As discussed above, suspended solid particles can strongly alter the effective rheology of a fluid, the same also applies
to gaseous inclusions such as bubbles \citep{mader_etal_2013}. Here again, for both liquid water and hydrocarbons,
gases initially dissolved in the fluid may generate bubbles during an eruption.\\

     Regardless of the composition of the cryofluid under consideration, its rheological behavior is governed by the Bingham parameters $\tau_0$ and
$\eta$; varying these parameters is therefore equivalent to varying the composition, without requiring detailed knowledge of the latter.
     Two potential approaches are therefore available to us: (1) for various examples in which the rheology of a mixture is known,
the corresponding numerical simulations may be carried out; or (2) starting from a prescribed rheology, the yield stress $\tau_0$ and
the plastic viscosity  (also named ``plastic viscosity'')  $\eta$ may be varied over
a given range, which is effectively equivalent to implicitly varying the composition.
    Among the mixtures available in the literature and relevant to the context of Titan, we found only the work of \cite{zhong_etal_2009}
to be formulated within a Bingham framework and, moreover, to provide the temperature dependence of the fluid behavior. We therefore chose
to follow the second approach described above and to adopt the constitutive laws proposed by \cite{zhong_etal_2009} as a proxy,
subsequently varying the Bingham parameters, as it will be reported in Sec.~\ref{param}, so as to mimic the influence of composition.\\

The work conducted by \cite{zhong_etal_2009} is an experimental study of a 40\% methanol--water mixture, whose rheology was described using a
Herschel–Bulkley law, $\tau = \tau_0 + \eta \, \dot{\gamma}^{n}$, which closely resembles a Bingham law.
The measured values are summarized in Tab.~1 of \cite{zhong_etal_2009}; as shown there, the exponent $n$ remains very close to unity,
justifying the approximation of these results with a Bingham law. The results of \cite{zhong_etal_2009} are shown in Fig.~\ref{visZhong09_a}-(a,b).
As noted by \cite{zhong_etal_2009}, the yield stress $\tau_0$ exhibits a maximum near $-180^{\circ}$C before decreasing
with further cooling. Since the number of crystals in the considered slurry is expected to increase, the yield stress would continue rising. 
To our knowledge, this anomalous behavior has not yet been physically explained.
The yield stress $\tau_0$ and the viscosity coefficient $\eta$ were determined only for temperatures between $169$~K 
($-104^{\circ}$C) and $208$~K ($-65^{\circ}$C). 
As shown in panels (a) and (b) of Fig.~\ref{visZhong09_a}, we initially used a constant extrapolation of $\tau_0$ and $\eta$ toward 
low temperatures; a more realistic approach was subsequently tested and is described in Sec.~\ref{param}.
The extrapolation is {\it a priori} necessary because Titan’s surface temperature is far below $169$~K: the temperature measured
by Huygens in a tropical region is about $94$~K \citep{fulchignoni_etal_2005}, whereas temperatures in polar regions are expected to be
a few degrees lower, around $90$~K \citep{jennings_etal_2016}. We emphasize that if, during the flow phase, the fluid
temperature does not fall below $169$~K, with the remainder of the cooling occurring after the flow has ceased, the proposed interpolation
will have no influence.

\begin{figure}[htbp]
\begin{center}
\includegraphics[angle=0, width=9cm]{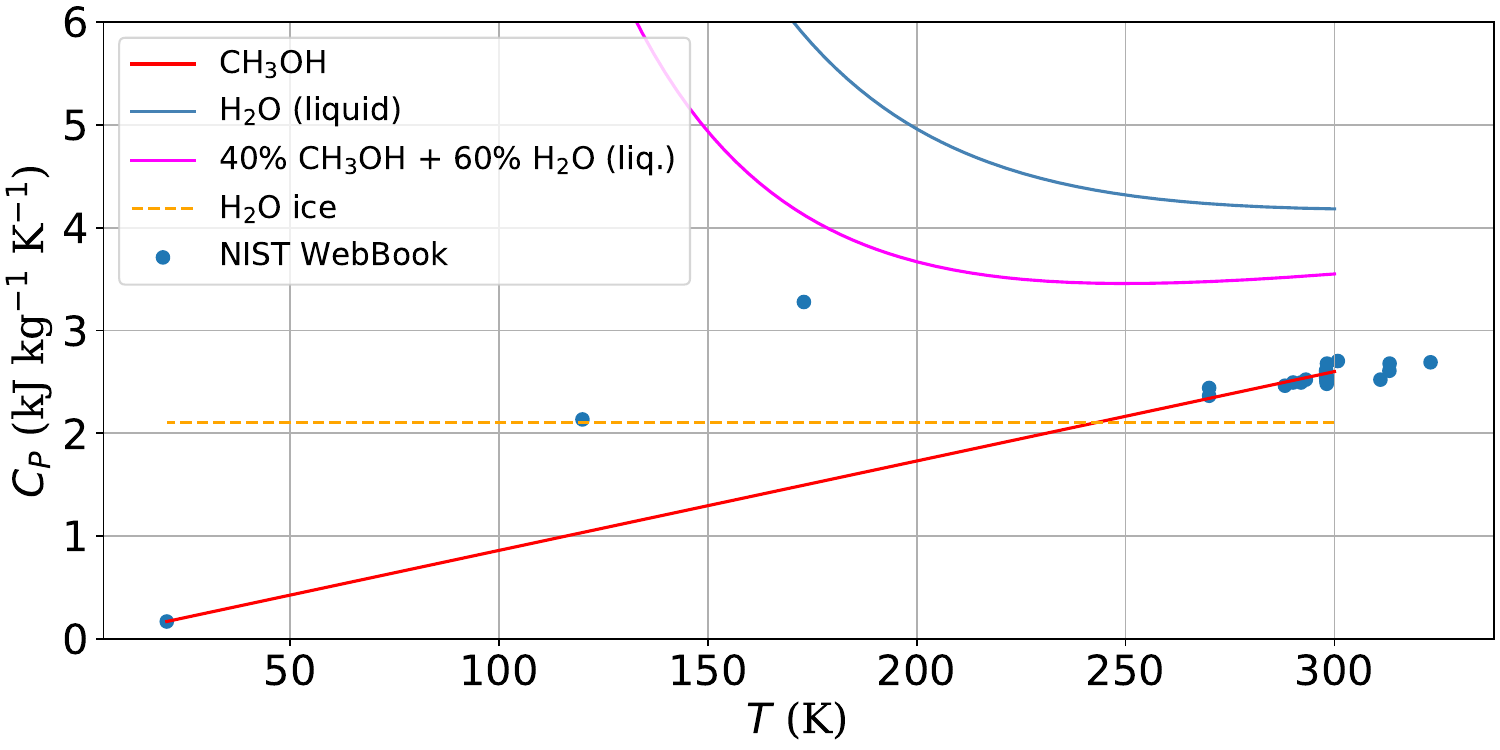}

\caption[]{\label{Cp_methanol_water}Estimated specific heat $C_p$ for methanol–water mixtures. Blue dots: methanol measurements from the NIST
           WebBook\footnotemark. Yellow line: water ice; light blue curve: liquid water; pink line: weighted 
           mean of methanol and water. Red line: extreme methanol values. The adopted average value is $2500$~J~K$^{-1}$~kg$^{-1}$.}
\end{center}
\end{figure}

Unfortunately, we did not find a highly accurate value for the density of the methanol–water mixture considered here, 
but a value around $930$~kg~m$^{-3}$ appears to be a reasonable estimate. This is consistent with the weighted mean of the densities 
of water ($10^3$~kg~m$^{-3}$) and methanol ($792$~kg~m$^{-3}$), which yields $917$~kg~m$^{-3}$. On another side, industrial
references\footnote{\url{https://methanol.org}} suggest a density of $970$~kg~m$^{-3}$ at $-40^{\circ}$C, differing by less than 
10\% from our estimate. For the specific heat at constant pressure $C_p$, we adopted $C_p = 2500$~J~K$^{-1}$~kg$^{-1}$ by comparing 
the values for methanol, liquid water, and ice (see Fig.~\ref{Cp_methanol_water}). A quick comparison with the values published 
by \cite{davies_etal_2010} (see their Table~3, p.~892) for water–ammonia mixtures shows that our estimates are reasonable. 
For our reference model, we chose to adopt these values of $\rho$ and $C_p$ and to keep them constant. However, in the case of a binary
mixture such as the water--methanol system, cooling leads to the formation of crystals within the fluid volume. The latent heat released 
during crystallization implies that the effective heat capacity may differ from the intrinsic value of $C_p$. This effect will be specifically 
investigated in Sec.~\ref{param}.\\

%
\footnotetext{\url{https://webbook.nist.gov}}
%
%
   Titan's eruption rates and durations are completely unknown. The only meaningful constraint is that the eruption rate should remain
consistent with the assumption of an effusive ({\it i.e.} not explosive) eruption.
In the context of Europa, an eruption mechanism has been proposed \citep{fagents_2003,lesage_etal_2020,lesage_etal_2022} based on the existence of a subsurface reservoir of liquid water \citep{vilella_etal_2020}
undergoing freezing. 
These authors found rather high fluid ejection velocities, on the order of $1$--$10$~m~s$^{-1}$, far too large to produce an effusive regime.
However, if we assume the existence of such ``magma chambers'', one may envisage significant pressure drop within the volcanic 
conduit \citep{bodin_etal_2026}, which would be sufficient to reduce the flow velocity to much lower values. While such an eruptive mechanism 
is qualitatively relevant in the present context, it does not allow the range of plausible eruption rates or eruption durations to be constrained.
For Ceres, \cite{sori_etal_2018} proposed an eruption rate of $10^4$~m$^3$~yr$^{-1}$, but this value is averaged over both time and space 
and therefore provides little information about the local conditions that may prevail on Titan.
Moreover, tidal forcing is approximately $10^4-10^5$~times
stronger on this Saturnian moon than on Ceres\footnote{A crude estimate of the tidal effect scales as ($\mathcal{M}_p R_s/d_{ps}^3$),
where $R_s$ is the radius of the object undergoing tides, $\mathcal{M}_p$ is the mass of the tide-raising body and $d_{ps}$ is the distance to it.
Ceres is primarily subject to solar tides, whereas Titan experiences tides raised by Saturn.},
implying a potentially very different cryovolcanic setting.
Finally, for a cryofluid draining from a crater lake, the ``eruption rate'' also remains poorly constrained, as it is highly sensitive to the 
geometry of the conduit and to the fluid level in the reservoir.
   Therefore, as a first approximation, we draw inspiration from values reported for terrestrial planets. For example, \cite{miyamoto_sasaki_2000} 
considered the lava flow associated with Xantippe Crater on Venus, whose extent is broadly comparable to that of Mohini Fluctus.
    The eruption parameters adopted for Xantippe Crater were an effusion rate of $10^4$~m$^3$~s$^{-1}$ and an eruption duration of $10^6$~s
($11.6$~terrestrial days). For our first Titan simulation, we selected comparable but slightly reduced values: an eruption duration of 
$3\times 10^5$~s (3.5 terrestrial days) and an effusion rate of $3.3 \times 10^3$~m$^3$~s$^{-1}$. The eruption rate is assumed to remain constant
throughout the eruption, corresponding to a time-averaged value.
These choices yield a total erupted volume of
$9.9 \times 10^8$~m$^3$, which is far smaller than the possible maximum volume of the entire Mohini Fluctus deposit \citep[$3.3 \times 10^{12}$~m$^3$, adopting the maximum thickness estimated by][]{lopes_etal_2013}.
If the cryofluid source is reduced to a single $500 \times 500$~m cell, the mean cryofluid velocity at the vent is $1.3$~cm~s$^{-1}$.
For a dynamic viscosity of the order of $\eta \sim 10$~Pa~s (see Fig.~\ref{visZhong09_a}-b), the fluid velocity at the outlet corresponds
to a Reynolds number of approximately $600$, indicating a regime where inertial forces remain small enough to ensure an effusive flow.
  In the following, we investigate the influence of the fluid discharge rate (eruption rate) and the duration of fluid emission (eruption duration) 
by varying their values around those adopted in the reference case.\\

\begin{figure*}[htbp]
\centering
\adjustbox{center}{%
\includegraphics[angle=0, width=16 cm]{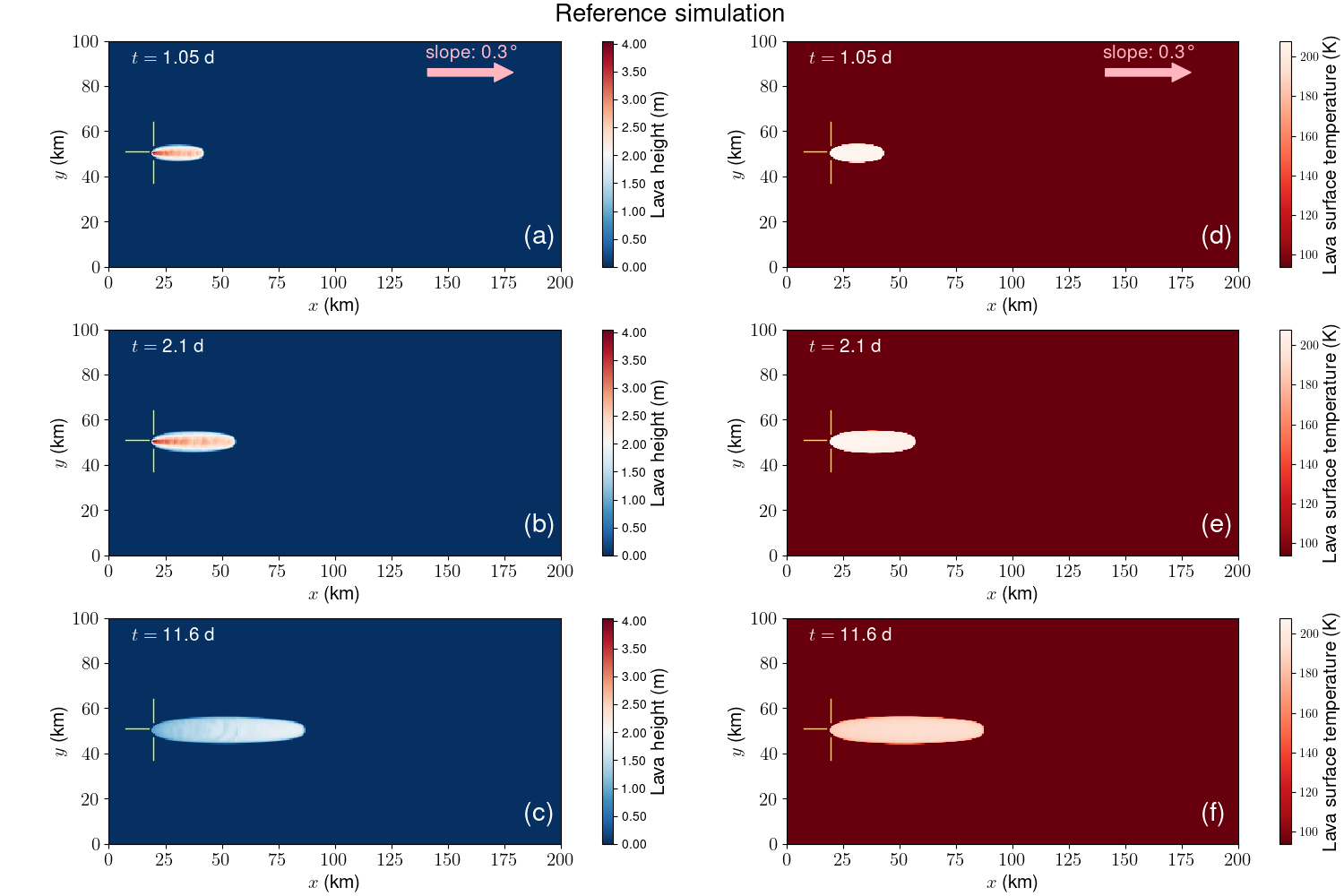}
}

\caption[]{\label{Z9V7ref}An example of cryofluid flow at the surface of Titan. The spatial coordinates $x$ and $y$
           are expressed in km, and the position of the cryofluid source is indicated with a yellow cross. The main slope runs
           parallel to the $x$-axis, with a gradient of $0.3$ degrees, as shown by the pinkish arrow in panel (a). 
           The color scale in panels (a)–(b)–(c) represents the cryofluid thickness $h$ (m), while panels (d)–(e)–(f) show the cryofluid
           temperature $T$ (K). 
           Panels (b) and (e) correspond approximately to the end of the eruption, whereas panels (c) and (f) represent the state of the 
           system at the end of the simulation. Thermal radiation is the only cooling process taken into account in this numerical experiment. 
           The adopted cryofluid rheology is based on laboratory measurements conducted on a water–methanol mixture \citep{zhong_etal_2009},
           extrapolated toward lower temperatures ({\it i.e.}, below $169$~K). The cell size is $0.5 \times 0.5$~km.
           This simulation (labelled ``Ref'') required approximately $3$~days and $20$~hours of computation on 
           $64$~CPUs\footnotemark, using the 
           anisotropy correction algorithm proposed by \cite{vicari_etal_2007}.}
\end{figure*}
\begin{figure}[!t]
\begin{center}
\includegraphics[angle=0, width=9 cm]{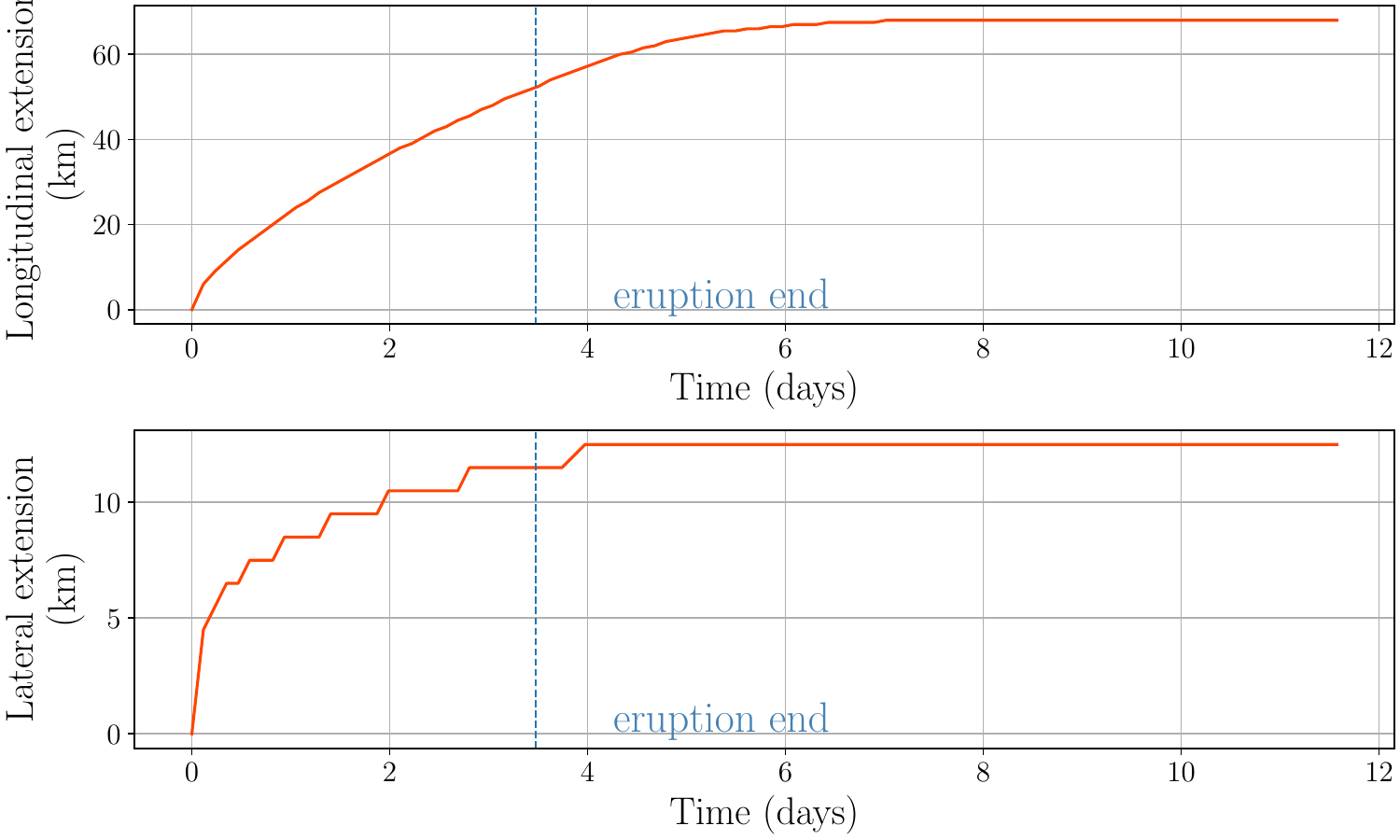}

\caption[]{\label{Ref_Xextension}(a) The longitudinal extension of our reference simulation (``Ref'') as a function of time, 
           the blue-dashed vertical line indicates the end of the eruption, while the red one shows the extension stops at around $7$~days.
           (b) The lateral extension of the cryofluid flow of the same simulation. In this case, the scalloping effect due to the
           $500$-meter-wide cells is clearly visible}
\end{center}
\end{figure}

    Using all the reference parameter values discussed above, we constructed our reference model, the results of which are shown in
Fig.~\ref{Z9V7ref} (simulation labelled ``Ref''). The eruption duration was set to $3 \times 10^5$~s
(3.5~days), the total simulation time is $10^6$~s (11.6~days), and the eruption rate 
is $e_{\rm rate} = 3.3 \times 10^3$~m$^3$~s$^{-1}$, with these values also being adopted as the reference model parameters.
%
\footnotetext{therefore corresponding to around 6000~hours of CPU-time.}
%
 As expected, the cryofluid flows downslope, forming a lobed structure symmetrical with respect to the axis containing the vent and
parallel to the $x$-axis. The flow reaches its maximum longitudinal extent at $t \simeq 7$~days, {\it i.e.}, $3.5$~days after the end of 
the eruption, as shown in Fig.~\ref{Ref_Xextension}, which depicts the temporal evolution of both the longitudinal and lateral extents 
of the flow. An internal flow remains active, {\it i.e.}, although the flow’s contour is unchanged, the thickness continues to evolve, 
resulting in an accumulation of material in the lowest part of the deposit (compare panels b and c of Fig.~\ref{Z9V7ref}), 
the maximum of lava thickness stops evolving at around $8.5$~days, {\it i.e.} $5$~days after the end of the eruption.
During its advance, the average velocity of the flow front was $9.7$~km/day, or $0.41$~km~h$^{-1}$, corresponding to $11$~cm~s$^{-1}$.
%
%
    The resulting flow is relatively thin, with an average thickness of $1.39$~m and a maximum thickness of $1.88$~m.
     The Reynolds number $\mathcal{R}e$ associated with the flow can be estimated by considering a dynamic viscosity on the order of 
$10$~Pa~s ({\it cf.} Fig.~\ref{visZhong09_a}) and a characteristic length scale of approximately one meter. This yields $\mathcal{R}e \sim 10$, 
indicating that the flow is expected to be laminar at large scale.
Its longitudinal extent reaches $68$~km, with a lateral width of $12.5$~km (see Fig.~\ref{Ref_Xextension}). 
The eruption temperature was set to $208$~K 
\citep[the maximum value in the range reported by][]{zhong_etal_2009}, which we adopt as a starting point to avoid introducing additional assumptions
concerning rheology at temperautre higher than $208$~K.
Since only radiative cooling was considered in this preliminary experiment, heat loss is relatively inefficient, leading to an average temperature of 
$186$~K at the end of the simulation, {\it i.e.}, the fluid is not fully cooled. More importantly, during the dynamic phase of the flow, 
the temperature remains close to the eruption temperature, yielding an average of $195$~K when the flow stops. This value lies well within the range of
validity of the results reported by \cite{zhong_etal_2009} (see Fig.~\ref{visZhong09_a}), suggesting that the extrapolated portion of the rheological law
does not significantly affect the flow behavior in this simulation. 
We also extended this simulation in time, keeping only the thermal phenomena. The average cryofluid temperature asymptotically tends toward the ambient temperature, namely $94$~K. After an evolution of $400$~days, the average temperature still reaches $100$~K. We did not go any further,
as extending this calculation has only limited value for the computational cost.
Overall, this first simulated lava flow appears physically consistent and resembles, 
to some extent, a terrestrial basaltic flow. In this initial scenario, the overall extent of the lava flow seems to be controlled primarily by its volume
rather than by its temperature. These results are also  compatible with the size of flows previously mentioned, observed by Cassini radar.
Indeed, the 68~km extent reached by the simulated flow lies, for instance, between the spatial extents observed for Ara Fluctus and Mohini Fluctus
(see Fig.~\ref{topoMohiniFluctus}).\\

  We emphasize that all the parameters discussed above are uncertain—such as those related to rheology—or may even have non-unique values,
potentially depending on the location or time considered, as is the case, for example, with the average sub-flow slope. In the remainder of
this paper, we will therefore use the reference model described above to study the influence of the various parameters, aiming to highlight
the corresponding trends in flow behavior. It appears beyond the scope of this work to attempt to reproduce the details of any specific flow, 
as certain parameters may have antagonistic effects: for instance, a steeper slope might lengthen a flow, while a stronger wind above the flow 
might tend to reduce its extent.\\
We have divided this parametric study into two sections: the first addresses the most physical aspects, while the second generically examines the effects of basic topographical characteristics. To facilitate our understanding of these different effects, the parameters are tested one at a time, with all others retaining the values assigned in the reference model.\\


\section{Influence of physical parameters}
\label{param}
%
\begin{figure}[htbp]
\begin{center}
\includegraphics[angle=0, width=9 cm]{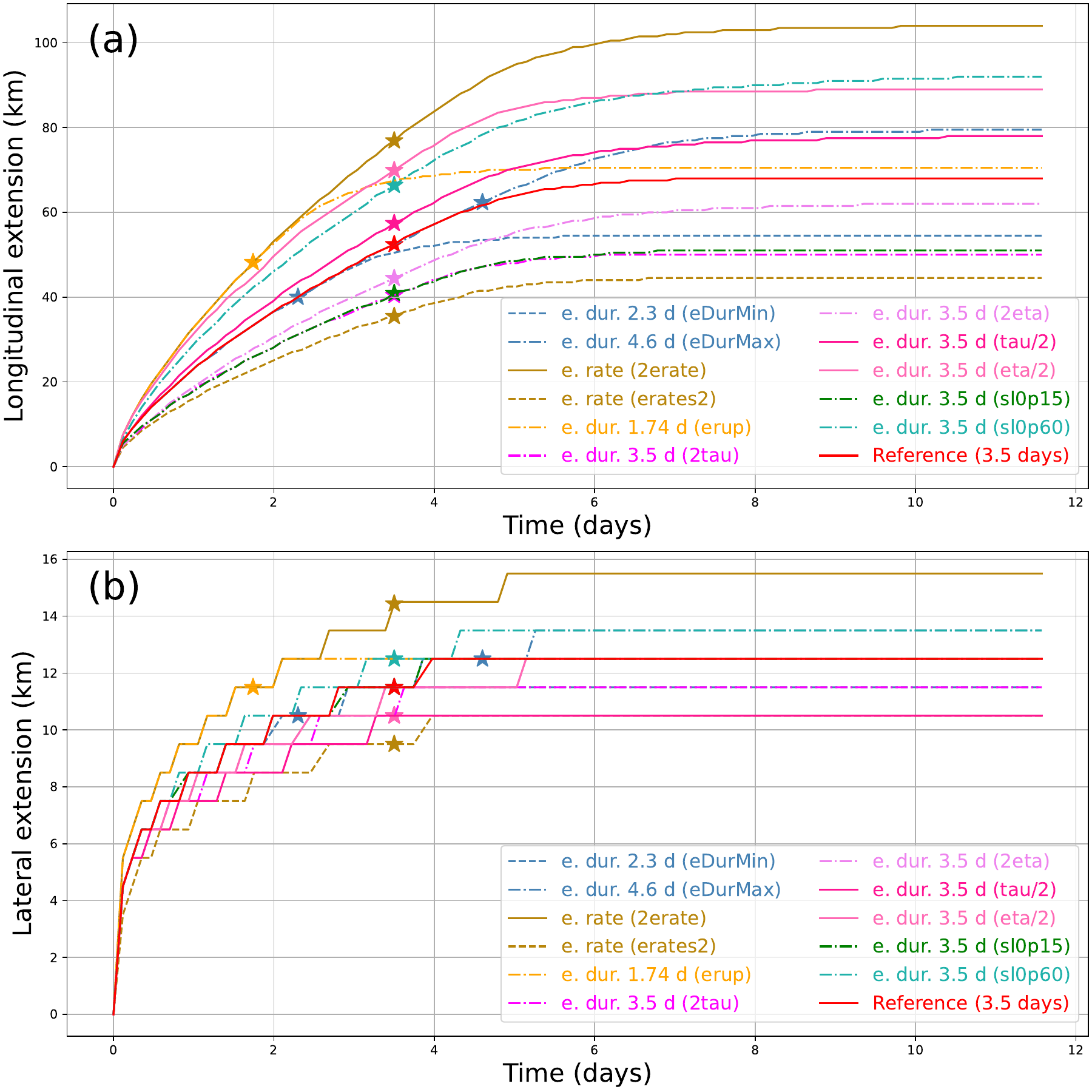}

\caption[]{\label{compa_phys_param}
           Panel (a) shows comparison of the longitudinal extension, as a function of time, for cryofluid flows. The eruption duration has been
           modified, ``Ref'' corresponds to the set of parameters used in Sect.~\ref{model} for which the eruption duration was $3\times 10^5$~s
           ($3.5$~terrestrial days), the simulation labeled ``eDurMin'' has a shorter (by $-33$\%) duration fixed at $2\times 10^5$~s ($2.3$~days).
           In the opposite, the simulation ``eDurMax'' used a longer (by $+33$\%) eruption duration set at  $4\times 10^5$~s ($4.6$~days).
           A simulation with a doubled eruption rate (``2erate''), together with a simulation corresponding to this rate divided
           by a factor of 2 (``erates2''), were also carried out.
           The computation labelled ``erup'' involved a total erupted volume equal to that of our standard model, but with an 
           eruption rate twice the standard one, leading to an eruption duration shorter by a factor of two.
           Experiments with the Bingham parameters, describing the fluid rheology, were conducted by multiplying either $\tau_0$ 
           (simulation ``2tau'') or $\eta$ (simulation 2eta'') by a factor of two.
           Similarly, simulations characterized by lower rheological strength were conducted by decreasing either $\tau_0$ (simulation ``taus2'')
           or $\eta$ (simulation ``etas2'') by a factor of two.
           The simulation corresponding to a slope reduced to $0.15^{\rm o}$ is ``sl0p15'', while that of slope set to $0.60^{\rm o}$ is 
           ``sl0p60''.
           In panel (b), we have displayed the lateral extension, {\it i.e.} parallel to the $y$-axis of Fig.~\ref{Z9V7ref}, of the flow as a 
           function of time. The crenellated appearance of the curve is due to combined effect of the spatial resolution ($500$~m here) and also
           to the Bingham nature of the fluid, which requires a threshold to be crossed before it can flow.
           In all cases, the moment when the eruption ends is indicated by a star ($\bigstar$).\\
           }
\end{center}
\end{figure}
%

   In this section, we examine the primary parameters controlling the eruption itself, namely the eruption rate and duration.
We then consider the role of cooling mechanisms by including wind-driven convection above the cryofluid flow, and finally, we assess the effects of the
Bingham parameters that govern the fluid’s rheology. In all cases, our focus will be on the spatial extent of the flow when it reaches its maximum.\\

   We begin by examining the effect of eruption duration, while keeping all other parameters identical to those used in the 
previous section. Two simulations were performed: one in which the duration was reduced by $33$\% (simulation ``eDurMin''), decreasing from
$3.5$~days in the reference simulation (``Ref'') to $2.3$~days, and another in which it was increased by $33$\% (simulation ``eDurMax''), 
reaching $4.6$~days. The temporal evolution of the longitudinal and lateral extents is presented in Fig.~\ref{compa_phys_param}.
We recall the for the reference simulation, the final longitudinal and lateral extents are $68$~km and $12.5$~km, respectively. 
In the reduced-duration case (``eDurMin''), these values decrease to $54.5$~km ($-20$\%) and $12.5$~km ($-8$\%), while in the 
increased-duration case (``eDurMax''), they increase to $79.5$~km ($+17$\%) and $13.5$~km ($+8$\%). 
Consistent with mass conservation, the final cryofluid thicknesses are $1.39$~m (``Ref''), $1.28$~m (``eDurMin''), and $1.48$~m (``eDurMax'').
These results indicate that the average final thickness is relatively insensitive to the total eruption duration, and thus to the total erupted 
cryofluid volume. Within the framework of the present scenario, it is primarily the longitudinal extent that is most sensitive to changes in the
total erupted volume.
%
\begin{figure}[htbp]
\begin{center}
\includegraphics[angle=0, width=9 cm]{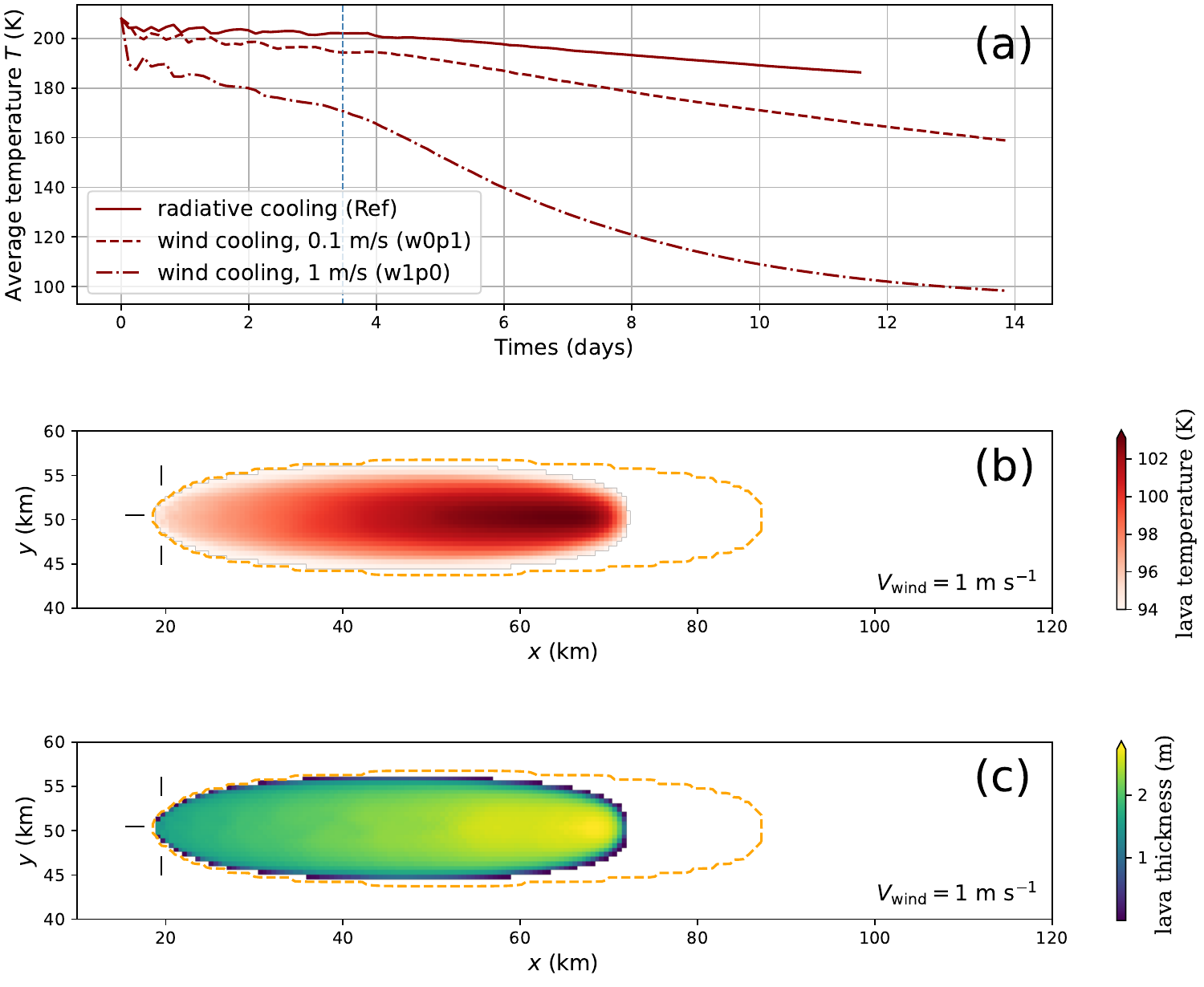}

\caption[]{\label{cooling}Effect of thermal cooling on a cryofluid flow.
Panel (a) shows the evolution of the average surface temperature of the lava as a function of time. 
The end of the eruption is represented by a vertical dashed blue line.
The solid line corresponds to radiative cooling only (``Ref'', our reference simulation); the dashed line includes convective cooling following 
the prescription of \cite{davies_etal_2010} with a wind speed of $0.1$~m~s$^{-1}$ (``w0p1''); the dash-dotted line corresponds to a wind speed of 
$1$~m~s$^{-1}$ (``w1p0''). In order to save computing time, all simulations were stopped before thermal equilibrium between the lava and the 
environment was reached; nevertheless, the two simulations that took wind cooling into account were run somewhat longer.
Panel (b) shows the temperature map for the simulation with a wind speed of $1$~m~s$^{-1}$, at $11.46$~days 
after the start of the eruption. Panel (c) shows the lava thickness map for the same simulation, at the same time.
In both panels (b) and (c), the vent position is marked with a black ``cross'', while the dashed orange contour indicates the final extent 
of the reference simulation (radiative cooling only).}
\end{center}
\end{figure}
%

  Regarding the eruption rate, we performed a sensitivity test by first multiplying it by a factor of two, increasing it from
$3.3 \times 10^{3}$~m$^3$~s$^{-1}$ to $6.6 \times 10^{3}$~m$^3$~s$^{-1}$, while keeping the eruption duration unchanged. This therefore corresponds 
to a doubling of the total erupted volume. Our model shows that the maximum longitudinal extent increases from $68$~km (reference model) to $104$~km 
(see Fig.~\ref{compa_phys_param}), whereas the increase in the maximum lateral extent is less pronounced, from $12.5$~km to $15.5$~km.
The final average thickness increases only slightly, from $1.39$~m to $1.49$~m. In a second simulation, we reduced the eruption rate by a factor of $2$,
decreasing its value from $3.3 \times 10^3$~m$^3$~s$^{-1}$ to $1.65 \times 10^3$~m$^3$~s$^{-1}$ (see the new Fig.~\ref{compa_phys_param}).
In this case, the total volume of cryofluid is reduced by a factor of two, leading to a maximum longitudinal extent of $44.5$~km
and a maximum lateral extent of $10.5$~km. The final average thickness of the flow is $1.30$~m.
   Similarly to the results obtained in the eruption-duration sensitivity test, increasing or decreasing the total volume of cryofluid through changes
in the eruption rate primarily affects the final longitudinal extent of the flow. In contrast, the final lateral extent and the average flow thickness
are only marginally affected.
     Finally, to avoid volume-related effects, we chose to keep the total erupted volume constant by adjusting the eruption duration
ccording to the eruption rate. In this way, we aimed to isolate the effect of the fluid ejection velocity at the vent.
Then, we have increased the  eruption rate from $3.3 \times 10^{3}$~m$^3$~s$^{-1}$ to $6.6 \times 10^{3}$~m$^3$~s$^{-1}$,
reducing in the same time the eruption duration by the same factor, {\it i.e.} $2$, leading to $1.5 \times 10^5$~s ($1.74$~days). 
As shown in Fig.~\ref{compa_phys_param}, the resulting extensions lead to a final state (simulation ``erup'') essentially similar 
to that of our reference model, with, for instance, an average thickness of $1.32$~m instead of $1.39$~m. In our eruption scenario, the eruption 
rate value seems to have only a marginal impact on final flow extension which appears more controlled by the total erupted volume.\\

      Up to this point, we have considered only radiative cooling through the upper surface of the fluid.
Given the low temperatures involved, radiative heat transfer is generally far less efficient than wind-driven forced convection.
For this reason, we conducted additional tests incorporating the formulation described by \cite{davies_etal_2010} and recalled in Sec.~\ref{model}.
The Huygens probe measured weak near-surface winds during its descent, with values around $1$~m~s$^{-1}$ \citep{bird_etal_2005}. 
Likewise, atmosphere General Circulation Model (GCM) models predict similarly low surface wind speeds; for example, \cite{tokano_lorenz_2015} 
and \cite{lefevre_etal_2025} report values typically in the range $0.1$--$1$~m~s$^{-1}$. We therefore performed two simulations using the 
convective cooling formulation of \cite{davies_etal_2010}: one with a wind speed of $0.1$~m~s$^{-1}$ (``w0p1’’) and another with a wind speed 
of $1$~m~s$^{-1}$ (``w1p0’’), representative of the conditions expected at Titan’s surface.
The results are presented in Fig.~\ref{cooling}. 
As can be seen at first glance, convective cooling under Titan’s conditions is much more efficient than radiative cooling alone, particularly 
at higher wind speeds (see Fig.~\ref{cooling}-a).\\

For the simulation including a wind of $1$~m~s$^{-1}$ (``w1p0''), after $13.9$~days the lava is almost in thermal equilibrium with its environment, 
whose temperature was set to $94$~K. At the end of the ``w1p0'' simulation, the average temperature of the lava was $98$~K.
In panels (b) and (c), the flow reaches a longitudinal extent of $53.5$~km 
(for an average thickness of $1.90$~m), significantly smaller than the $68$~km obtained for the radiative-only cooling case. 
This behavior results from the dependence of rheology on temperature, particularly that of $\eta$, with constant extrapolation as depicted in 
Fig.~\ref{visZhong09_a}-a and b.\\
%
\begin{figure}[htbp]
\begin{center}
\includegraphics[angle=0, width=9 cm]{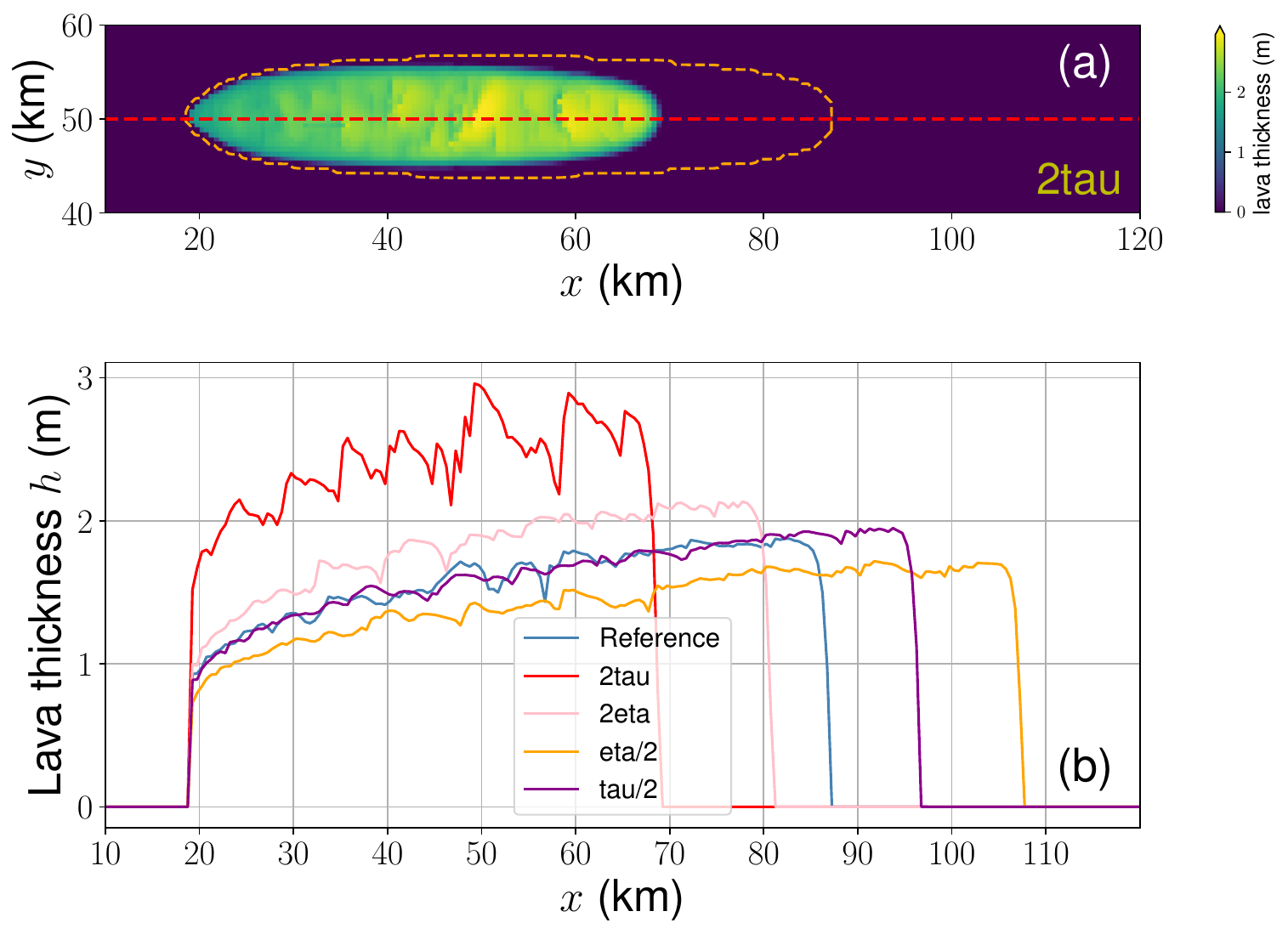}

\caption[]{\label{rheo_surface}(a) Cryolava thickness map of the final state obtained in simulation ``2tau'', for which the Bingham parameter $\tau_0$
           has been increased by a factor of two. The yellow dashed line indicates the final extent of the reference simulation (``Ref'').
           (b) Cryolava thickness along the longitudinal flow axis for the reference simulation (``Ref'') and the enhanced-rheology
           simulations ``2tau'' and ``2eta'' (see main text), together with reduced-viscosity simulations ``tau/2'' and ``eta/2''.}
\end{center}
\end{figure}
%

      Although Titan’s surface is globally an arid environment, rainfall events may occur, as suggested by several observations 
\citep{turtle_etal_2009,barnes_etal_2013}. Upon contact with cryofluid, at temperatures for instance between $200$ and $300$~K, liquid methane
vaporizes instantaneously, inducing cooling of the cryofluid associated with the latent heat of vaporization of methane.
The thickness of the cryofluid,
$e_{\rm lava}$, experiencing a temperature decrease $\Delta T$, can be estimated using an expression of the form
\begin{equation}
   e_{\rm lava} = \frac{\rho_{\rm CH_4} \, p_{\rm rain} \, \Delta t_{\rm rain} \, \Delta_{\rm vap}H_{\rm CH_4}}{\rho_{\rm lava} \,
                        C_{\rm p, lava} \, \Delta T \, M_{\rm CH_4}}
\end{equation}
where $\rho_{\rm CH_4}$ is the liquid methane density (kg~m$^{-3}$), $p_{\rm rain}$ the rain precipitation rate (m~s$^{-1}$),
$\Delta_{\rm vap}H_{\rm CH_4}$ the liquid methane enthalpy of vaporization (J~mol$^{-1}$), $\rho_{\rm lava}$ the cryofluid density (kg~m$^{-3}$),
$C_{\rm p, lava}$ the lava specific heat (J~K$^{-1}$~kg$^{-1}$), $\Delta T$ the cryofluid temperature decrease (K) and $M_{\rm CH_4}$ the methane
molecular weight. In the NIST database\footnote{\url{https://webbook.nist.gov/chemistry}} we found 
$\Delta_{\rm vap}H_{\rm CH_4} \simeq 8.6$~kJ~mol$^{-1}$ \citep{stock_henning_2006}, the Global Circulation Models predicts relatively 
modest liquid methane precipitation rates, according to \cite{lora_etal_2019} a rate $p_{\rm rain} \simeq 10$~mm~Titan-day$^{-1}$ is a rather 
large value, we adopted a shower duration of $1$~Titan-day, {\it i.e.} around $15$~terrestrial days, the density and specific heat of the considered
cryofluid have been given previously in the text. Taking into account all these quantities, we found for a cooling of only $\Delta T= 10$~K, the vaporization
the falling liquid methane lead to a cryofluid thickness of $10$~cm cooled down.
     This evaluation suggests that rain-induced cooling of cryofluid is likely a secondary process, owing primarily to the low precipitation
rates and the limited latent heat of vaporization of liquid methane.\\

   As described in Sec.~\ref{model}, the composition of the working fluid is primarily characterized in our model by its rheological parameters.
The influence of cryofluid composition was investigated by varying the parameters of the Bingham rheology with respect to the reference case considered here, 
namely the rheological model derived by \cite{zhong_etal_2009}.
We then conducted four simulations: two simulations were performed with $\tau_0$ and $\eta$ increased by a factor of two, denoted ``2tau'' and ``2eta'', respectively,
together with two additional simulations in which these parameters were decreased by a factor of two, denoted ``taus2'' and ``etas2'', respectively.
An increase in the Bingham parameters may, for instance, represent a higher load of suspended solid particles, or a chemical composition with higher concentrations 
of certain solutes. Conversely, a less viscous cryofluid may simulate a situation in which these solid or chemical loads are reduced.
The results pertaining to the spatial extent of the flow are presented in Fig.~\ref{compa_phys_param}.
    The effects of $\tau_0$ and $\eta$ are broadly comparable. As expected, an increase in viscosity tends to reduce the longitudinal extent of 
the flow, whereas a decrease in viscosity enhances its axial extension.
With the increased viscosity, the final average lava thickness is larger, reaching $2.01$~m (for ``2tau'' and $1.56$~m for ``2eta'') instead of $1.39$~m for ``Ref''. 
Since viscosity tends to increase as temperature decreases (see Fig.~\ref{visZhong09_a}), these results are fully consistent with those of the ``w0p1'' and ``w1p0'' 
simulations, which include efficient wind cooling. Conversely, a decrease in viscosity leads to a substantially greater longitudinal extent of the flow.
Consistently, the final mean thickness decreases to $1.28$~m in the ``eta/2'' simulation and to $1.43$~m in the ``tau/2'' simulation.
These behaviours are also clearly evident in Fig.~\ref{rheo_surface}-b.\\

Notably, on the surface of the flows with increased viscosity, one observes the appearance of small ripples, particularly visible in 
the ``2tau'' simulation (see Fig.~\ref{rheo_surface}-a). These patterns are approximately aligned perpendicular to the flow direction and are almost absent 
in the transverse direction. Such a phenomenon is not surprising, given that a Bingham fluid only begins to move once the applied stresses 
exceed a certain threshold (the yied stress $\tau_0$). 
   Instabilities at the surface of a visco-plastic fluid have been studied both theoretically and experimentally for a long time 
\citep[\textit{e.g.}][]{liu_mei_1994,mounkailaNoma_etal_2021}, and similar phenomena are observed in simulations using other numerical 
techniques such as SPH\footnote{Smoothed-particle hydrodynamics} \citep{li_etal_2022} or LBM\footnote{Lattice-Boltzmann Method} 
\citep{leonardi_etal_2015}. 
We furthermore note slight asymmetries of the flow with respect to its longitudinal axis (red line in Fig.~\ref{rheo_surface}-a).
This is easily explained by the limited spatial resolution of $500$ m and by the use of the anisotropy-correction algorithm 
of \cite{vicari_etal_2007}, which relies on a Monte Carlo–type approach.\\

In Fig.~\ref{rheo_surface}-b, for the same instant, we plot the cryofluid thickness $h$ along the longitudinal axis for different flows.
As can be seen, the reference simulation (``Ref'') exhibits small ripples with amplitudes on the order of a few centimeters, which, over the distances 
considered --{\it i.e.}, a total flow length of about $70$~km-- is negligible compared to the uncertainties inherent to this type of model 
and to our potential detection capabilities. For the two high-viscosity simulations, the amplitude of ripples is larger, reaching up to several 
tens of centimeters. These structures have a ``quasi-wavelength'' of approximately $5-10$~km, making them more significant; however, in practice, 
it is probably unrealistic to expect that such features could be observed at these scales on Titan.\\


   In our reference model, in order to address the lack of experimental data below $169$~K, we used a constant extrapolation of 
the viscosity coefficient $\eta(T)$ (see Fig.~\ref{visZhong09_a}-a,b); {\it i.e}., we adopted the value of $\eta$ at $169$~K for all 
temperatures below this threshold. 
However, there is no physical reason to assume that $\eta$ remains constant below $169$~K. We therefore performed an additional test using an 
extrapolation based on a cubic law  (see Fig.~\ref{visZhong09_a}-d).
Regarding the yield stress $\tau_0$, following a suggestion by \cite{pinkerton_wilson_1994}, we took advantage of the property $(\tau_0/\eta)^{1/3} \sim 1$ 
({\it cf.} Fig.~\ref{visZhong09_a}-c) to estimate the yield stress by adopting the approximation $\tau_0(T) \sim \eta(T)$. 
Of course, the experimental points at the lowest temperatures show some deviation from this law, but, as previously noted, the measurements reported by \cite{zhong_etal_2009} at these temperatures remain open to question.
We obtained a final longitudinal extension of $83.5$~km ($68$~km for the reference simulation), with a lateral one of $13.5$~km and 
an average thickness of $1.07$~m ($1.39$~m for the reference simulation). This behavior is not surprising since, at ``high'' temperature, the employed 
interpolation underestimates the yield stress by approximately $25$\% (see Fig.~\ref{visZhong09_a}-a), leading to a more fluid flow at the beginning 
of the process.\\

%
\begin{figure}[htbp]
\begin{center}
\includegraphics[angle=0, width=9 cm]{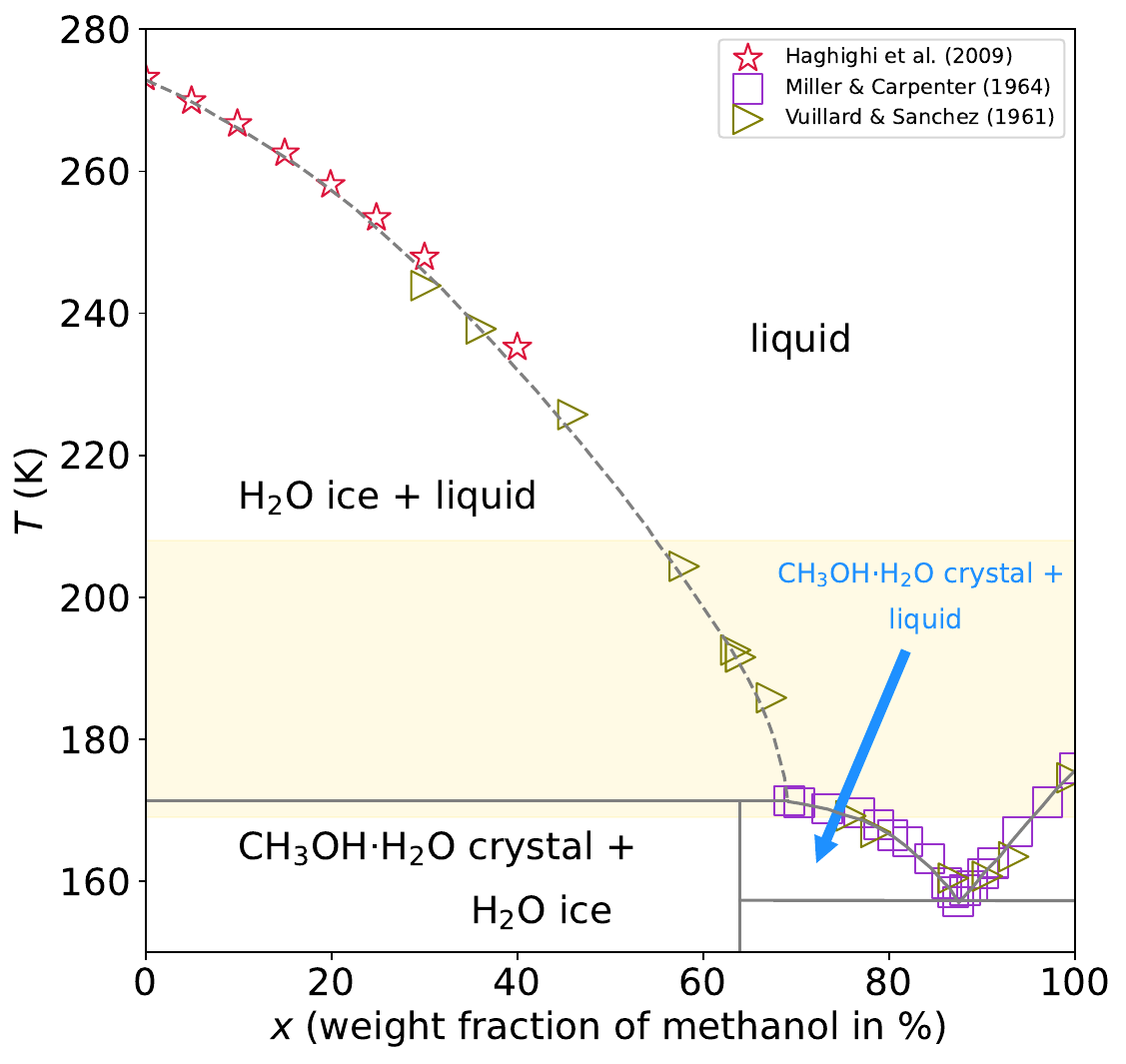}

\caption[]{\label{H2OCH3OH_bindiag}Binary phase diagram \citep[inspired by Fig.~1 in][]{dougherty_etal_2018} of the water--methanol 
           system: temperature $T$ (K) as a function of $x$, 
           the methanol weight fraction in the liquid phase. Experimental data are represented by symbols, and the corresponding references 
           are indicated. The different phase domains are delineated by grey lines, and the nature of the phases is specified within 
           each domain. Among these grey lines, the liquidus curve has been distinguished by using a dashed line rather than a solid one.
           The shaded yellow region corresponds to the temperature range considered by \cite{zhong_etal_2009}.
          In Zhong's study, the system is dominated by the coexistence of two phases, as illustrated here: a solid phase 
          composed of water ice and a liquid phase consisting of a water--methanol mixture, in which the methanol fraction ranges from 
          approximately 50\% to 70\%.}
\end{center}
\end{figure}
%

     The model developed in this work does not explicitly include phase transitions, in particular the formation of ice crystals
within the cryofluid. The effect on rheology is implicitly accounted for. As temperature decreases, the number of ice crystals per unit
volume increases, which enhances crystal–crystal interactions and consequently leads, at the macroscopic scale, to an increase in viscosity, 
as reported by \cite{zhong_etal_2009}.
            Partial crystallization of the fluid may also have thermal consequences: the formation of ice crystals releases latent 
heat $L_{\rm ice}$, and the heat capacity of ice, $C_{p,{\rm ice}}$, is slightly different from that of liquid water. 
Fig.~\ref{H2OCH3OH_bindiag} shows the phase diagram of the water--methanol system. As can be seen, within the domain explored by 
\cite{zhong_etal_2009}, two phases coexist: water-ice crystals and a liquid phase that becomes progressively enriched in methanol as 
temperature decreases.

\begin{figure}[htbp]
\begin{center}
\includegraphics[angle=0, width=9 cm]{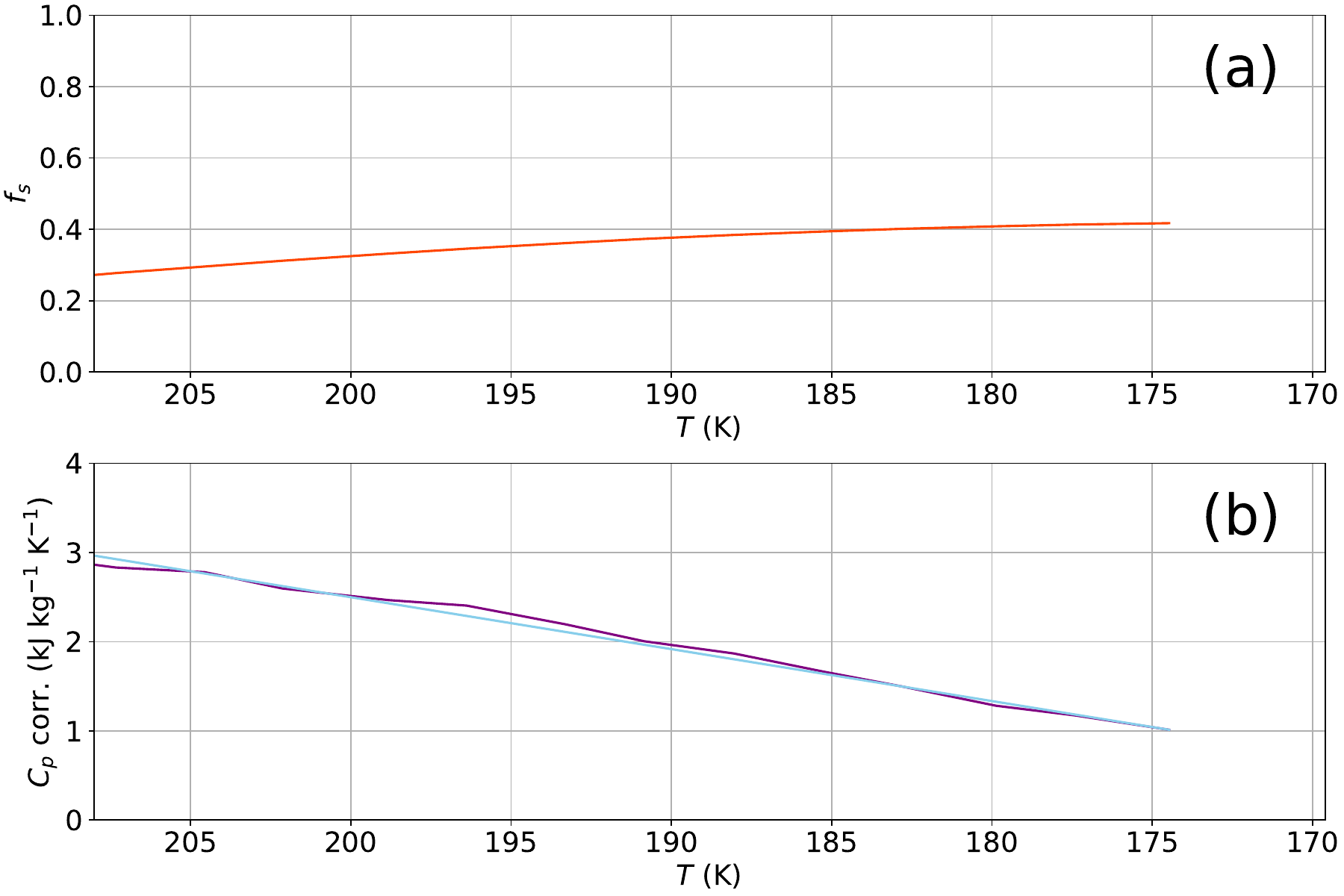}
\caption[]{\label{fs_Cpcorr}(a) The solid fraction $f_s$ in the binary water--methanol system during cooling of the mixture ($x$-axis reversed) 
          over the temperature range considered by \cite{zhong_etal_2009}. As the mixture cools, $f_s$ increases and eventually reaches 
          approximately $0.4$.
          (b) The corrective term $C_{\rm p, corr}$ (violet curve), numerically derived from the liquidus curve shown in Fig.~\ref{H2OCH3OH_bindiag}, 
          together with its linear approximation (sky-blue line). The slightly undulating appearance of the purple curve is likely a numerical
          artifact arising from the evaluation of the derivative $\mathrm{d}f_s/\mathrm{d}T$ using a finite-difference scheme along the 
          liquidus curve.}
\end{center}
\end{figure}

If $f_s$ denotes the solid fraction (here, water ice) present in the system at a given temperature, the effective heat 
capacity, $C_{p, {\rm eff}}$, can be expressed as follows:
       
\begin{equation}
   C_{p, {\rm eff}} = (1-f_s) \, C_{p, {\rm liq}} + f_s \, C_{p, {\rm ice}} + L_{\rm ice} \, \frac{\mathrm{d}f_s}{\mathrm{d} T}
\end{equation}
       
where $C_{p, {\rm liq}}$ is the specific heat of the liquid phase, $C_{p, {\rm ice}}$ the specific heat of ice and 
$L_{\rm ice}$ the freezing latent heat of water. The derivative $\mathrm{d}f_s/\mathrm{d} T$ has to be computed along the 
liquidus line in Fig.~\ref{H2OCH3OH_bindiag} according to the equation
       
\begin{equation}\label{dfsdT}
   \frac{\mathrm{d}f_s}{\mathrm{d} T} = \frac{\displaystyle  x_0}{\displaystyle x_L^2(T) \; \mathrm{d}f_s/\mathrm{d} T}
\end{equation}
       
where $x_0$ is the global fraction of methanol (here $40$\%) and $x_L$ the fraction of methanol in the liquid phase (given the $x$-axis
in Fig.~\ref{H2OCH3OH_bindiag}). The fraction of solid $f_s$ is given by
\begin{equation}\label{fs}
   f_s(T) = 1 - \frac{x_0}{x_L(T)}
\end{equation}
Fig.~\ref{fs_Cpcorr} shows the solid fraction $f_s$ as a function of temperature along the liquidus curve displayed in Fig.~\ref{H2OCH3OH_bindiag}, 
together with the correction term $C_{p, {\rm corr}} = f_s , C_{p, {\rm ice}} + L_{\rm ice} , \mathrm{d}f_s/\mathrm{d} T$. As can be seen, 
the correction $C_{p, {\rm corr}}$ is largest at the highest temperatures and decreases as the amount of crystals increases. This higher degree of 
crystallinity therefore leads to an increase in the effective heat capacity by a few kJ~K$^{-1}$~kg$^{-1}$. One may thus expect a delayed 
cooling of the cryofluid during an eruption.\\
%
%

\begin{figure}[htbp]
\begin{center}
\includegraphics[angle=0, width=12 cm]{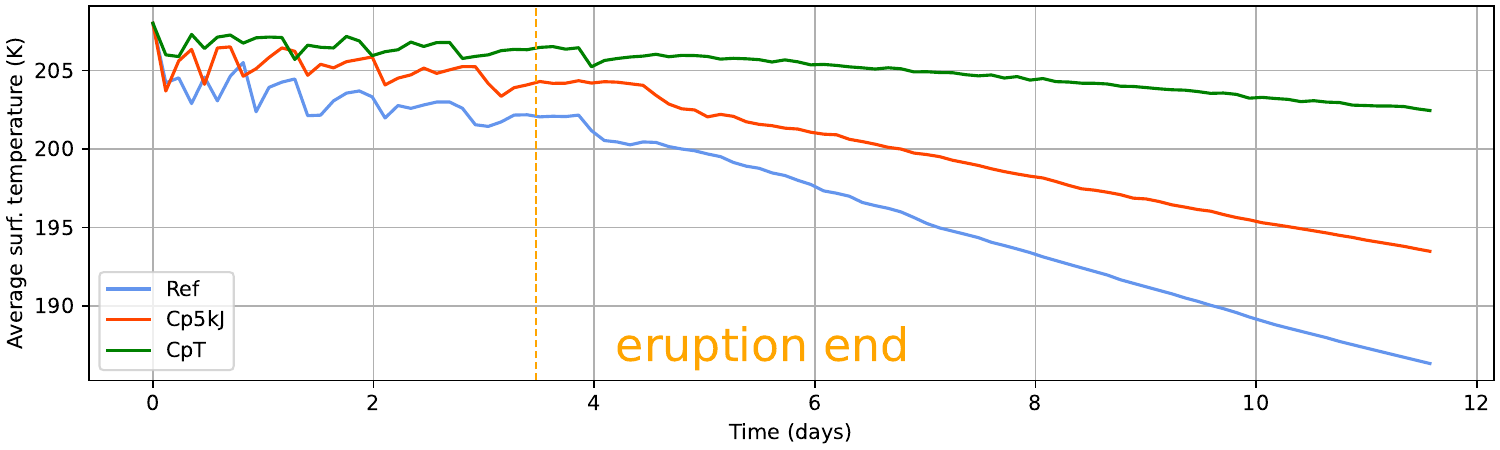}

\caption[]{\label{compa_Tav_Ref-C}Comparison of the temporal evolution of the mean flow temperature between the reference simulation 
          (``Ref''), for which the heat capacity is assumed constant and equal to $2.5$~kJ~K$^{-1}$~kg$^{-1}$, and a simulation (``Cp5kJ'') 
          in which the heat capacity is also assumed constant but set to $5.0$~kJ~K$^{-1}$~kg$^{-1}$. We also performed an additional simulation 
          (``CpT'') in which the heat capacity was allowed to vary with temperature according to a $C_p(T)$ relationship derived from the 
          liquidus curve displayed in Fig.~\ref{H2OCH3OH_bindiag}. The results reveal an even stronger delay in cooling than in the 
          previous cases.}
\end{center}
\end{figure}

      As a first step, we performed a simulation in which the heat capacity of the cryofluid was doubled, from $2.5$~kJ~K$^{-1}$~kg$^{-1}$ to 
$5.0$~kJ~K$^{-1}$~kg$^{-1}$, while remaining constant throughout the simulation. Figure~\ref{compa_Tav_Ref-C} compares the temporal evolution 
of the mean temperature of the reference flow (``Ref'') with that obtained for the simulation using $C_p = 5.0$~kJ~K$^{-1}$~kg$^{-1}$. As expected, 
the increase in $C_p$ leads to a delay in cooling, although the resulting temperature difference remains limited to only a few degrees.\\
%
%

We further extended our investigation by incorporating a temperature-dependent heat-capacity law. This relationship was derived from the liquidus 
curve of the phase diagram shown in Fig.~\ref{H2OCH3OH_bindiag}, using Eqs.~\ref{dfsdT} and \ref{fs}, which allow the calculation of $C_{p,{\rm eff}}(T)$. 
The resulting effective heat capacity was then directly implemented in the model, employing the linearized version of $C_{p, {\rm corr}}$ shown
in Fig.~\ref{fs_Cpcorr}-b. The results, presented in Fig.~\ref{compa_Tav_Ref-C}, show that the cooling delay is even more pronounced in this case, 
reaching approximately $16$~K after $11.6$~days relative to the reference simulation.

    A fluid that remains at a relatively high temperature for a longer period of time also remains less viscous than one whose temperature 
decreases more rapidly. One therefore expects a greater spatial extent when the $C_p(T)$ law is included. At the end of the simulation, 
{\it i.e.} after $11.6$~days, the reference case (``Ref'') exhibits a longitudinal extent of $\Delta x = 68$~km and a lateral extent of 
$\Delta y = 12.5$~km, with a mean thickness of $\bar{h} = 1.39$~m. When the above-mentioned $C_p(T)$ law is incorporated, the corresponding values 
become $\Delta x = 101$~km, $\Delta y = 11.5$~km, and $\bar{h} = 1.00$~m. This numerical experiment clearly demonstrates the potentially important 
role of the thermal behaviour of the cryofluid heat capacity. Consequently, determining $C_p(T)$ through laboratory measurements is just as important 
as characterizing the rheological properties of the fluid.

   It should be noted that the effect highlighted here is primarily driven by the crystallization of water and is therefore particularly 
relevant to aqueous binary mixtures. Nevertheless, other types of mixtures, including those containing solid organic particles or gas bubbles, 
may also exhibit significant temperature-dependent variations in heat capacity, potentially leading to similar effects on flow evolution.

\section{Influence of the terrain topography}
\label{subs}

      Up to this point, the modeled cryofluid flow has been simulated over a domain consisting of a planar inclined surface, assumed to be
perfectly smooth at the adopted spatial resolution. The rationale for this idealization is to isolate the factors that may influence 
the final horizontal extent and thickness of the flow. The interactions between the substrate and the 
flow itself can be categorized into two types:
(1) interactions governed purely by fluid mechanics, resulting from the local topography, or micro-topography; and
(2) interactions involving the exchange of matter or heat between the substrate and the fluid.
Both types of interactions can significantly alter the final state of the liquid--either by directly modifying its trajectory or by affecting 
its rheological properties, which in turn influence the flow dynamics.
In the present study, we focus exclusively on interactions of topographic origin. Specifically, we investigate 
(i) the contribution of the slope of the inclined plane,   
(ii) the impact of terrain ``rugosity'', and (iii) the impact of recessed or protruding obstacles with characteristic 
dimensions comparable to the flow scale itself.
   In the absence of a digital elevation model representative of the substrate over which the cryofluid flows, we therefore adopted an approach in
which the effects of simple topographic features are isolated. The objective is to identify general trends in the flow behaviour, rather than 
to reproduce specific morphological features of Titan's surface in detail.\\

    Regarding the influence of slope, we considered two values: one half of the reference value ({\it i.e.}, $0.15^\circ$, simulation
``sl0p15'') and one twice as large ({\it i.e.}, $0.60^\circ$, simulation ``sl0p60''). The corresponding results, expressed in terms of the spatial 
extent of the lava flow, are shown in Fig.~\ref{compa_phys_param}. As expected, increasing the slope angle enhances the longitudinal extent of 
the cryofluid flow, whereas lateral variations remain relatively modest. For reference, our baseline simulation (``Ref'' in Fig.~\ref{compa_phys_param}),
which assumes a slope of $0.3^\circ$, produced a longitudinal extent of $68$~km and an average thickness of $1.39$~m. Consistent with mass conservation, 
the final average flow thickness decreases from $1.85$~m for the lowest slope ($0.15^\circ$, sl0p15'') to $0.96$~m for the highest slope ($0.60^\circ$, 
``sl0p60''). The resulting longitudinal extents were $51$~km for the smaller slope and $92$~km for the larger one, respectively.\\

   In our context, the terrain roughness has to be understood as topography at scales smaller than the adopted spatial resolution.
The existence of such roughness is supported by observational evidence. For example, the study of RADAR backscattering in the Selk crater 
region revealed high-amplitude scattering, which could be explained by several surface properties, among which is the presence of roughness with 
a scale ranging from centimeters to kilometers \citep{bonnefoy_etal_2022}. We acknowledge that surface roughness can vary substantially between 
locations and may arise from different processes in the surroundings of impact and cryovolcanic craters.\\

\begin{figure}[htbp]
\begin{center}
\includegraphics[angle=0, width=9 cm]{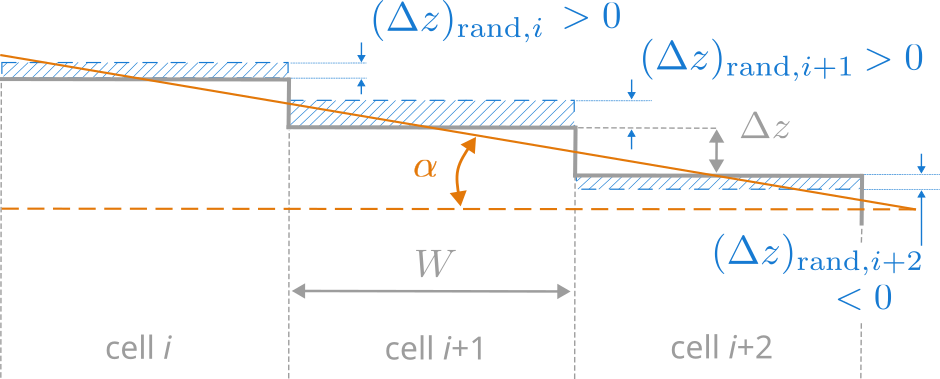}

\caption[]{\label{slope_rugo_def}Schematic view of a vertical cross-section of the inclined plane, with a slope angle $\alpha$ (in orange) 
           relative to the horizontal. This slope defines the altitude difference $\Delta z$ (in grey) between the surfaces of adjacent 
           cells. To simulate surface ``roughness'', each cell $i$ is assigned a random elevation component, 
           denoted ($\Delta z$)$_{\mathrm{rand}, i}$ (in blue), which could be positive or negative.}
\end{center}
\end{figure}

\begin{figure}[htbp]
\begin{center}
\includegraphics[angle=0, width=9 cm]{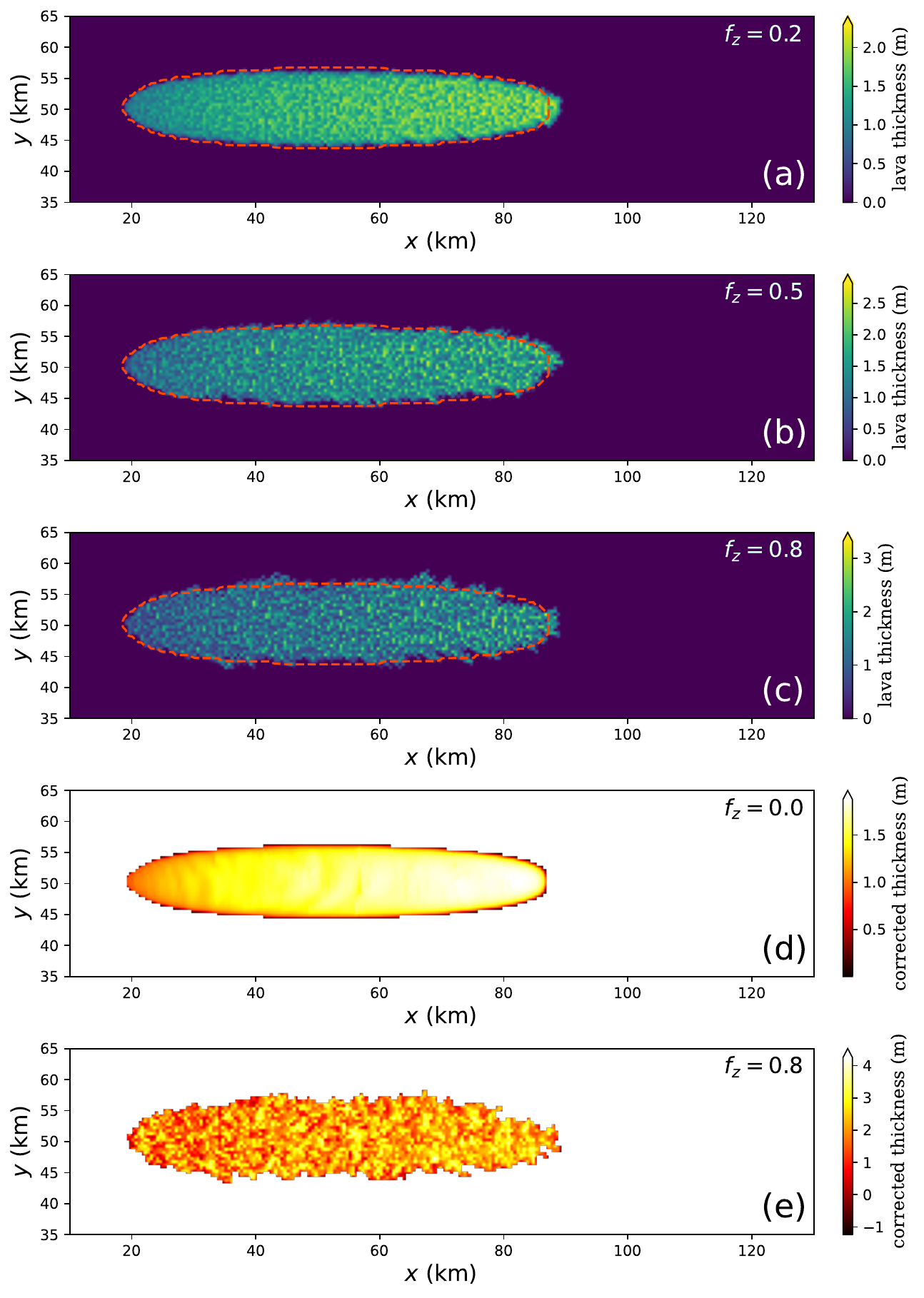}

\caption[]{\label{slope_rugo_influ}On panels (a), (b), and (c), we present the map of cryofluid flow thickness
for three values of the roughness coefficient $f_z$: $0.2$, $0.5$, and $0.8$,
corresponding to the final state of the flow. The red dashed line represents the outline
of the reference flow, in which the terrain is assumed to be perfectly smooth. Panels (d)
and (e) compare the surface morphology of the reference flow (panel d) with that
of the flow accounting for the maximum roughness $0.8$. In these last two panels, we used
a corrected lava thickness, as defined in the main text.}
\end{center}
\end{figure}

In the cellular automata framework, the inclined plane is approximated by cells of width $W$, arranged in a ``stair-step'' 
configuration (see Fig.~\ref{slope_rugo_def}). The slope angle $\alpha$ with respect to the horizontal defines the elevation difference 
$\Delta z$ between adjacent cells. For a perfectly smooth inclined plane, $\Delta z$ is uniform across the grid. To simulate surface 
roughness at the spatial scale $W$, a random elevation component ($\Delta z$)$_{\mathrm{rand}, i}$ is added to each cell 
(see Fig.~\ref{slope_rugo_def}). In order to preserve the average slope, the magnitude of the random perturbation is bounded by a fraction 
of $\Delta z$, through the parameter $f_z$ such that $\max($$|\Delta z|$$_{\mathrm{rand}, i}) = f_z \, \Delta z$. The surface elevation of each 
cell is thus perturbed by a value randomly drawn from a uniform distribution in the range $[-D_m, +D_m]$, where $D_m = f_z \, \Delta z$.\\
%
%
 In addition to $f_z=0.0$ corresponding to the reference simulation (``Ref''), we have considered three values of $f_z$: $0.2$,
$0.5$ and $0.8$.
    The results are shown in Fig.~\ref{slope_rugo_influ}, where it can be observed that the higher the roughness parameter $f_z$, 
the more the cryofluid tends to spread; however, the effect appears to be relatively weak.
Beyond these aspects, it is noteworthy that dendritic structures 
emerge along the flow margins, a feature that becomes increasingly pronounced with higher surface roughness.
It is also instructive to assess the effect of terrain roughness on the surface state of the final flow: does the 
lava surface reproduce the roughness of the underlying terrain? To this end, we introduce a ``corrected thickness''
$h_{\rm corr}$ of cryofluid, defined as:

\begin{equation}
h_{\rm corr} = (h+z) - z_{\rm smooth}
\end{equation}

where $h$ and $z$ denote, respectively, the lava thickness and the terrain elevation within a given cell, while $z_{\rm smooth}$ 
represents the elevation of the corresponding perfectly smooth inclined plane ({\it i.e.} corresponding to $f_z=0.0$) in the same cell.
With this definition of $h_{\rm corr}$, if the cryofluid tended to smooth out the ``asperities'' of the terrain, approaching a situation
similar to that of a flow over a smooth inclined plane, the values of $h_{\rm corr}$ would be expected to be similar regardless of the surface 
roughness considered.
Figures~\ref{slope_rugo_influ}-(d) and \ref{slope_rugo_influ}-(e) show $h_{\rm corr}$ for two extreme cases: the reference simulation ($f_z=0.0$) 
and the case with the highest roughness ($f_z=0.8$). 

Notably, terrain roughness induces a corresponding roughness in the lava flow surface. 
Computing the standard deviations of $h_{\rm corr}$ confirms this behavior: for the reference simulation, the value is $0.428$~m, whereas 
the high-roughness simulation ($f_z=0.8$) yields $0.885$~m. 
The non-zero standard deviation obtained for the reference simulation corresponds to variations in the value of $h_{\rm corr}$ from one cell to 
another due to the mean terrain slope, here $0.3^{\rm o}$. Since the mean slopes in the reference simulation and in the one with $f_z = 0.8$ are the same, 
the increase in the computed standard deviation indeed corresponds to a quantification of the roughness of the cryofluid surface.
However, if one computes the standard deviation of the terrain elevation within the area covered by the flow footprint, one obtains $1.198$~m, a 
value higher than $0.885$~m. One can therefore conclude that the terrain roughness induces a roughness in the lava surface, although the latter remains
smoother than the terrain it covers.
   It is quite clear that the rheology of the cryofluid has a direct and significant influence on the smoothing of the underlying terrain 
roughness. Indeed, in the limiting case of a low-viscosity Newtonian fluid, the fluid would conform closely to the terrain, following the shape 
of its surface irregularities, and its ability to smooth the underlying topography would therefore be negligible. Conversely, a highly viscous fluid 
flows less readily and may more effectively smooth out surface irregularities.
More indirectly, parameters that affect the rheological properties of the cryofluid will also influence its smoothing capacity. Temperature is 
perhaps the clearest example of this effect: higher temperatures generally correspond to lower viscosities and therefore to a weaker smoothing effect, 
whereas lower temperatures lead to higher viscosities and, consequently, more pronounced smoothing.
In addition, processes unrelated to the intrinsic properties of the cryofluid may also affect the final surface morphology. For example, during flow, 
the cryofluid may entrain sedimentary blocks which, if buoyant, could contribute to the development of surface roughness. At smaller scales, the 
solidification of the cryofluid, particularly if it consists of a complex mixture, may lead to the formation of different types of crystals, which 
can likewise generate surface roughness.
     The surface morphology of any potential cryofluid flows could be investigated at small scales using the optical instruments aboard {\it Dragonfly}, 
in particular the {\it DragonCam} microscopic camera \citep{clark_etal_2025}. At larger scales, the topography of the terrains encountered could be 
mapped using the navigation cameras and, most likely, the LIDAR system \citep{lorenz_etal_2018b}.\\

One could therefore potentially expect cryofluid flows to decrease the apparent roughness of the surface, which could be observable with a
RADAR such as the one carried by CASSINI. Perhaps paradoxically, \cite{bonnefoy_etal_2022} observed, using CASSINI, an increased RADAR 
backscatter on the slopes of the Selk crater possibly covered by ancient cryofluid deposits; however, this effect could have many
causes other than scattering strictly from the surface. 
Around Selk, the observed effect may be unrelated to cryofluid and instead correspond to a simple ejecta deposit; moreover,
since RADAR waves penetrate below the surface up to few times the wavelength 
({\it i.e.} few centimenters to few tens of centimeters), one can readily envision a volume-scattering effect produced 
by internal heterogeneities \citep{janssen_etal_2011} within the cryofluid, or simply by the underlying terrain.
The scattering heterogeneities could originate from centimetre-sized sediments entrained by the flow during its propagation or already present
at the time of emergence from the vent. Gaseous inclusions could also develop as a result of the exsolution of previously dissolved gases. 
On Titan, centimetre-sized bubbles may form more readily than on Earth \citep{cordier_etal_2017a,cordier_ligerbelair_2018,cordier_etal_2024}.\\

%
%
   We now turn to the question of partially guided flow.
A lava flow can experience the effect of large-scale topographic singularities, {\it i.e.}, features whose size is comparable
to that of the flow itself. Nevertheless, we propose two cases that are as idealized as they are extreme: one corresponding to
an excavation of the terrain, while in the other the flow is stopped by a physical barrier.
In the first case, we placed along the flow path a basin of
$40 \times 6$~km, with a depth of $1$~m, carved into the initial inclined plane. In the second case, we placed along the free trajectory of 
the cryolava an $30$~m high wall, $40$~km long in the direction perpendicular to the axis of the flow.
%
\begin{figure}[htbp]
\begin{center}
\includegraphics[angle=0, width=9 cm]{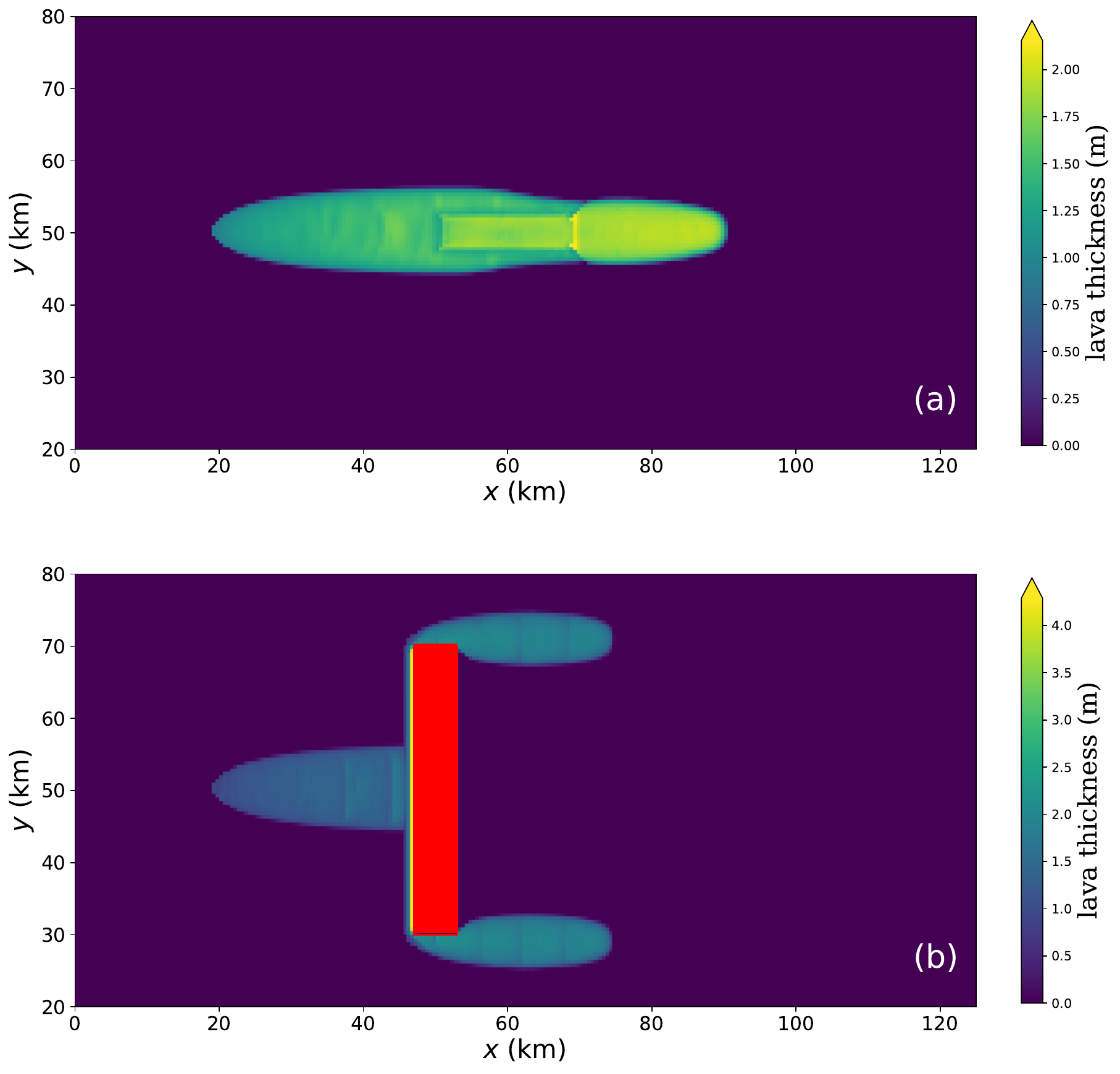}

\caption[]{\label{chanWall}(a) An example of the behavior of a cryolava flow encountering a depression measuring $20$x$6$~km and $1$~m deep 
           along its path. (b) Another example in which an impassable wall ($30$~m high and $40$~km long) is erected across the fluid’s trajectory.
           This wall is shown in red in this panel.}
\end{center}
\end{figure}
%

The results are shown in Fig.~\ref{chanWall}. In the first case, the channelling of part of the material flow led to a slight increase in the 
maximum extent of the flow, which then reaches $71.5$~km instead of $68$~km for the reference simulation, while the width remains unchanged. 
A narrowing of the flow can be noted around $x \sim 70$~km, as well as an accumulation of cryolava at the right end--{\it i.e.}, also the lower en--of 
the channel. The wall induces a substantial decrease in the extent along the $x$-axis, due to the division of the flow into two ``daughter flows``,
the amount of material available for each of them being obviously less than that mobilized for the unobstructed flow. Against the upstream face of the wall, 
a strong accumulation of material is visible, reaching a thickness of $\sim 4$~m, while the rest of the flow has a thickness on the order of one meter, 
not very different from the average thickness of the reference flow. These results, which could be expected, show the sensitivity of a cryolava flow 
to the presence of significant obstacles. They also highlight the need for a relatively realistic digital terrain model if one wishes to perform 
simulations that come closer to observations.\\

  On a realistic digital elevation model, the effects discussed above would combine in a more or less complex manner, depending on the topological
characteristics of the terrain. Of course, additional processes may also come into play, in particular those neglected in this study, such as mass exchange 
between the substrate and the flow.

\section{Discussion}
\label{discu}
  The rheological behavior of the model liquid considered in this study, a methanol–water mixture, has been
characterized over a limited temperature range, with temperatures below $169$~K and above $208$~K not investigated. 
In addition, the two measurements obtained at the lowest temperatures are associated with significant uncertainties. 
An extension of the investigated temperature range is therefore required, particularly toward higher temperatures, where prebiotic 
chemical activity is expected to be more favorable. Although aqueous solutions containing methanol or ammonia, as well as 
saline solutions, are of considerable relevance for icy satellites, the specific case of Titan allows for the consideration 
of a wider range of possible liquid compositions. Indeed, as already mentioned, during their flow over Titan's surface, cryofluid may become loaded
with organic sediments produced by atmospheric photochemistry. These materials—likely composed of particles spanning a wide range 
of sizes, from molecular scales to centimeters or larger—may either dissolve in water or remain in suspension and be transported 
by the flowing fluid. Such a mixture would necessarily lead to a substantial modification of the cryofluid's rheological properties.
Therefore, any experimental characterization of the rheological properties of such complex fluids is of great interest.
%
   In the present study, we have only considered a simplified representation of the effect of composition on rheology, by varying the Bingham 
yield stress $\tau_0$ and the dynamic viscosity $\eta$.
 Our work, although it includes thermal losses at the surface of cryofluid flows, does not account for the possible
formation of a crust or the details of vertical transport processes occurring within the fluid layer. 
Depending on the exact chemical composition of the lava, the formation of a solid phase at the surface 
may lead to the development of a crust or, for example in aqueous ammonia solutions \citep{kargel_1995}, 
to the formation of crystals denser than the liquid phase. These crystals are therefore distributed 
throughout the cryofluid volume, producing an ice slurry and leading to an evolution of its rheological
properties. For compositions in which a crust forms, this crust exerts a mechanical effect over the entire flow, 
tending to slow the flow, which must then apply sufficient stress to break the crust. In general, 
rheology describes the redistribution of momentum within the fluid volume. In the cellular automata–based 
approach, the adopted rheological laws represent, in an implicit and highly macroscopic manner, 
the processes occurring at smaller scales across the thickness of the flow. In sum, laboratory experiments
on candidate cryofluids, and complex mixtures of fluids and solid phases, possibly organic, are still very important to determine their rheological properties.\\

 In our approach, we have neglected thermal and mechanical interactions between the cryofluid and the underlying substrate.
Vertical transport mechanisms and crust formation could be explicitly addressed using a multilayer cellular automaton approach 
\citep[\textit{e.g.}][]{hidaka_etal_2005} or other numerical paradigme like SPH\footnote{Smoothed-particle hydrodynamics} method 
\citep[see \textit{e.g.}][]{bodin_etal_2026}, which could be adapted to incorporate sediment entrainment by the advancing flow or 
thermal erosion leading to partial melting of the substrate \citep[see for instance][and references therein]{gallant_etal_2020}. 
However, if the flow is laminar as shown by the estimated Reynolds number and the velocity of the fluid is small (see Sec.~\ref{model}),
the mechanical erosion of the substrate, perhaps consisting of a layer of organic material could be limited. In a similar vein, 
thermal erosion could be unefficient due to the drop of the temperature at the interface 
between the erupting fluid and the substrate \citep{davies_etal_2016}.\\


   In the scenarios we explored, the complete cooling time of a cryofluid flow, {\it i.e.} its thermal equilibration with the environment,
is longer than the flow duration. For our model of reference, because of the relatively modest thickness of the deposit, the cooling time is on the order of a few
tens of days.
%
   However, duration up to a few months depending of the adopted scenario, may be expected. This is the case for instance if we account for the 
effective heat capacity $C_{P, {\rm eff}}$ that
can significantly polong the cooling time. Unlike the case of Enceladus, there are currently no direct observations on Titan of the temporal evolution
of cryovolcanic activity. However, indirectly, after correcting VIMS infrared data for atmospheric effects, \cite{solomonidou_etal_2016} suggested 
that the spectral albedo of the Sotra Patera region may have increased between April 2005 and February 2006. If this observation is real and is related 
to cryovolcanic activity, it would impose a temporal constraint on the order of one year. This is fully compatible with our simulations, as a sequence 
of flow and cooling episodes, each lasting several months, could readily extend over a timescale of about one year.
In any case, the thermal timescales obtained in our simulations are consistent with those suggested by \cite{solomonidou_etal_2016}, and
are much shorter than those associated with impact melt pools. The latter 
%
    fully freeze after $10^2$ to $10^4$ years \citep{artemieva_lunine_2003,artemieva_lunine_2005,obrien_etal_2005,hedgepeth_etal_2022,wakita_etal_2023}
due to their initial depths reaching several hundred meters. Cryolava flows, however, have the advantage of being able to cover larger surface areas than
craters and having depths that are more accessible to {\it in situ} analysis.
It is very likely that cryofluid encounters organic material produced by atmospheric photochemistry, or transported from the interior
\citep{cordier_etal_2024,goncalves_etal_2025}, along its path. One of the most important 
questions concerning Titan’s environment is that of the interaction between liquid water, or any aqueous solution, and the ubiquitous organic 
matter present in this context. Such interactions could constitute the precursors of prebiotic chemistry. Water indeed has the ability to produce 
new chemical species through hydrolysis of the material it comes into contact with. The higher the temperature, the more this alteration is favored; 
the same holds for long contact durations.
%
  We emphasize that the evolution of the cryofluid composition during the flow is consistent with the evolution of the shape of the spectral signature 
of the Sotra Patera region suggested by \cite{solomonidou_etal_2016}.\\
%
   Regarding the important astrobiological aspect of Titan, several laboratory experiments have studied the hydrolysis products of 
tholins and possible molecular sediment—analogues of Titan’s organic matter—after exposure 
on the order of a year \citep{neish_etal_2010, cleaves_etal_2014}, while some studies have performed chemical analyses after only a few days or weeks
\citep{raulin_etal_2007, ramirez_etal_2010, poch_etal_2012, taniuchi_etal_2013, brasse_etal_2017}. Recently, \cite{farnsworth_etal_2024} showed, 
for example, the production of glycine and alanine after a few weeks, through alkaline hydrolysis of aminonitriles. 
All these experiments were carried out near room temperature,
{\it i.e.} between $253$~K for the lowest reported value \citep{neish_etal_2010,ramirez_etal_2010,poch_etal_2012,brasse_etal_2017,farnsworth_etal_2024} 
and $383$~K \citep{taniuchi_etal_2013}. One exception has to be noted: \cite{ramirez_etal_2010}, who performed measurements at $96$~K.
The ``reference'' eruption temperature adopted here corresponds to the highest  value considered in \cite{zhong_etal_2009}, namely $208$~K,
which is well below the temperatures investigated in the hydrolysis experiments. Temperature is a key parameter in chemical reaction kinetics, 
most of which follow an Arrhenius-type law, whereby rate constants vary exponentially with temperature.\\

At temperatures higher than the maximum value considered by \cite{zhong_etal_2009}, one can expect the fluid to become progressively less viscous 
as temperature increases, potentially leading to the disappearance of its non-Newtonian behavior. This could place the system outside the domain of
applicability of the cellular automaton method. Nevertheless, the encounter between a cryofluid flow, initially considered as an aqueous solution,
and the organic matter sedimented at the surface may, through mixing, yield a fluid much more viscous than water, and possibly even non-Newtonian. 
Indeed, soluble organic matter, as well as insoluble matter remaining as solid particles, can induce a very significant change in the rheological 
properties of the resulting complex fluid. Such behavior is not uncommon; one may cite, for instance, beverages of the ``milkshake'' type
\citep{dogaru_etal_2014}.
To mimic the flow of this type of fluid on Titan’s surface, we considered a fluid that retains the maximal Bingham parameters of 
\cite{zhong_etal_2009} at an ambient temperature, {\it i.e.} $300$~K, to which we added a $0.1$~m/s wind for increased realism. 
Regarding rheology, for the sake of simplification we adopted a constant rheology, for which we chose $\tau_0 = 15$~Pa and $\eta = 30$~Pa~s. 
These values correspond to the maxima obtained by \cite{zhong_etal_2009}, even though, under the assumption of an ``aqueous/organic matter'' 
mixture, the fluid composition would be very different. 
To fix ideas, these Bingham parameter values are comparable to these typically encountered for ketchup or honey \citep{steffe_1996}.
High values of the rheological parameters slow down cooling by reducing the surface area in contact with the air. Here, our goal is to evaluate how long 
a given volume of cryofluid remains near ambient temperature, at which the majority of hydrolysis experiments have been conducted.
    With this simulation setup, the average temperature of the cryofluid remained above $250$~K (the temperature considered in four papers
previously mentioned) for $10$~days, while the maximum temperature stayed above this value for $13$~days.
   Most hydrolysis experiments at $250$~K were conducted over durations on the order of $10$~weeks, in alkaline environments. Although the 
flow simulation proposed here shows temperatures above $250$~K for $1$--$2$~weeks, hydrolysis products should be detectable, albeit in small 
quantities. Indeed, Fig.~4 of \cite{farnsworth_etal_2024} suggests that aniline could be detected after such a short duration and at such
a low temperature.
   Of course, the simulations reported in this paragraph, do not account for the behavior of $C_{P, {\rm eff}}$, which may result in conditions that are even more
favorable for chemical reactions. \\

\begin{figure}[htbp]
\begin{center}
\includegraphics[angle=0, width=9 cm]{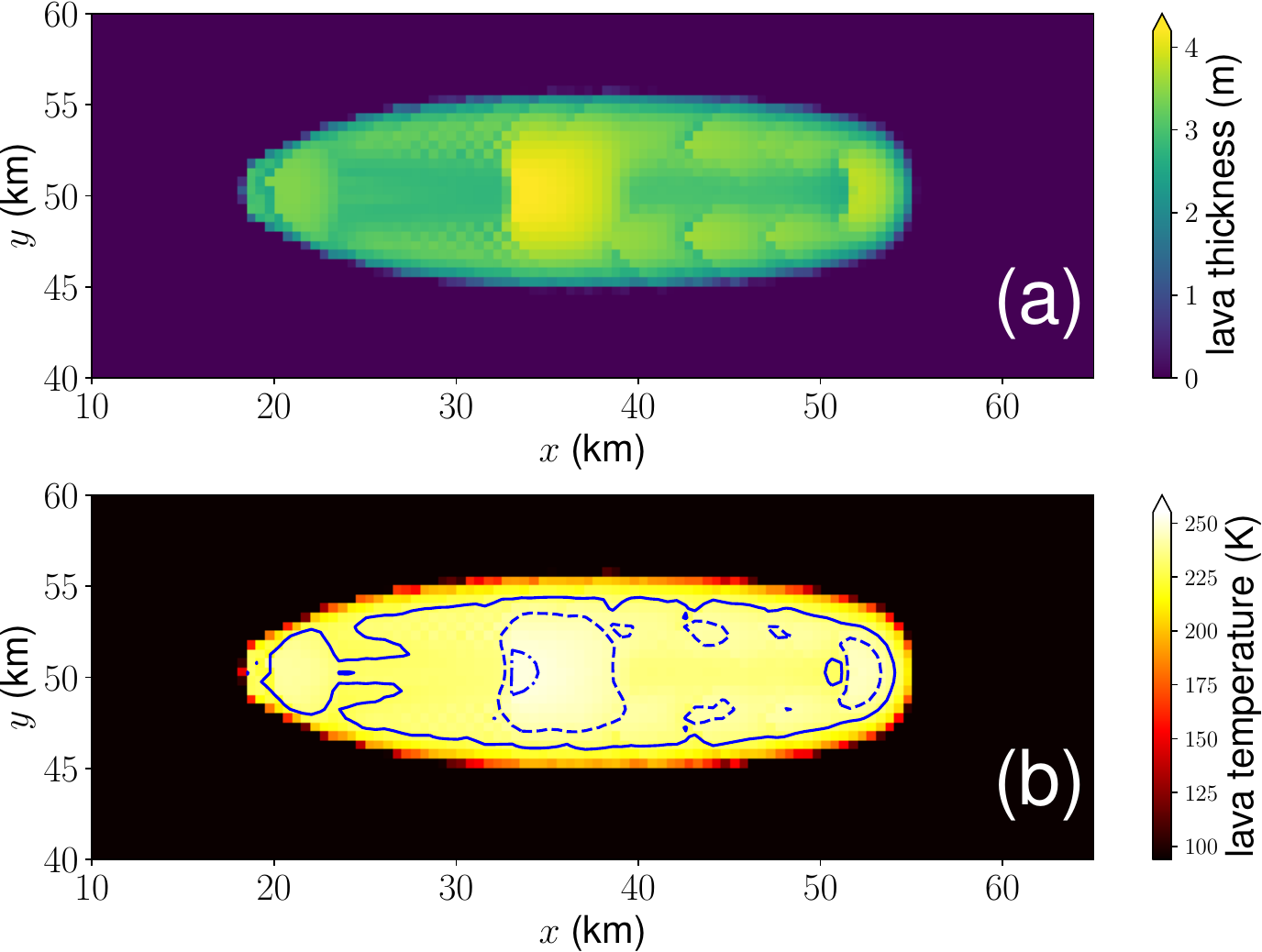}

\caption[]{\label{hot_and_viscous}Cryolava flow simulation for a hot ($T_{\rm eruption}= 350$~K), and high yield stress 
           fluid ($\tau_0= 20$~Pa) and $eta= 30$~Pa~s, kept contant. This snapshot corresponds to $t= 21.6$~days.
           (a) The map of lava thickness. (b) The map of cryofluid temperature, with countour at $T= 230$, $240$ and $249$~K (respectively
           solid, dashed and dashdot line).}
\end{center}
\end{figure}

   The initial temperature of the effused cryofluid is a function of the
geophysical mechanisms that might give rise to cryovolcanic eruptions, and which are currently poorly constrained. Any
initial temperature assumption remains speculative. Nonetheless, in the specific case of a fluid derived from an impact-generated lava lake, 
initial temperatures on the order of $350$~K may not be unreallistic.
Indeed, temperatures up to $500$~K can be inferred from Fig.~6 of \cite{artemieva_lunine_2003}. In addition, Fig.~3 of \cite{artemieva_lunine_2005}
shows temperatures above $273$~K within the impact melt. In the introduction of \cite{lorenz_etal_2021}, which discusses the rationale
for selecting Selk crater as the landing site for the {\it Dragonfly} mission, temperatures higher than $273$~K are considered, based on
\cite{neish_etal_2018}.
      Then, we revisited our latter simulation and set the eruption temperature to $350$~K, the mean flow temperature remains above $250$~K for $13.5$~days,
whereas the maximum temperature stays above this threshold for $19.5$~days, {\it i.e.} about $3$~weeks. The effect can be further enhanced by increasing
$\tau_0$ from $15$ to $20$~Pa, which results in a mean temperature above $250$~K for $14.8$~days and a maximum temperature above $250$~K for $21.5$~days.
As already shown, rheology plays a key role in the cooling of a cryofluid flow: a more viscous fluid tends to form a thicker flow, thereby reducing the
surface area in contact with the atmosphere. Figure~\ref{hot_and_viscous} presents the thickness and temperature maps of the simulation corresponding
to $\tau_0 = 20$~Pa and $T_{\rm erup} = 350$~K at $t = 21.6$~days. The temperature is clearly correlated with the cryofluid thickness.
Determining the rheological properties of cryofluid is therefore a crucial aspect in the search for extraterrestrial prebiotic activity.\\

 In our analysis of thermal properties, we did not consider certain very specific, but plausible, circumstances that could nevertheless
significantly increase the cooling timescale. One such example is the formation of a foam layer above the cryofluid, produced by the degassing of
a gas initially dissolved in the liquid ({\it e.g}., methane) and by the presence of surfactant molecules supplied by the encountered organic material. 
Because foam has a thermal conductivity close to that of air, our estimates indicate that its conductivity could be on the order of 40 times lower 
than that of liquid water, leading to a comparable increase in the cryofluid cooling timescale.\\


   Regarding the thermo-mechanical coupling between a cryofluid flow and its immediate environment, we have already discussed its interaction 
with the ground, for example through possible thermal erosion. We have also investigated the influence of wind on the flow behavior: 
as expected, wind enables very efficient cooling, leading to a fast temperature drop, which in turn results in an 
increase in the fluid viscosity and in the thickness of the final flow.
    On the other hand, the occurrence of an eruption could also trigger convection currents 
in the atmosphere in contact with the cryofluid flow, in turn generating an accelerated cooling effect. This type of effect could be
modeled using a ``mesoscale'' atmospheric model, as has already been done for Titan's atmosphere above hydrocarbon lakes
\citep{chatain_etal_2022,chatain_etal_2024}.

\section{Conclusion}
\label{concl}

   By considering laboratory-characterized fluids as analogues for cryofluids, we show that our cellular-automaton-based
model can produce, to first order, cryofluid flows with spatial extents of several tens of kilometers, depending on the mobilized fluid volume.
The mean deposit thickness remains on the order of one meter, assuming a single eruptive episode. 
This value seems more conservative than the inferred spatial extent.
        The main result of this study is the compatibility between the spatial extent predicted by our model and the observations. 
    The sensitivity of this spatial extent was investigated by examining the effects of both the mean terrain slope and the terrain 
``roughness'' at the scale of the spatial resolution, {\it i.e}., $500$~m. Naturally, the average terrain slope affects how far a flow extends: 
steeper slopes allow the fluid to travel farther, while gentler  slopes restrict its reach. Our findings revealed that, although cell-to-cell irregularitie 
create some unevenness on the cryofluid flow’s surface, the flow itself remains smoother than the terrain beneath it.
  We also tested the presence of large obstacles, {\it i.e.} of a size comparable to that of the flow itself, which can very significantly affect 
the trajectory of the cryofluid.
The agreement between observations and model outputs could be refined further, but this would not be of great interest because the observed extents are quite variable 
\citep{lopes_etal_2013}.
%
   Indeed, reproducing the morphology of a particular flow: Mohini Fluctus as a whole, larger structures such as the Rohe Fluctus-Winia Fluctus 
complex, or smaller ones such as Ara Fluctus (see Fig.~\ref{topoMohiniFluctus}), or any other feature identified in the literature 
\citep{lopes_etal_2010,lopes_etal_2013}, would be of limited value given the insufficient observational constraints currently available. 
Indeed, as we have shown, the morphology of the simulated flows depends on numerous parameters: the total extruded volume, the rheological and 
thermal properties of the cryofluid, the terrain properties (average slope, presence of obstacles, roughness at several scales, etc.), the wind 
regime above the flow, and the interaction with soil material that could be mobilized by the flow. For most of these parameters, observational 
constraints are either weak or entirely lacking for the moment. Given the resulting freedom in the choice of these parameters, it would therefore always be 
possible to reproduce a given morphology using several different sets of parameters, meaning that the solutions would not be unique.
The non-uniqueness of the solutions consequently limits the extent to which the simulation can provide meaningful insight.\\

      However, the diversity of the observed morphologies can itself be regarded as valuable information, as it implies variability in the local
conditions governing cryofluid flows. These parameters, which may vary from one location to another, are readily identifiable 
and can be grouped into four categories :
(1) the parameters controlling the total volume of extruded cryofluid and the duration of the eruption (as well as the number of
    eruptions in the same region, which may lead to interactions between successive flows); 
(2) the parameters related to the local topography: the mean terrain slope; the number, location, and morphology of potential obstacles;
    and any surface ``roughness'',
(3) the nature of the terrain, which may interact with the cryofluid in various ways, for example through thermal erosion or 
    mixing with organic sediments, thereby modifying the rheological or chemical properties of the cryofluid;
(4) the parameters related to atmospheric activity: the presence and strength of winds, and possible liquid methane precipitation.
    Conversely, one may expect the physical properties of the cryofluid to be less variable at the vent, particularly if the physical 
processes giving rise to the eruptions are similar.\\
%
%
%
%
    Finally, one of the main limitations of the numerical method employed in this work is the absence of an explicit
treatment of vertical heat transport across the thickness of the flow.
  Similarly to what we have already suggested for the vertical transport of material,
a more detailed investigation of this thermal behavior is needed, and would require a more sophisticated model,
at the expense of increased computational cost since it requires a 3D approach.\\

   To briefly summarize this work, we have established a connection between laboratory experiments on cryolava/impact-melt analogs, showing that,
within a reasonable scenario, the spatial extent of the observed flows can be approximately reproduced. On the other hand, from an astrobiological
perspective, we have shown that the thermal behavior of the simulated flows allows for chemical alteration of the organic sediments that could be
encountered by the flow.\\


   Many of the uncertainties identified in this study may be resolved through the in situ exploration that {\it Dragonfly}
is expected to undertake in the mid-2030s.
  This instrumented octocopter may indeed help characterize possible impact-melt flows in the Selk crater region in two main ways:
(1) through imaging and (2) through the analysis of the chemical composition of surface materials. The Dragonfly Camera Suite (DragonCam)
\citep{barnes_etal_2021} will include two panoramic cameras capable of imaging the large-scale morphology of the terrains encountered. 
Wide-angle cameras will provide imaging at smaller spatial scales, while two microscopic cameras will deliver information on surface textures 
at resolutions as fine as $60$~$\mu$m per pixel \citep{clark_etal_2025}.
With regard to compositional analyses, the Dragonfly Mass Spectrometer (DraMS) and the Dragonfly Gamma-Ray and Neutron Spectrometer (DraGNS) 
are the two primary instruments designed for this purpose. A drilling system will be capable of collecting samples from depths of up to 
approximately $6$~cm \citep{zacny_etal_2022}. These samples will then be analysed using one of two operating modes: 
the Laser Desorption Mass Spectrometer (LDMS) \citep{grubisic_etal_2021} and the Gas Chromatograph Mass Spectrometer (GCMS) 
\citep{stern_etal_2023}. Together, these instruments will be able to measure molecular masses up to approximately $2000$ Daltons \citep{trainer_etal_2018,barnes_etal_2021}.
  As Dragonfly makes new discoveries in the field, whether physical or chemical in nature, our model, or any derivative thereof, could provide 
a useful tool for interpreting these observations.\\

\noindent{\bf Data availability} The lava flow simulation code based on cellular automata, \texttt{CALava}, developed as part of 
this project, along with post-processing 
tools, is publicly available on GitLab server of the University of Nantes\footnote{\url{https://gitlab.univ-nantes.fr/cordier-d/calava}} 
and archived on the Zenodo platform
\url{https://doi.org/10.5281/zenodo.23038363}.
The \texttt{CALava} program is written in modern \texttt{FORTRAN}, while the post-processing tools were developed in \texttt{Python}. 
\texttt{CALava} is parallelized and is best compiled using the GNU \texttt{gfortran} compiler.\\

\noindent{\bf Acknowledgements} 
This work was supported by the Programme National de Planétologie (PNP) of CNRS-INSU, co-funded by CNES.
Computations were performed using resources from CNES, ROMEO\footnote{\url{https://romeo.univ-reims.fr}}
and CRIANN\footnote{\url{https://www.criann.fr}} HPC Centers. We warmly thank Jean-Christophe Malapert for access to
CNES computing facilities. Part of this research was carried out at the Jet Propulsion Laboratory,
California Institute of Technology, under a contract with the National Aeronautics and Space Administration (80NM0018D0004).

\appendix
\section{The algorithm of CA anisotropy correction}
\label{appAnisoCorr}

%
\begin{figure}[!h]
\begin{center}
\includegraphics[angle=0, width=9 cm]{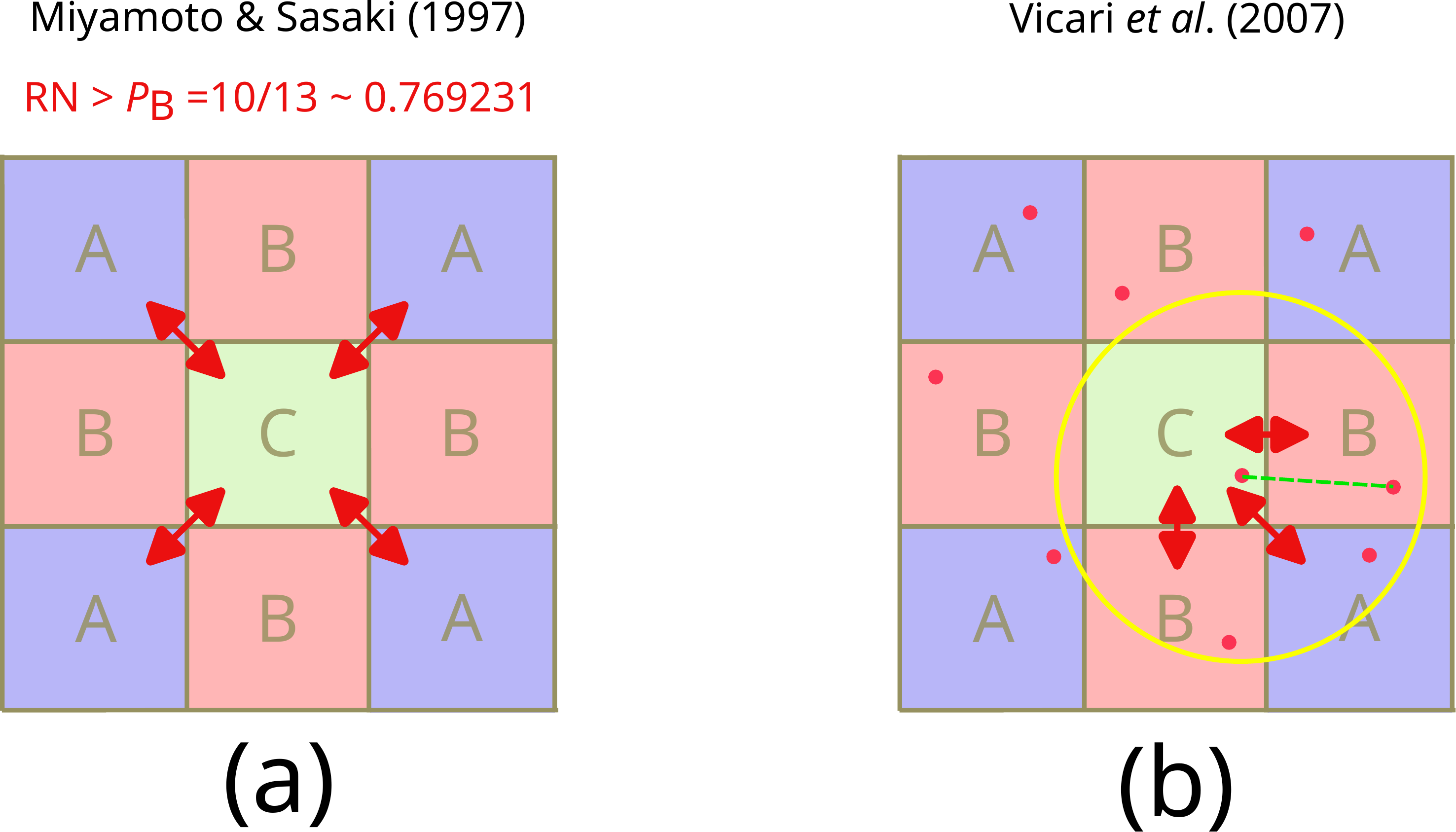}

\caption[]{\label{princip_m97_v07}(a) Schematic representation of the anisotropy-correction algorithm proposed by \cite{miyamoto_sasaki_1997}.
           When the pseudo-random number RN is greater than $P_{\rm B}$, all type-A cells can exchange mass and energy with the central cell C.
           Regardless of the value of RN, any possible fluxes between the B cells and cell C are taken into account.
           (b) Schematic representation of the anisotropy-correction algorithm proposed by \cite{vicari_etal_2007}. Before computing the fluxes, 
           a representative point, selected at random, is assigned to each cell (red points); here, only the cells whose representative point lies 
           at a distance smaller than $W$ can exchange mass with cell C. The yellow circle has radius 
           $W$, and its center is the representative point of cell C. The green dashed segment in panel (b) illustrates an example of the distance 
           between the representative points of neighboring cells.}
\end{center}
\end{figure}
%

\begin{figure*}[!t]
\centering
\adjustbox{center}{%
\includegraphics[angle=0, width=15 cm]{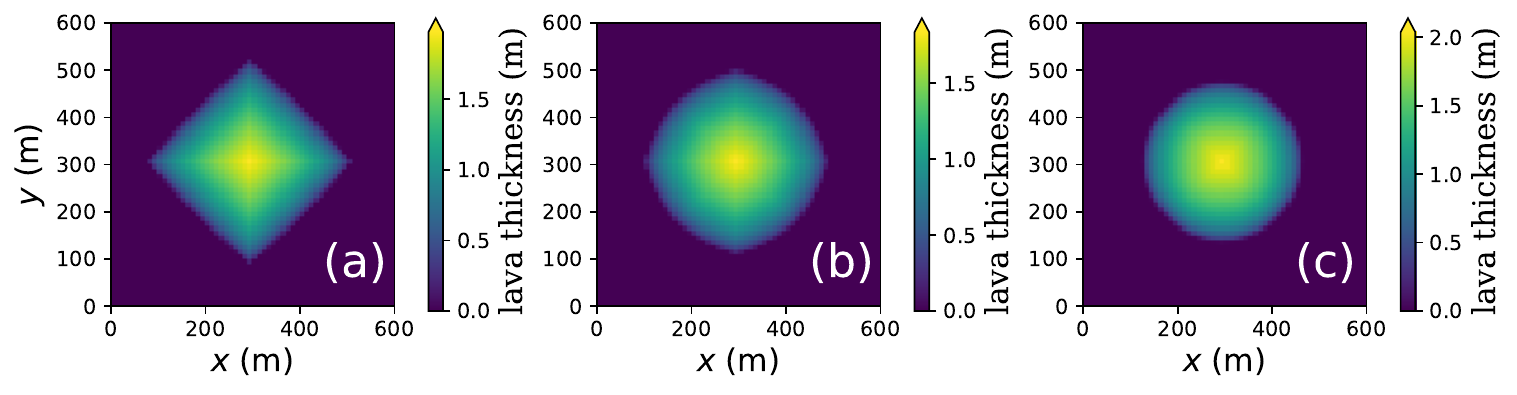}
}
\caption[]{\label{compaAnisoM97V07}(a) Terrestrial lava eruption without correcting for the anisotropy inherent in the cellular automaton method, 
           (b) the same simulation with an anisotropy correction suggested by \cite{miyamoto_sasaki_1997}, 
           (c) a third version of the same physical eruption using a correction algorithm inspired by \cite{vicari_etal_2007}. 
           The physical, geometrical, and temporal parameters are inspired by the simulations performed in \cite{miyamoto_sasaki_1997}. 
           The parameter values are recalled in the text of the appendix. These simulations were carried out 
           on a PC with a computation time of less than $5$~min.}
\end{figure*}

    As has long been noted by various authors \citep[see][]{kroc_etal_2010}, the cellular automaton method generally introduces anisotropies 
related to the grid used. This is the case for flow simulations performed on a square, flat, and horizontal grid (see Fig.~\ref{compaAnisoM97V07}-a). 
In this situation, one would expect to obtain an axisymmetric result centered on the location of the volcanic vent. To correct this issue at low 
computational cost, \cite{miyamoto_sasaki_1997} proposed a method based on a Monte Carlo algorithm. The principle of the Miyamoto~\&~Sasaki's 
method is recalled in Fig.~\ref{princip_m97_v07}-a. This method does not strictly conserve mass and energy, 
since the fluxes of mass and energy at a given time --for example, from cell $(i,j)$ to its neighbor $(i+1,j+1)$-- are not necessarily equal. 
However, in practice it has been observed that conservation laws are satisfied to within better than 1\%, which is sufficient at least for preliminary 
tests or for simulations with prohibitive computation times. 
An example of a lava flow simulation, using of the algorithm of \cite{miyamoto_sasaki_1997} is shown in Fig.~\ref{compaAnisoM97V07}-b.
Compared to the results obtained without this algorithm, as depicted in Fig.~\ref{compaAnisoM97V07}-a, the improvement is noticeable.
We have also implemented in \vb{CALava} an anisotropy-correction algorithm inspired by \cite{vicari_etal_2007}, in which at each time step, each cell is 
assigned a representative point whose position is randomly selected; only cells whose representative points are separated by a distance smaller 
than $W$ exchange mass and energy. For a given time step, this Monte Carlo-type procedure is repeated $N_{\rm mc}$ times and averaged. 
The principle of our Vicari~{\it et al.}–like algorithm is summarized in Fig.~\ref{princip_m97_v07}-b.
An example of lava flow simulation, obtained with $N_{\rm mc} = 10$ is shown in Fig.~\ref{compaAnisoM97V07}-c, which—despite the small 
value of $N_{\rm mc}$--provides a good correction, as can be seen.
The higher the value of $N_{\rm mc}$, the better the correction, but the computation time increases proportionally, so a compromise must be found.
We emphasize that the quality of the anisotropy correction also depends on the time step used, since a smaller time step implies more Monte Carlo iterations.
However, it can be noted that if one only wishes to obtain the maximum runout of a flow, even an uncorrected calculation may prove sufficient 
as a first approximation; indeed, the maximum runouts observed in Figs.~\ref{compaAnisoM97V07}-a, b, c are very similar.
The geometrical, temporal, and physical characteristics of mentioned simulations, inspired by the values used in \cite{miyamoto_sasaki_1997}, are recalled below:

\begin{itemize}
   \item geometry used in the simulation:
        \begin{itemize}
           \item dimensions of the simulation box: $600$x$600$~m
           \item number of cells in the $x$-axis direction: $60$
           \item number of cells in the $y$-axis direction: $60$
           \item lenght of a cell edge: $10$~m
           \item elevation: flat horizontal terrain.
           \item position of the vent: $(i,j)=(30,30)$\\
        \end{itemize}
        
  \item eruption parameters:
        \begin{itemize}
           \item eruption rate: $100$~m$^3$~s$^{-1}$
           \item eruption duration: $1000$~s
           \item lava temperature at the vent: $1320$~K\\
        \end{itemize}
 
  \item temporal parameters:
        \begin{itemize}
           \item simulation total duration: $5000$~s
           \item time step: $0.1$~s (kept constant).
           \item number of time records in the output file: $100$\\
        \end{itemize}

  \item physical parameters:
        \begin{itemize}
           \item gravity: $9.81$~m~s$^{-2}$
           \item lava density: $2.5 \times 10^{3}$~kg~m$^{-3}$
           \item lava specific heat $C_p = 840$~J~K$^{-1}$~kg$^{-1}$
           \item lava emissivity $\epsilon= 0.9$
           \item lava viscosity $\eta = 850$~Pa~s
           \item lava yield stress $\tau_0 = 225$~Pa
           \item cooling by black body radiations.
        \end{itemize} 
\end{itemize}

In a manner related to the treatment of the anisotropy problem, one must also correctly estimate the derivative $\partial h/\partial x$
that appears in the equation giving the critical thickness $h_{\rm c}$ (see Eq.~\ref{hc}). For terrain that is only slightly inclined
with respect to the horizontal, {\it i.e.} with $\alpha \simeq 0$, we have
\begin{equation}
h_{c, (i,j) - (k,l)} \simeq \frac{\tau_0}{\rho g} \, \frac{\Delta x}{\Delta h}
\end{equation}
where $\Delta x$ represents, in our context, the distance between the representative points of neighboring cells $(i,j)$ and $(k,l)$.
In the case of the algorithm of \cite{miyamoto_sasaki_1997}, this distance is taken to be equal to $W$ between a cell C and a cell of
type B, and equal to $\sqrt{2} W$ between a cell C and a cell of type A. When using the algorithm inspired by \cite{vicari_etal_2007},
the distance $\Delta x$ must not be taken to be the distance between two randomly selected representative points. Indeed, these points
may be arbitrarily close to each other, leading to a distance $\Delta x \simeq 0$, which results in $h_{\rm c} \simeq 0$
and therefore an arbitrarily large mass flux between cells $\Delta V$ (see Eq.~\ref{DeltaV}). Computed in this way, the slope
$\partial h/\partial x$ of the lava surface may be very different from physical reality; the volume variations $\Delta V$ may exceed
the amount of lava initially available in the cell losing material; and ultimately, the occurrence of arbitrarily large values of $\Delta V$
means that a flow would never stop, regardless of the rheological properties of the fluid. To correct this problem, we have adopted,
in the case of the algorithm of \cite{vicari_etal_2007}, an average distance calculated using the normalized probabilities
$P_{\rm A} = 3/13$ and $P_{\rm B} = 10/13$ \citep[see][]{miyamoto_sasaki_1997}.
This average distance is therefore:
\begin{equation}
\frac{3\sqrt{2} + 10}{13} \, W \simeq 1.09559 \times W
\end{equation}
%

%
%

\bibliographystyle{elsarticle-harv} 
\def\sciam{Sci.
  Am.}\def\nature{Nature}\def\nat{Nature}\def\science{Science}\def\natastro{Nat.
  Astron.}\def\natgeo{Nat. Geosci.}\def\natcom{Nat.
  Commun.}\def\pnas{PNAS}\def\AnnderPhys{‎Ann. Phys.
  (Berl.)}\def\icarus{Icarus}\def\pss{Planet. Space
  Sci.}\def\psj{PSJ}\def\planss{Planet. Space Sci.}\def\ssr{Space Sci.
  Rev.}\def\solsr{Sol. Syst. Res.}\def\ApSSP{ApSSP}\def\psj{Planet. Sci.
  J.}\def\jqsrt{J. Quant. Spectrosc. Radiat. Transfer}\def\expastro{Exp.
  Astron.}\def\jcis{‎J. Colloid Interface
  Sci.}\def\aap{A\&A}\def\apj{ApJ}\def\apjl{ApJL}\def\apjs{ApJS}\def\aj{AJ}\def\mnras{MNRAS}\def\araa{Annu.
  Rev. Astron. Astrophys.}\def\areps{Annu. Rev. Earth Planet.
  Sci.}\def\pasj{Publ. Astron. Soc. Jpn.}\def\apss{Astrophys. Space
  Sci.}\def\expastro{Exp. Astron.}\def\pasp{Publ. Astron. Soc.
  Pac.}\def\expastron{Exp. Astron.}\def\asr{Adv. Space
  Res.}\def\galaxies{Galaxies}\def\astrobiol{Astrobiology}\def\areps{Annu. Rev.
  Earth Planet. Sci.}\def\georl{Geophys. Res. Lett.}\def\grl{Geophys. Res.
  Lett.}\def\jgr{J. Geophys. Res.}\def\jgrp{J. Geophys. Res.:
  Planets}\def\gca{Geochim. Cosmochim. Ac.}\def\epsl{Earth Planet. Sci.
  Lett.}\def\ess{Earth Space Sci.}\def\plasci{Planet. Sci.}\def\ggg{Geochem.
  Geophys. Geosyst.}\def\rmg{Rev. Mineral. Geochem.}\def\gji{Geophys. J.
  Int.}\def\tpm{Transport Porous Med.}\def\philtrans{Phil.
  Trans.}\def\faradis{Farad. Discuss.}\def\jcis{‎J. Colloid Interface
  Sci.}\def\jfm{J. Fluid Mech.}\def\physflu{Phys. Fluids}\def\pachem{Pure Appl.
  Chem.}\def\jpcA{J. Phys. Chem. A}\def\chemrev{Chem. Rev.}\def\AppOpt{Appl.
  Opt.}\def\sciam{Sci.
  Am.}\def\nature{Nature}\def\nat{Nature}\def\science{Science}\def\natastro{Nat.
  Astron.}\def\natgeo{Nat. Geosci.}\def\natcom{Nat.
  Commun.}\def\pnas{PNAS}\def\AnnderPhys{‎Ann. Phys.
  (Berl.)}\def\icarus{Icarus}\def\pss{Planet. Space Sci.}\def\planss{Planet.
  Space Sci.}\def\ssr{Space Sci. Rev.}\def\solsr{Sol. Syst.
  Res.}\def\expastro{Exp. Astron.}\def\jcis{‎J. Colloid Interface
  Sci.}\def\aap{A\&A}\def\apj{ApJ}\def\apjl{ApJL}\def\apjs{ApJS}\def\aj{AJ}\def\mnras{MNRAS}\def\araa{Annu.
  Rev. Astron. Astrophys.}\def\pasj{Publ. Astron. Soc.
  Jpn.}\def\apss{Astrophys. Space Sci.}\def\pasp{Publ. Astron. Soc.
  Pac.}\def\expastron{Exp. Astron.}\def\asr{Adv. Space
  Res.}\def\astrobiol{Astrobiology}\def\areps{Annu. Rev. Earth Planet.
  Sci.}\def\georl{Geophys. Res. Lett.}\def\jgr{J. Geophys. Res.}\def\jgrse{J.
  Geophys. Res. Solid Earth.}\def\jgrp{J. Geophys. Res.
  Planets}\def\gca{Geochim. Cosmochim. Ac.}\def\epsl{Earth Planet. Sci.
  Lett.}\def\plasci{Planet. Sci.}\def\ggg{Geochem. Geophys.
  Geosyst.}\def\rmg{Rev. Mineral. Geochem.}\def\tpm{Transport Porous
  Med.}\def\philtrans{Phil. Trans.}\def\faradis{Farad. Discuss.}\def\jcis{J.
  Colloid Interface Sci.}\def\jfm{J. Fluid Mech.}\def\physflu{Phys.
  Fluids}\def\pachem{Pure Appl. Chem.}\def\jpcA{J. Phys. Chem.
  A}\def\chemrev{Chem. Rev.}\def\sciam{Sci.
  Am.}\def\nature{Nature}\def\nat{Nature}\def\science{Science}\def\natastro{Nat.
  Astron.}\def\natgeo{Nat. Geosci.}\def\natcom{Nat.
  Commun.}\def\pnas{PNAS}\def\AnnderPhys{‎Ann. Phys.
  (Berl.)}\def\icarus{Icarus}\def\pss{Planet. Space
  Sci.}\def\psj{PSJ}\def\planss{Planet. Space Sci.}\def\ssr{Space Sci.
  Rev.}\def\solsr{Sol. Syst. Res.}\def\ApSSP{ApSSP}\def\psj{Planet. Sci.
  J.}\def\jgla{J. Glaciol.}\def\jqsrt{J. Quant. Spectrosc. Radiat.
  Transfer}\def\expastro{Exp. Astron.}\def\jcis{‎J. Colloid Interface
  Sci.}\def\aap{A\&A}\def\apj{ApJ}\def\apjl{ApJL}\def\apjs{ApJS}\def\aj{AJ}\def\mnras{MNRAS}\def\araa{Annu.
  Rev. Astron. Astrophys.}\def\pasj{Publ. Astron. Soc.
  Jpn.}\def\apss{Astrophys. Space Sci.}\def\pasp{Publ. Astron. Soc.
  Pac.}\def\expastron{Exp. Astron.}\def\asr{Adv. Space
  Res.}\def\galaxies{Galaxies}\def\astrobiol{Astrobiology}\def\areps{Annu. Rev.
  Earth Planet. Sci.}\def\georl{Geophys. Res. Lett.}\def\grl{Geophys. Res.
  Lett.}\def\jgr{J. Geophys. Res.}\def\jgrp{J. Geophys. Res.
  Planets}\def\gca{Geochim. Cosmochim. Ac.}\def\epsl{Earth Planet. Sci.
  Lett.}\def\ess{Earth Space Sci.}\def\plasci{Planet. Sci.}\def\ggg{Geochem.
  Geophys. Geosyst.}\def\rmg{Rev. Mineral. Geochem.}\def\gji{Geophys. J.
  Int.}\def\tpm{Transport Porous Med.}\def\philtrans{Phil.
  Trans.}\def\faradis{Farad. Discuss.}\def\jcis{‎J. Colloid Interface
  Sci.}\def\jfm{J. Fluid Mech.}\def\physflu{Phys. Fluids}\def\pachem{Pure Appl.
  Chem.}\def\jpcA{J. Phys. Chem. A}\def\chemrev{Chem. Rev.}\def\AppOpt{Appl.
  Opt.}\def\pmpes{Proc. Math. Phys. Eng. Sci.}\def\cpc{Comput. Phys.
  Commun.}\def\araa{Annu. Rev. Astron. Astrophys.}\def\repprogphys{Rep. Prog.
  Phys.}\def\jcp{J. Comput. Phys.
  }\def\nature{Nature}\def\nat{Nature}\def\science{Science}\def\natastro{Nat.
  Astron.}\def\natgeo{Nat. Geosci.}\def\natcom{Nat. Commun.}\def\scirep{Sci.
  Rep.}\def\sciad{Sci.
  Adv.}\def\aap{A\&A}\def\apj{ApJ}\def\apjl{ApJL}\def\apjs{ApJS}\def\aj{AJ}\def\mnras{MNRAS}\def\araa{Annu.
  Rev. Astron. Astrophys.}\def\pasj{Publ. Astron. Soc.
  Jpn.}\def\apss{Astrophys. Space Sci.}\def\pasp{Publ. Astron. Soc.
  Pac.}\def\expastron{Exp. Astron.}\def\asr{Adv. Space
  Res.}\def\galaxies{Galaxies}\def\science{Sci}\def\jced{J. Chem. Eng.
  Data}\def\fpe{Fluid Phase Equilibria}\def\iecr{Ind. Eng. Chem.
  Res.}\def\aichej{AIChE J.}\def\pt{Powder Technol.}\def\etfs{Exp. Therm. Fluid
  Sci.}\def\jgr{J. Geophys. Res.}\def\gca{Geochim. Cosmochim.
  Acta}\def\chemgeol{Chem Geol.}\def\jcp{J. Chem. Phys.}\def\jcis{‎J. Colloid
  Interface Sci.}\def\jcsft{J. Chem. Soc. Faraday Trans.}\def\jpcB{J. Phys.
  Chem. B}\def\jsf{J. Supercrit. Fluids}\def\enerp{Energy
  Procedia}\def\aichej{AlChE J.}\def\IECPDD{Ind. Eng. Chem. Process Des.
  Dev.}\def\EF{Energy Fuels}\def\jacs{J. Am. Chem. Soc.}\def\sciam{Sci.
  Am.}\def\nature{Nature}\def\nat{Nature}\def\science{Science}\def\natastro{Nat.
  Astron.}\def\natgeo{Nat. Geosci.}\def\natcom{Nat.
  Commun.}\def\pnas{PNAS}\def\AnnderPhys{‎Ann. Phys.
  (Berl.)}\def\icarus{Icarus}\def\pss{Planet. Space
  Sci.}\def\psj{PSJ}\def\planss{Planet. Space Sci.}\def\ssr{Space Sci.
  Rev.}\def\solsr{Sol. Syst. Res.}\def\ApSSP{ApSSP}\def\psj{Planet. Sci.
  J.}\def\jqsrt{J. Quant. Spectrosc. Radiat. Transfer}\def\expastro{Exp.
  Astron.}\def\jcis{‎J. Colloid Interface
  Sci.}\def\aap{A\&A}\def\apj{ApJ}\def\apjl{ApJL}\def\apjs{ApJS}\def\aj{AJ}\def\mnras{MNRAS}\def\araa{Annu.
  Rev. Astron. Astrophys.}\def\pasj{Publ. Astron. Soc.
  Jpn.}\def\apss{Astrophys. Space Sci.}\def\pasp{Publ. Astron. Soc.
  Pac.}\def\expastron{Exp. Astron.}\def\asr{Adv. Space
  Res.}\def\galaxies{Galaxies}\def\astrobiol{Astrobiology}\def\areps{Annu. Rev.
  Earth Planet. Sci.}\def\georl{Geophys. Res. Lett.}\def\grl{Geophys. Res.
  Lett.}\def\jgr{J. Geophys. Res.}\def\gca{Geochim. Cosmochim.
  Ac.}\def\epsl{Earth Planet. Sci. Lett.}\def\ess{Earth Space
  Sci.}\def\plasci{Planet. Sci.}\def\ggg{Geochem. Geophys.
  Geosyst.}\def\rmg{Rev. Mineral. Geochem.}\def\gji{Geophys. J.
  Int.}\def\tpm{Transport Porous Med.}\def\philtrans{Phil.
  Trans.}\def\faradis{Farad. Discuss.}\def\jcis{‎J. Colloid Interface
  Sci.}\def\jfm{J. Fluid Mech.}\def\physflu{Phys. Fluids}\def\pachem{Pure Appl.
  Chem.}\def\jpcA{J. Phys. Chem. A}\def\chemrev{Chem. Rev.}\def\AppOpt{Appl.
  Opt.}\def\nature{Nature}\def\nat{Nature}\def\science{Science}\def\natastro{Nat.
  Astron.}\def\natgeo{Nat. Geosci.}\def\natcom{Nat. Commun.}\def\scirep{Sci.
  Rep.}\def\sciad{Sci.
  Adv.}\def\aap{A\&A}\def\apj{ApJ}\def\apjl{ApJL}\def\apjs{ApJS}\def\aj{AJ}\def\mnras{MNRAS}\def\araa{Annu.
  Rev. Astron. Astrophys.}\def\pasj{Publ. Astron. Soc.
  Jpn.}\def\apss{Astrophys. Space Sci.}\def\pasp{Publ. Astron. Soc.
  Pac.}\def\expastron{Exp. Astron.}\def\asr{Adv. Space
  Res.}\def\galaxies{Galaxies}\def\science{Sci}\def\jced{J. Chem. Eng.
  Data}\def\fpe{Fluid Phase Equilibria}\def\iecr{Ind. Eng. Chem.
  Res.}\def\aichej{AIChE J.}\def\pt{Powder Technol.}\def\etfs{Exp. Therm. Fluid
  Sci.}\def\jgr{J. Geophys. Res.}\def\gca{Geochim. Cosmochim.
  Acta}\def\chemgeol{Chem Geol.}\def\jcp{J. Chem. Phys.}\def\jcis{‎J. Colloid
  Interface Sci.}\def\jcsft{J. Chem. Soc. Faraday Trans.}\def\jpcB{J. Phys.
  Chem. B}\def\jsf{J. Supercrit. Fluids}\def\enerp{Energy
  Procedia}\def\aichej{AlChE J.}\def\IECPDD{Ind. Eng. Chem. Process Des.
  Dev.}\def\EF{Energy Fuels}\def\jacs{J. Am. Chem. Soc.}\def\sciam{Sci.
  Am.}\def\nature{Nature}\def\nat{Nature}\def\science{Science}\def\natastro{Nat.
  Astron.}\def\natgeo{Nat. Geosci.}\def\natcom{Nat.
  Commun.}\def\pnas{PNAS}\def\AnnderPhys{‎Ann. Phys.
  (Berl.)}\def\icarus{Icarus}\def\pss{Planet. Space
  Sci.}\def\psj{PSJ}\def\planss{Planet. Space Sci.}\def\ssr{Space Sci.
  Rev.}\def\solsr{Sol. Syst. Res.}\def\ApSSP{ApSSP}\def\psj{Planet. Sci.
  J.}\def\jqsrt{J. Quant. Spectrosc. Radiat. Transfer}\def\expastro{Exp.
  Astron.}\def\jcis{‎J. Colloid Interface
  Sci.}\def\aap{A\&A}\def\apj{ApJ}\def\apjl{ApJL}\def\apjs{ApJS}\def\aj{AJ}\def\mnras{MNRAS}\def\araa{Annu.
  Rev. Astron. Astrophys.}\def\pasj{Publ. Astron. Soc.
  Jpn.}\def\apss{Astrophys. Space Sci.}\def\pasp{Publ. Astron. Soc.
  Pac.}\def\expastron{Exp. Astron.}\def\asr{Adv. Space
  Res.}\def\galaxies{Galaxies}\def\astrobiol{Astrobiology}\def\areps{Annu. Rev.
  Earth Planet. Sci.}\def\georl{Geophys. Res. Lett.}\def\grl{Geophys. Res.
  Lett.}\def\jgr{J. Geophys. Res.}\def\gca{Geochim. Cosmochim.
  Ac.}\def\epsl{Earth Planet. Sci. Lett.}\def\ess{Earth Space
  Sci.}\def\plasci{Planet. Sci.}\def\ggg{Geochem. Geophys.
  Geosyst.}\def\rmg{Rev. Mineral. Geochem.}\def\gji{Geophys. J.
  Int.}\def\tpm{Transport Porous Med.}\def\philtrans{Phil.
  Trans.}\def\faradis{Farad. Discuss.}\def\jcis{‎J. Colloid Interface
  Sci.}\def\jfm{J. Fluid Mech.}\def\physflu{Phys. Fluids}\def\pachem{Pure Appl.
  Chem.}\def\jpcA{J. Phys. Chem. A}\def\chemrev{Chem. Rev.}\def\AppOpt{Appl.
  Opt.}\def\sciam{Sci.
  Am.}\def\nature{Nature}\def\nat{Nature}\def\science{Science}\def\natastro{Nat.
  Astron.}\def\natgeo{Nat. Geosci.}\def\natcom{Nat. Commun.}\def\sciad{Sci.
  Adv.}\def\scirep{Sci. Rep.}\def\AnnderPhys{‎Ann. Phys.
  (Berl.)}\def\icarus{Icarus}\def\pss{Planet. Space Sci.}\def\planss{Planet.
  Space Sci.}\def\ssr{Space Sci. Rev.}\def\solsr{Sol. Syst.
  Res.}\def\expastro{Exp. Astron.}\def\jcis{‎J. Colloid Interface
  Sci.}\def\aap{A\&A}\def\apj{ApJ}\def\apjl{ApJL}\def\apjs{ApJS}\def\aj{AJ}\def\mnras{MNRAS}\def\aapr{A\&ARv}\def\araa{Annu.
  Rev. Astron. Astrophys.}\def\pasj{Publ. Astron. Soc.
  Jpn.}\def\apss{Astrophys. Space Sci.}\def\pasp{Publ. Astron. Soc.
  Pac.}\def\expastron{Exp. Astron.}\def\astrobiol{Astrobiology}\def\areps{Annu.
  Rev. Earth Planet. Sci.}\def\georl{Geophys. Res. Lett.}\def\jgr{J. Geophys.
  Res.}\def\gca{Geochim. Cosmochim. Ac.}\def\epsl{Earth Planet. Sci.
  Lett.}\def\plasci{Planet. Sci.}\def\ggg{Geochem. Geophys.
  Geosyst.}\def\tpm{Transport Porous Med.}\def\philtrans{Phil.
  Trans.}\def\faradis{Farad. Discuss.}\def\jcis{‎J. Colloid Interface
  Sci.}\def\jfm{J. Fluid Mech.}\def\physflu{Phys.
  Fluids}\def\pnas{PNAS}\def\pachem{Pure Appl. Chem.}\def\jpcA{J. Phys. Chem.
  A}\def\chemrev{Chem.
  Rev.}\def\nature{Nature}\def\nat{Nature}\def\science{Science}\def\natastro{Nat.
  Astron.}\def\natgeo{Nat. Geosci.}\def\natcom{Nat. Commun.}\def\scirep{Sci.
  Rep.}\def\science{Sci}\def\jced{J. Chem. Eng. Data}\def\fpe{Fluid Phase
  Equilibria}\def\iecr{Ind. Eng. Chem. Res.}\def\aichej{AIChE J.}\def\pt{Powder
  Technol.}\def\etfs{Exp. Therm. Fluid Sci.}\def\gca{Geochim. Cosmochim.
  Acta}\def\chemgeol{Chem Geol.}\def\jcp{J. Chem. Phys.}\def\jcis{‎J. Colloid
  Interface Sci.}\def\jcsft{J. Chem. Soc. Faraday Trans.}\def\jpcB{J. Phys.
  Chem. B}\def\jsf{J. Supercrit. Fluids}\def\enerp{Energy
  Procedia}\def\aichej{AlChE J.}\def\IECPDD{Ind. Eng. Chem. Process Des.
  Dev.}\def\EF{Energy Fuels}\def\jacs{J. Am. Chem. Soc.}\def\jqsrt{J. Quant.
  Spectrosc. Radiat. Transfer}\def\astrobiol{Astrobiology}\def\areps{Annu. Rev.
  Earth Planet. Sci.}\def\georl{Geophys. Res. Lett.}\def\grl{Geophys. Res.
  Lett.}\def\jgr{J. Geophys. Res.}\def\gca{Geochim. Cosmochim.
  Ac.}\def\epsl{Earth Planet. Sci. Lett.}\def\ess{Earth Space
  Sci.}\def\plasci{Planet. Sci.}\def\ggg{Geochem. Geophys.
  Geosyst.}\def\rmg{Rev. Mineral. Geochem.}\def\gji{Geophys. J.
  Int.}\def\ssr{Space Sci. Rev.}\def\jas{J. Atmos. Sci.}


\end{document}